\documentclass[12pt,a4paper]{article} 
\usepackage{style}
\usepackage{amsmath}
\usepackage{amssymb}
\usepackage{amsthm}
\usepackage{psfrag}
\usepackage{graphicx}
\usepackage{hyperref}
\usepackage{feynmp}
\usepackage{color}
\usepackage{bm}
\usepackage{booktabs}

\newcommand{\be}{\begin{equation}}
\newcommand{\ee}{\end{equation}}
\newcommand{\beqa}{\begin{eqnarray}}
\newcommand{\eeqa}{\end{eqnarray}}
\newcommand{\bsm}{\begin{smallmatrix}}
\newcommand{\esm}{\end{smallmatrix}}

\newcommand\G{\mathcal{G}}

\newcommand{\hMpc}{h\text{Mpc}^{-1}}

\newcommand{\bseq}{\begin{subequations}}
\newcommand{\eseq}{\end{subequations}}

\renewcommand{\ln}{\mathop{\rm ln}\nolimits}

\newcommand\vk{\varkappa}

\newcommand{\kmax}{k_{\rm max}}

\newcommand{\true}{{\rm true}}

\newcommand{\astfootnote}[1]{%
\let\oldthefootnote=\thefootnote%
\setcounter{footnote}{0}%
\renewcommand{\thefootnote}{\fnsymbol{footnote}}%
\footnote{#1}%
\let\thefootnote=\oldthefootnote%
}

\usepackage{color}
\usepackage[caption=false]{subfig}

\makeatletter
\def\@makecaption#1#2{%
  \vskip\abovecaptionskip
  \sbox\@tempboxa{{\footnotesize #1: #2}}
  \ifdim \wd\@tempboxa >\hsize
    {\footnotesize #1: #2\par}
  \else
    \global \@borderpane\true
    \hbox to\hsize{\hfil\box\@tempboxa\hfil}%
  \fi
  \vskip\belowcaptionskip}
\makeatother

\renewcommand{\vec}[1]{\boldsymbol{#1}}

\renewcommand{\hMpc}{h\,\mathrm{Mpc}^{-1}}
\newcommand{\delD}[1]{(2\pi)^3\delta_\mathrm{D}\left({#1}\right)}
 
\renewcommand{\vk}{\vec k}

\newcommand{\vq}{{\vec q}}
\newcommand{\vx}{\vec x}

\newcommand{\vn}{\vec n}

\newcommand{\hn}{\hat{\vec n}}
\newcommand{\hq}{\hat{\vec q}}

\usepackage[dvipsnames]{xcolor}

\def\cG{\mathcal{G}}
\newcommand{\kd}{k_{12}}

\usepackage{color}
\def\beq{\begin{eqnarray}}
\def\eeq{\end{eqnarray}}

\relax

\title{Full-Shape Analysis of the Redshift-Space Galaxy Trispectrum in the EFTofLSS}

\author[a,b]{Beatriz Tucci\footnote{\href{mailto:beatucci@stanford.edu}{beatucci@stanford.edu}}}
\author[a,b]{and Oliver H.\,E. Philcox\footnote{\href{mailto:beatucci@}{ohep2@cantab.ac.uk}}}

\affiliation[a]{Leinweber Institute for Theoretical Physics at Stanford, 382 Via Pueblo, \\Stanford, CA 94305, USA}
\affiliation[b]{Kavli Institute for Particle Astrophysics and Cosmology, 382 Via Pueblo, \\Stanford, CA 94305, USA}

\abstract{\small
The effective field theory of large-scale structure (EFTofLSS) program for spectroscopic galaxy surveys has been remarkably successful for cosmological inference from the galaxy power spectrum and bispectrum. However, a full-shape cosmological analysis including the galaxy trispectrum has not yet been performed. In this work, we develop the tree-level EFTofLSS model of the galaxy trispectrum, including all relevant ingredients: galaxy bias, stochasticity, infrared resummation, Fingers-of-God (FoG) counterterms, Alcock-Paczynski distortions, binning effects, and covariance. We validate the model by performing cosmological inference using the one-loop galaxy power spectrum, tree-level bispectrum, and tree-level trispectrum measured from dark-matter halos in 25 \textsc{AbacusSummit} $N$-body simulations, whose large effective volume ($V=200h^{-3}\text{Gpc}^3$) allows us to test the precision of our model and determine scale cuts in both real-space and redshift-space. We also study the cosmological information content of the galaxy trispectrum in a DESI-like volume, finding that its inclusion can improve constraints e.g., on $\omega_{\rm cdm}$ by up to $15\%$. We further find that the maximum scale cut can be extended to $k_{\max}=0.12h\mathrm{Mpc}^{-1}$ depending on the tracer,  thanks to our improved treatment of FoG effects. This provides a path towards using the three-dimensional large-scale structure to constrain new-physics scenarios with distinctive signatures in the trispectrum, in particular primordial non-Gaussianity.
}

\normalsize

\begin{document}

\maketitle

\section{Introduction}
Cosmological information in large-scale structure is distributed across a hierarchy of $n$-point correlation functions, each sensitive to a different level of complexity in the galaxy density field \citep[e.g.,][]{1980lssu.book.....P,Bernardeau:2001qr}. At leading order, this information is encoded in the power spectrum, while the bispectrum captures the first departure from Gaussianity. The connected four-point function, or trispectrum, is the next step in this hierarchy, opening a window into interactions that cannot be fully reconstructed from lower-order statistics alone, including nonlinear mode coupling generated by gravitational evolution and primordial non-Gaussianity (PNG), namely departures from Gaussian initial conditions imprinted by early Universe physics. These two sources of non-Gaussian structure are mixed in the late-time galaxy field and must be carefully disentangled to isolate primordial physics. In this sense, the trispectrum is not merely a noisier extension of the three-point function, but a genuinely new observable, sensitive to aspects of both the early Universe and late-time structure formation that are otherwise inaccessible.

Many inflationary models produce a non-vanishing four-point function. Local primordial non-Gaussianity contains two independent trispectrum amplitudes, conventionally denoted $g_{\rm NL}$ and $\tau_{\rm NL}$. The former describes a cubic local interaction (e.g., a transformation of the form $\zeta\to \zeta+g_{\rm NL}\zeta^3$ for curvature $\zeta$), while the latter controls a collapsed contribution that arises commonly when inflation contains multiple fields. Their effects on halo clustering through scale-dependent and stochastic bias were studied by \citep{Smith:2011gnl,Baumann:2012stochastic,Ferraro:2014multifield}, and the resulting scale-dependent bias has also been used to place observational constraints on $g_{\rm NL}$ \citep{Desjacques:2009jb,Xia_2011,Leistedt:2014zqa}. Beyond these local templates, inflationary self-interactions and the exchange of massive particles generate characteristic contact and exchange trispectra \citep[e.g.,][]{Arkani-Hamed:2015bza,Arkani-Hamed:2018kmz,Senatore:2010jy,Chen:2009bc,Baumann:2017jvh,Liu:2019fag,2015arXiv150200635S}. The latter are a central target of cosmological collider models, whose momentum and angular dependence can encode the masses and spins of particles present during inflation \citep{Arkani-Hamed:2015bza,Lee:2016vti,Chen:2009zp}. A long-term goal of large-scale structure analyses is therefore to measure $g_{\rm NL}$, $\tau_{\rm NL}$, and more general cosmological collider signals directly from the three-dimensional tracer trispectrum.

To date, almost all searches for the primordial trispectrum have used the cosmic microwave background (CMB). A recent series of papers \citep{Philcox:2025cmbtrispectrum,Philcox:2025cmbcode,Philcox:2025cmbplanck} developed separable primordial templates and optimal estimators and applied them to \textit{Planck} PR4 temperature and polarization data (building on a number of previous works \citep[e.g.,][]{2015arXiv150200635S,Fergusson:2010gn,Marzouk:2022utf}). They constrained a broad set of local, effective-field-theory, direction-dependent, and particle-exchange templates, including several cosmological collider signals. These works demonstrate the sensitivity of cosmological four-point analysis to inflationary physics. 

Large-scale structure (LSS) provides a complementary three-dimensional probe of primordial non-Gaussianity, allowing us to access many more modes than with the CMB \citep[e.g.,][]{Sailer:2021yzm,Kalaja:2020mkq}. As current and upcoming galaxy surveys extend over increasingly large volumes, it becomes possible to use the trispectrum to test primordial physics with higher precision. The main challenge is that the primordial contribution must be disentangled from signals generated by nonlinear gravitational evolution, tracer bias, stochasticity, and redshift-space distortions (see \citep[e.g.,][]{Cabass:2022epm,MoradinezhadDizgah:2020whw,Baumann:2021ykm} for the bispectrum). A reliable analysis therefore requires accurate modeling of these late-time effects and a clear understanding of how they may be distinguished from the primordial signal. In practice, the high dimensionality of the trispectrum introduces further challenges, since the theoretical model must be consistently projected and averaged over finite bins, account for Alcock-Paczynski (AP) distortions, and be combined with an accurate covariance model. It is also necessary to determine the range of scales over which the perturbative description remains reliable and does not bias the inferred cosmological parameters.

The theoretical modeling of the late-time trispectrum has progressed substantially in recent years. \citep{Bertolini:2016bmt} computed the one-loop matter trispectrum in Standard Perturbation Theory (SPT) and within the effective field theory of large-scale structure (EFTofLSS), and \citep{Steele:2021lnz} later calibrated the associated EFT counterterms against $N$-body simulations. More recently, \citep{Bakx:2026galaxybias} derived the complete one-loop trispectrum of biased tracers in real space as part of a fully renormalized treatment of galaxy bias (building on \citep{Bakx:2025cvu}), including stochasticity and all relevant higher-derivative terms. The redshift-space galaxy trispectrum has also been studied at tree level. In particular, \citep{Leonard:2024xvy} derived the corresponding galaxy four-point correlation function in SPT, including redshift-space distortions and deterministic galaxy bias through third order. However, their calculation does not include ingredients such as AP distortions and an EFT treatment of biased tracers.

Other works have focused on particular limits of the trispectrum. \citep{Giri:2023modulation} combined the squeezed bispectrum and collapsed trispectrum through the large-scale modulation of small-scale power to constrain local primordial non-Gaussianity. Along similar lines, \citep{Goldstein_2024} developed nonperturbative models for cosmological-collider signals in the squeezed matter bispectrum and collapsed matter trispectrum, and tested them using $N$-body simulations. The collapsed trispectrum is also directly related to power-spectrum covariance. Recently, \citep{Goldstein:2026covariance} showed that squeezed-bispectrum and collapsed-trispectrum estimators can reconstruct the complete non-Gaussian matter power-spectrum covariance using far fewer simulations than a sample-covariance estimate.

The observational study of the large-scale structure four-point function has also advanced, particularly with the advent of new high-precision data. \citep{Gualdi:2020mattertrispectrum} introduced and measured the integrated matter trispectrum in real- and redshift-space simulations, and showed through forecasts that it can improve constraints on PNG. \citep{Gualdi:2021joint} then performed a joint MCMC analysis of the power spectrum, bispectrum, and integrated trispectrum of redshift-space matter, though this used modified tree-level SPT kernels for the bispectrum and trispectrum rather than a full EFT model.

In galaxy data, \citep{Philcox:2021hbm} presented the first detection of the connected configuration-space four-point function in the SDSS-BOSS CMASS sample, while \citep{Gualdi:2022kwz} detected the monopole and quadrupoles of the integrated galaxy trispectrum. More recently, \citep{Hou:2025desi4pcf} measured the connected four-point function of DESI Data Release 1 luminous red galaxies, finding a $14.7\sigma$ detection of the gravitationally induced signal. These measurements establish that the galaxy four-point function is observable at high significance, but they have not yet been used for full-shape cosmological inference. 

A particularly distinctive application of the trispectrum is the search for parity violation. For a scalar density field, the four-point function is the lowest-order correlator that can carry a handedness, since four points form a tetrahedron that cannot in general be rotated onto its mirror image \citep{Cahn:2021ltp}. Using this idea, \citep{Hou:2022parityodd} and \citep{Philcox:2022parityodd} reported evidence for a (mildly) nonzero parity-odd four-point function in BOSS. This motivated more model-specific studies of parity violation during inflation \citep{Cabass:2022colliding,Bao:2025onc,Cabass:2022rhr}, as well as the construction of parity-violating $N$-body simulations \citep{Coulton_2024}. The interpretation of the BOSS signal has since become less clear. \citep{Philcox:2024mmz} found that its significance is strongly reduced when the covariance is estimated from a different full $N$-body simulation suite, while \citep{Krolewski:2024parity} showed that a mismatch in the parity-even eight-point function between the data and mocks can bias the inferred significance, finding no compelling evidence for parity violation. Searches in other datasets and with different statistics have so far not given a clear confirmation either. No signal was found in the SDSS Ly$\alpha$ forest \citep{Adari:2024vkf}, while the DESI DR1 galaxy analyses found an excess in some auto-correlation tests but no convincing signal in cross-correlations or after accounting for covariance uncertainties \citep{Slepian:2025kbb,Hou:2025cey}. A complementary analysis using parity-odd compressed trispectrum statistics in BOSS and DESI likewise found no evidence for parity violation \citep{Gao:2026zaz}. Regardless of the status of the original BOSS anomaly, these studies show that the galaxy trispectrum gives access to symmetry tests that are simply not available to the power spectrum or bispectrum.

Incorporating the full shape of the galaxy trispectrum into a cosmological analysis is far from straightforward, since its large dimensionality complicates both the theoretical modeling and the covariance estimation. Additionally, several late-time and observational effects contribute at the level of accuracy required for cosmological inference and must therefore be modeled consistently. In this work, we construct a complete tree-level EFTofLSS model for the galaxy trispectrum in both real- and redshift-space, carefully accounting for all effects needed to compare theory and data: galaxy bias, stochastic (shot-noise) contributions, infrared resummation of long-wavelength displacements, Fingers-of-God counterterms, nonlinear redshift-space distortions, AP distortions, angular projection, and binning effects. Importantly, this is built upon the same framework as the power spectrum and bispectrum, facilitating robust joint analyses. We additionally derive and validate the covariance structure required for a joint power-spectrum, bispectrum, and trispectrum likelihood. 

Our model is tested on halo catalogs from 25 \textsc{AbacusSummit} simulations, whose combined effective volume of $V_{\rm eff}=200h^{-3}\mathrm{Gpc}^3$ allows for high precision tests of the theory. Our analytical covariance is further validated using $15{,}000$ \textsc{Quijote} simulations. In the future, this pipeline will provide a way to perform full-shape galaxy power-spectrum, bispectrum and trispectrum analyses of spectroscopic surveys such as DESI~\cite{Aghamousa:2016zmz} and Euclid~\cite{Euclid:2024yrr}. Moreover, it will finally make it possible to constrain beyond-$\Lambda$CDM parameters such as $g_{\rm NL}$, $\tau_{\rm NL}$, and more general cosmological collider signals using the large-scale distribution of galaxies.

To our knowledge, this is the first full-shape cosmological analysis of the galaxy trispectrum using $N$-body simulations. Our work therefore allows us to study the cosmological information content of the galaxy trispectrum and determine whether its inclusion can further tighten the $\Lambda$CDM constraints obtained from the power spectrum and bispectrum alone (of particular importance given the evolving debate on the information content of the perturbative field \citep[e.g.,][]{Nguyen:2024yth,Spezzati_2025,Akitsu:2025boy,Stadler:2026ufu}). In contrast to the integrated analysis by \citep{Gualdi:2021joint}, we retain the five-dimensional dependence of the binned trispectrum monopole and use an EFTofLSS description of biased tracers, fitting the tree-level trispectrum monopole jointly with the one-loop power-spectrum multipoles and tree-level bispectrum monopole. We sample the cosmological parameters $\{h,\omega_{\rm cdm},A_s\}$ together with the leading-order galaxy bias parameters, while analytically marginalizing over the remaining parameters. This allows us to quantify both the independent information encoded in the trispectrum and the range of scales over which the tree-level redshift-space model remains reliable, setting the stage for a future survey analyses for both $\Lambda$CDM and PNG.

The remainder of this paper is organized as follows. In Sec.~\ref{sec: model}, we present the underlying galaxy trispectrum model, as well as its estimator, projection, binning, and covariance. Sec.\,\ref{sec: data} describes the simulations and likelihood analysis. Results in real and redshift space are given in Secs.~\ref{sec: results-real} and \ref{sec: results-rsd}, followed by implications for a DESI-like survey analyses in Sec.~\ref{sec: desi}. We conclude in Sec.~\ref{sec: discussion}. Additional details on perturbation-theory kernels, stochastic terms, FoG countertem model, discreteness weights, covariance derivation, detection significance and fisher forecasts are discussed in the appendices.

\begin{figure}
    \centering
    \includegraphics[width=0.5\linewidth]{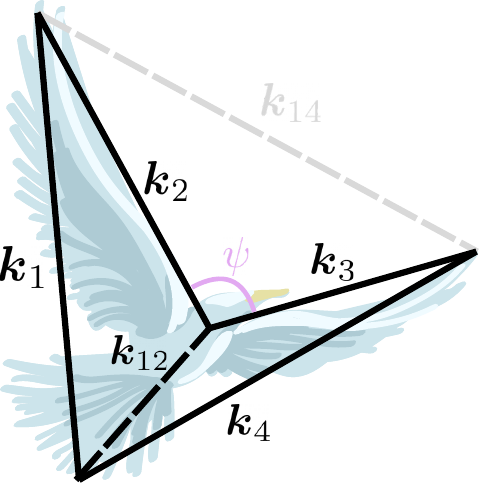}
    \caption{A flying bird provides an intuitive picture of the trispectrum geometry. The four external sides $k_1, k_2, k_3, k_4$ form a tetrahedron, with diagonals $k_{12}$ and $k_{14}$. The trispectrum estimator used in this work fixes a $\{k_1, k_2, k_3, k_4, k_{12}\}$ bin, and the wing flapping movement represents our integration over the diagonal $k_{14}$ (or equivalently the folding angle $\psi$).}
    \label{fig:bird}
\end{figure}

\section{The Galaxy Trispectrum}\label{sec: model}

The connected four-point correlation function describes configurations with one galaxy at each vertex of a tetrahedron, while its Fourier transform is the trispectrum. More precisely, the connected trispectrum of biased tracers, $T$, is defined as
\begin{equation}\label{eq: tri-def}
    \langle \delta_g(\vk_1)\delta_g(\vk_2)\delta_g(\vk_3) \delta_g(\vk_4)\rangle_c \equiv (2\pi)^3 \delta_D(\vk_1+\vk_2+\vk_3+\vk_4) \, T(\vk_1,\vk_2,\vk_3,\vk_4).
\end{equation}
In redshift space, after imposing momentum conservation, the trispectrum is a function of eight variables. We take the six lengths to be $\{k_1,k_2,k_3,k_4,k_{12},k_{14}\}$, where $k_{12}\equiv|\boldsymbol{k}_1+\boldsymbol{k}_2|=|\boldsymbol{k}_3+\boldsymbol{k}_4|$ and $k_{14}\equiv|\boldsymbol{k}_1+\boldsymbol{k}_4|=|\boldsymbol{k}_2+\boldsymbol{k}_3|$. The remaining two variables specify the orientation of the tetrahedron with respect to the line-of-sight (hereafter LoS). We parametrize it using the direction of the diagonal $\boldsymbol{k}_{12}$, defining $\mu\equiv\mu_{12}=\hat{\boldsymbol{k}}_{12}\cdot\hat{\boldsymbol{n}}$, together with an azimuthal angle $\varphi$ describing rotations about $\boldsymbol{k}_{12}$. For each external leg, we define $\mu_i\equiv\hat{\boldsymbol{k}}_i\cdot\hat{\boldsymbol{n}}$. In real space, rotational invariance removes the dependence on $\{\mu,\varphi\}$, leaving only the six lengths.

The integrated trispectrum studied by \citep{Gualdi:2020mattertrispectrum,Gualdi:2021joint} averages over both diagonal degrees of freedom. Here, following \citep{Goldstein_2024} and \citep{Coulton_2024}, we keep the dependence on $k_{12}$, while integrating over $k_{14}$, or equivalently over the folding angle $\psi$, which gives the angle between $\vk_1\times\vk_2$ and $\vk_3\times \vk_4$. This is necessary to obtain a factorizable estimator (Sec.\ \ref{subsec: estimator}). The bird in Fig.~\ref{fig:bird} provides a useful picture of this geometry. As the bird flies and flaps its wings, the two triangular faces rotate about the common diagonal $k_{12}$, sweeping through different values of $k_{14}$ (or $\psi$). Our estimator averages over this wing motion while preserving the dependence of the tetrahedron on $k_{12}$. 

In Sec.~\ref{subsec: theory}, we discuss the deterministic and stochastic contributions to the theoretical model of the galaxy trispectrum, together with other necessary ingredients such as Fingers-of-God counterterms and infrared resummation. We then present the trispectrum estimator in Sec.~\ref{subsec: estimator}. Finally, in Sec.~\ref{subsec: model-data}, we show how the theoretical model is mapped into the observable measured by the estimator, including diagonal averaging, angular integration, Alcock–Paczynski distortions, and binning effects.

\subsection{Theoretical Model}\label{subsec: theory}

To obtain the tree-level galaxy trispectrum model, we use Eulerian perturbation theory and the galaxy bias expansion \cite{Desjacques:2016bnm}. This expresses the real-space galaxy density field, $\delta_g$, as a sum over all allowed bias operators at each order in perturbation theory, themselves constructed from $\delta$ and $\theta$ (density and velocity fields). The galaxy bias expansion allows us to separate the trispectrum model as
\begin{equation}
    T(\vk_1,\vk_2,\vk_3,\vk_4) = T_{\rm det} + T_{\rm stoch},
\end{equation}
where $T_{\rm stoch}$ encodes stochasticity: large-scale fluctuations that do not correlate with the linear density field, also known as shot-noise \cite{Assassi:2014fva,Angulo:2015eqa,Senatore:2014eva,Mirbabayi:2014zca,Desjacques:2016bnm}. 

\paragraph{Deterministic Contributions.}
At third order in perturbation theory the deterministic galaxy density field in real-space is
\begin{equation}\label{eq: bias-expan}
\delta_{g,\rm det} =
b_1 \delta
+ \frac{b_2}{2} \delta^2
+ b_{\mathcal{G}_2} \mathcal{G}_2
+ b_{\Gamma_3} \Gamma_3
+ \frac{b_3}{6} \delta^3
+ b_{\mathcal{G}_3} \mathcal{G}_3
+ b_{\delta \mathcal{G}_2} \delta \mathcal{G}_2,
\end{equation}
using the Galiean basis defined in \citep{Assassi:2014fva} (matching \textsc{ClassPT} \citep{Chudaykin:2020aoj}). The Galileon operators are defined by
\beq
    \cG_2(\Phi_g) &\equiv& \nabla_i\nabla_j \Phi_g\nabla^i\nabla^j\Phi_g-(\nabla^2\Phi_g)^2 , \\\nonumber
    \cG_3(\Phi_g) &\equiv& -\frac{1}{2}[2\nabla_i\nabla_j\Phi_g\nabla^j\nabla_k\Phi_g\nabla^k\nabla^i\Phi_g-3\nabla_i\nabla_j\Phi_g\nabla^i\nabla^j\Phi_g\nabla^2\Phi_g+(\nabla^2\Phi_g)^3],
\eeq
where $\Phi_g\equiv \nabla^{-2}\delta$ is the scaled gravitational potential. At leading order in perturbations, this is equivalent to the velocity potential $\Phi_v \equiv \nabla^{-2}\delta$, with the differences encoded by $\Gamma_3 \equiv \G_2(\Phi_g)-\G_2(\Phi_v)$. We do not include the deterministic higher-derivative term $b_{\nabla^2\delta}\nabla^2\delta$ that appear at same perturbative order, since it does not contribute to the tree-level bispectrum or trispectrum and is degenerate with the leading-order power-spectrum deterministic counterterm. 

Expanding the operators in Eq.~\eqref{eq: bias-expan} perturbatively in the linear density field $\delta^{(1)}$, the $n$-th order contribution to the galaxy density contrast can be written as
\beq\label{eq: delta-n-exp}
    \delta^{(n)}_g(\vk) = \int_{\vq_1\ldots\vq_n}\delD{\vq_1+\ldots+\vq_n-\vk}K_n(\vq_1,\ldots,\vq_n)\delta^{(1)}(\vq_1)\ldots\delta^{(1)}(\vq_n),
\eeq
where the real-space kernels $K_n$ given in Appendix \ref{app:kernels-real} depend on the bias parameters above. 

To describe observations in redshift space, we further account for the mapping induced by LoS peculiar velocities. In the plane-parallel approximation, this mapping takes the form \citep[e.g.,][]{Kaiser:1987qv}
\beq\label{eq: rsd-mapping}
\delta_g^{(s)}(\vk)
=
\delta_g(\vk)
+
\int d\vx\,e^{-i\vk\cdot\vx}
\left[e^{-ik_z f u_z(\vx)}-1\right]
\left[1+\delta_g(\vx)\right],
\eeq
where $f$ denotes the logarithmic growth rate and
$u_z(\vq)=(i\mu_{\vq}/q)\theta(\vq)$ is the Fourier-space LoS velocity field, with
$\mu_{\vq}\equiv\hat{\vq}\cdot\hn$ and $k_z\equiv\vk\cdot\hn$ for LoS direction $\hn$. Expanding this expression perturbatively leads to a decomposition analogous to Eq.~\eqref{eq: delta-n-exp}, with the real-space kernels $K_n$ replaced by the redshift-space kernels $Z_n$, and the expressions up to third order are given in Appendix~\ref{app:kernels-rsd}.

The deterministic terms of the tree-level trispectrum can be decomposed as
\begin{equation}
    T_{\rm det}(\vk_1,\vk_2,\vk_3,\vk_4) = T_{1122} + T_{1113},
\end{equation}
with
\begin{equation}
\begin{aligned}
    \big\langle\delta_g^{(1)}(\vk_1)\delta^{(1)}_g(\vk_2)\delta^{(2)}_g(\vk_3)\delta^{(2)}_g(\vk_4)\big\rangle_c &\equiv \delD{\vk_1+\vk_2+\vk_3+\vk_4}T_{1122} 
    , \\
    \big\langle\delta_g^{(1)}(\vk_1)\delta_g^{(1)}(\vk_2)\delta^{(1)}_g(\vk_3)\delta^{(3)}_g(\vk_4) \big\rangle_c &\equiv \delD{\vk_1+\vk_2+\vk_3+\vk_4}T_{1113}
    .
\end{aligned}
\end{equation}
Using the redshift-space kernels from Appendix~\ref{app:kernels-rsd}, we obtain \citep{Gualdi:2018pyw,Sugiyama:2019ike,Leonard:2024xvy,Spezzati_2025}
\beq\label{eq: Tk-det}
    T_{1122}(\vk_1, \vk_2, \vk_3, \vk_4) &=& 4Z_1(\vk_1)Z_1(\vk_2) \big[
    Z_2(\vk_1, -\vk_{13}) Z_2(\vk_2, \vk_{13}) P(k_{13})
    \\\nonumber
    &&
    +Z_2(\vk_1, -\vk_{14}) Z_2(\vk_2, \vk_{14}) P(k_{14})\big]P(k_{1})P(k_{2}) + \text{5 perm.},
    \\\nonumber
    T_{1113}(\vk_1, \vk_2, \vk_3, \vk_4) &=& 6Z_1(\vk_1)Z_1(\vk_2)Z_1(\vk_3) Z_3(\vk_1, \vk_2, \vk_3) P(k_1)P(k_2)P(k_3) + \text{3 perm.},
\eeq
where $P\equiv P_L(k)$ denotes the linear power spectrum, with the modifications due to infrared resummation discussed below. The real-space model can be obtained simply by replacing the redshift-space kernels $Z_n$ to the real-space ones, $K_n$ (or simply setting $f=0$). The redshift-space trispectrum depends on seven bias parameters
$\{b_1,b_2,b_{\cG_2},b_{\Gamma_3},b_3,b_{\cG_3},b_{\delta\cG_2}\}$,
as well as on the logarithmic growth rate $f(z)$. 

For completeness, we also list here the galaxy power spectrum and bispectrum used in this work. The deterministic one-loop redshift-space galaxy power-spectrum is \cite{Ivanov:2019pdj,Nishimichi:2020tvu}
\be\label{eq:pk-model}
    P_{\rm det}(\vk) = P_g^{\rm tree}(\vk)
    + P_g^{\rm 1\text{-}loop}(\vk) + P_g^{\rm ctr}(\vk).
\ee
Although the one-loop power spectrum is also evaluated consistently to third order in the bias expansion, it depends only on the subset $\{b_1,b_2,b_{\cG_2},b_{\Gamma_3}\}$ due to renormalization \citep{Assassi:2014fva}. The counterterm contribution $P_g^{\rm ctr}$ is parametrized by the coefficients $\{c_0,c_2,c_4\}$ at leading derivative order, together with the higher-order FoG coefficient $\tilde c$. The deterministic tree-level redshift-space galaxy bispectrum is
\be\label{eq:bk-model}
    B_{\rm det}(\vk_1,\vk_2,\vk_3) = 2\,Z_1(\vk_1)\,Z_1(\vk_2)\,Z_2(\vk_1,\vk_2)\,
    P(k_1)P(k_2) + \text{2 perm.}.
\ee
In the below, we use the one-loop power spectrum together with the tree-level bispectrum and trispectrum. Although we restrict the bispectrum to tree level, complete one-loop models of the redshift-space galaxy bispectrum are available in the literature \citep[e.g.,][]{Angulo:2014tfa,Eggemeier:2021cam,Philcox:2022b1loop,DAmico:2022ukl,Bakx:2025pop} and we plan to incorporate it in future work.

\paragraph{Stochasticity.}
Galaxy formation involves short-scale fluctuations that are not captured by the deterministic bias expansion. Their large-scale effects are described by stochastic fields, often referred to as shot noise, which are uncorrelated with the long-wavelength deterministic operators in the low-$k$ limit. Up to third order, the real-space stochastic expansion is
\begin{equation}
\delta_{g,\text{stoch}}=
\epsilon+\epsilon_\delta\delta+\frac{1}{2}\epsilon_{\delta^2}\delta^2+\epsilon_{\mathcal{G}_2}\mathcal{G}_2,
\end{equation}
following \cite{Senatore:2014eva,Desjacques:2016bnm,Rubira:2024tea,Rubira:2025rqo,Bakx:2026galaxybias}. Here, $\epsilon$ is the leading stochastic field, while $\epsilon_\delta\delta$ describes the modulation of stochasticity by the local density. The fields $\epsilon_{\delta^2}$ and $\epsilon_{\mathcal{G}_2}$ similarly describe stochastic modulation by the quadratic density and tidal operators. The stochastic fields are specified by their correlation functions, whose amplitudes are independent parameters in the EFTofLSS. For Poisson sampling, these correlators reduce to the familiar shot-noise hierarchy, with amplitudes scaling as inverse powers of the tracer number density $\bar n$ \cite{Desjacques:2016bnm}.
After the redshift-space mapping given in Eq.~\eqref{eq: rsd-mapping}, the third-order galaxy stochastic field becomes
\begin{equation}
\begin{aligned}
\delta_{g,\text{stoch}}^{(s)}
&=\,
\epsilon+
[\epsilon_\delta \delta]+[\epsilon_\delta \delta^{(2)}]
-if k_z[\epsilon u_z^{(2)}]
\\
&+
\frac{1}{2}[\epsilon_{\delta^2}\delta^2]
+
[\epsilon_{\mathcal{G}_2}\mathcal{G}_2]
-if k_z[\epsilon_\delta\delta\,u_z]
-\frac{1}{2}f^2k_z^2[\epsilon u_z^2].
\end{aligned}
\end{equation}
A systematic derivation of these is presented in Appendix~\ref{app:stoch}, and leads to the following decomposition for the stochastic trispectrum:
\be
T_{\rm stoch}= T^{(\bar n^{-1})}_{\rm stoch} + T^{(\bar n^{-2})}_{\rm stoch} + T^{(\bar n^{-3})}_{\rm stoch}\,,
\ee
where 
\be \label{eq:stoch_pieces}
\begin{split}
& T^{(\bar n^{-1})}_{\rm stoch}=
\frac{Z_1(\mathbf{k}_1)Z_1(\mathbf{k}_2)P(k_1)P(k_2)}{\bar{n}}
    \Bigg[
    \frac{b_1B_{\rm shot}}{2}
    \Big(
    4F_2(\mathbf{k}_1,\mathbf{k}_2)
    +2f
    \Big(
    \mu_1^2+\mu_2^2
    +
    \Big[
    \frac{k_1}{k_2}
    +
    \frac{k_2}{k_1}
    \Big]\mu_1\mu_2
    \Big)
    \Big)
    \\
    &\quad\quad\quad+
    (1+P_{\rm shot})
    \Big(
    2f\mu_{12}^2G_2(\mathbf{k}_1,\mathbf{k}_2)
    +
    f^2
    \frac{k_{12}^2}{k_1k_2}
    \mu_1\mu_2\mu_{12}^2
    \Big)
    +
    b_2C_{\rm shot}
    +
    2b_{\mathcal G_2}D_{\rm shot}\,
    \kappa(\mathbf{k}_1,\mathbf{k}_2)
    \Bigg]
    \\
    &\quad\quad\quad+
    \frac{Z_1(\mathbf{k}_1)
    Z_2(\mathbf{k}_1,-\mathbf{k}_{12})P(k_1)P(k_{12})}{\bar{n}}
    \Big[
    2b_1B_{\rm shot}
    +
    2f\mu_{12}^2(1+P_{\rm shot})
    \Big]
    \\
    &\quad\quad\quad+
    \frac{Z_1(\mathbf{k}_2)
    Z_2(\mathbf{k}_2,-\mathbf{k}_{12})P(k_2)P(k_{12})}{\bar{n}}
    \Big[
    2b_1B_{\rm shot}
    +
    2f\mu_{12}^2(1+P_{\rm shot})
    \Big]
    \\
    &\quad\quad\quad+\text{5 perm.},\\
& T^{(\bar n^{-2})}_{\rm stoch}=\frac{Z_1(\mathbf{k}_1)P(k_1)}{\bar n^2}
    \Big[
    b_1E_{\rm shot}
    +
    f\mu_1^2A_{\rm shot}
    \Big]
    +\text{3 perm.}
    \\
    &\quad\quad\quad+
    \frac{P(k_{12})}{\bar n^2}
    \Big[
    b_1B_{\rm shot}
    +
    f(1+P_{\rm shot})\mu_{12}^2
    \Big]^2
    +(12\leftrightarrow13)+(12\leftrightarrow14),
    \\
& T^{(\bar n^{-3})}_{\rm stoch}=\frac{F_{\rm shot}}{\bar n^3},
\end{split}
\ee
where $P= P_L(k)$ (up to infrared resummation effects), and the permutations are over the external Fourier modes. For example, one of the five permutations in $T^{(\bar n^{-1})}_{\rm stoch}$ is $(\vk_1,\vk_2)\leftrightarrow(\vk_1,\vk_3)$, while the first permutation of $T^{(\bar n^{-2})}_{\rm stoch}$ sends $\vk_1\leftrightarrow \vk_2$. In $T^{(\bar n^{-2})}_{\rm stoch}$, the $(12\leftrightarrow13)$ and $(12\leftrightarrow14)$ exchanges correspond to a diagonal permutation, \textit{i.e.}\ $\vk_{12}\leftrightarrow\vk_{13}$ and $\vk_{12}\leftrightarrow\vk_{14}$, respectively. The real-space result, which is recovered by setting $f=0$, agrees with \cite{Bakx:2026galaxybias}, but contains one additional stochastic contribution compared to \cite{Spezzati_2025}.

For completeness, stochastic contribution to the galaxy power-spectrum is \cite{Nishimichi:2020tvu}
\be\label{eq:pk-stoch}
    P_{\rm stoch} = \frac{1}{\bar n}\left[(1+P_{\rm shot})
    + a_0\left(\frac{k}{k_{\rm NL}}\right)^2 + a_2\,\mu^2\left(\frac{k}{k_{\rm NL}}\right)^2\right],
\ee
with $k_{\rm NL}=0.45\,h\,{\rm Mpc}^{-1}$, while the bispectrum stochastic contribution is \cite{Ivanov:2021kcd}
\be\label{eq:bk-stoch}
\begin{aligned}
    B_{\rm stoch} &= \frac{b_1^2}{\bar n}
    \Big[B_{\rm shot}
    + \beta\mu_i^2\big[B_{\rm shot}+(1+P_{\rm shot})\big]
    + \beta^2\mu_i^4\,(1+P_{\rm shot})\Big]P(k_1) + \text{2 perm.} \\
    &+ \frac{A_{\rm shot}}{\bar n^2},
\end{aligned}
\ee
with $\beta\equiv f/b_1$, and the Poisson term $(1+P_{\rm shot})$ appears because the data are not shot-subtracted. Therefore, the trispectrum stochasticity in Eq.~\eqref{eq:stoch_pieces} contains the same $P_{\rm shot}$ parameter that enters the power-spectrum and the same $\{A_{\rm shot},B_{\rm shot}\}$ parameters that appear in the bispectrum, together with four additional shot noise amplitudes, $\{C_{\rm shot},D_{\rm shot},E_{\rm shot},F_{\rm shot}\}$. In the Poisson limit, $A_{\rm shot}=B_{\rm shot}=C_{\rm shot}=D_{\rm shot}=E_{\rm shot}=F_{\rm shot}=1$ and $P_{\rm shot}=a_0=a_2=0$. As elsewhere in this work, the inverse power of $\bar n$ factors are absorbed into the analytically marginalized templates.

\paragraph{FoG Counterterms.}\label{sec: fog}
The deterministic bispectrum and trispectrum models used in this work are evaluated at tree level. Strictly speaking, a tree-level calculation does not require ultraviolet counterterms, since there are no loop integrals whose small-scale sensitivity must be renormalized. In redshift space, however, small-scale velocities lead to the nonperturbative Fingers-of-God (FoG) effect \citep{Kaiser:1987qv,Jackson:2008yv}. In the EFT description, the leading FoG corrections are one-loop redshift-space counterterms and enter as a derivative expansion in powers of the LoS momentum~\citep{Perko:2016puo,Ivanov:2021kcd}. Following conventional tree-level bispectrum analyses \citep{Ivanov:2021kcd,Philcox:2021kcw}, we include leading terms of this expansion to partially resum the effects of small-scale velocities, which become nonperturbative on relatively large scales. They should therefore be regarded as a phenomenological extension of the tree-level prediction, rather than as a complete one-loop renormalization of the bispectrum or trispectrum.

In the tree-level redshift-space bispectrum prescription of \citep{Ivanov:2021kcd}, the FoG counterterms are implemented by means of a shift in the linear redshift-space kernel,
\be
\label{eq:Z1fog}
Z_1(\vk)
\to
Z_1^{\rm FoG}(\vk)
=b_1+f\mu^2-
c_1\mu^2
\left(\frac{k}{k^r_{\rm NL}}\right)^2 .
\ee
For our fiducial treatment, we instead apply the FoG correction coherently to the full deterministic bispectrum and trispectrum. Multiplicative damping factors of this kind have long been used as phenomenological models of small-scale velocities in redshift-space higher-order statistics, with both Lorentzian-like and Gaussian forms considered in the literature \citep{Verde:1998zr,Scoccimarro:1999ed,Gil-Marin:2014sta,Hashimoto:2017klo,Yankelevich:2018uaz}. Rather than using the full damping function, we keep only its leading contribution in the derivative expansion,
\be
\begin{aligned}
\label{eq:cbct}
B_{\rm det}(\vk_1,\vk_2,\vk_3)&\to B_{\rm det}
\left[1-c_B\sum_{i=1}^{3}\frac{k_i^2\mu_i^2}{(k^r_{\rm NL})^2}\right],
\\
T_{\rm det}(\vk_1,\vk_2,\vk_3,\vk_4)&\to T_{\rm det}
\left[1-c_T\sum_{i=1}^{4}\frac{k_i^2\mu_i^2}{(k^r_{\rm NL})^2}\right],
\end{aligned}
\ee
where we introduced the new FoG counterterms $c_B$ and $c_T$, which will be used throughout this work, and $k^r_{\rm NL}$ is the redshift-space velocity scale (which we fix to $k^r_{\rm NL}=0.3\hMpc$). This structure can be understood as the large-scale limit of the commonly used Gaussian FoG ansatz \citep{Yankelevich:2018uaz,Ivanov:2021kcd},
\be
\label{eq:gaussian_fog}
\exp\left[
-\frac{1}{2}\sigma_v^2
\sum_i k_i^2\mu_i^2
\right],
\ee
expanded to leading order in $k^2\mu^2$. Here, $\sigma_v$ is an effective one-dimensional dispersion of the line-of-sight redshift-space displacement and therefore has units of length, parametrizing the impact of small-scale peculiar velocities.

If the same effective velocity dispersion described both statistics, the coefficients would satisfy
\be
c_B\simeq c_T\simeq
\frac{1}{2}\sigma_v^2(k^r_{\rm NL})^2 .
\ee
We do not impose this relation in our analysis. The Gaussian damping model is only used to motivate the leading momentum dependence of the correction, while $c_B$ and $c_T$ are treated as effective coefficients that can also absorb higher-order contributions missing from the tree-level model. Since the bispectrum and trispectrum receive different loop corrections and, in general, have different counterterm structures, there is no reason for these effective coefficients to be equal. We therefore vary $c_B$ and $c_T$ independently.

Eq.~\eqref{eq:cbct} captures the leading $k^2\mu^2$ dependence expected from FoG corrections while treating all external legs symmetrically. We also find that it describes the simulations better than applying the counterterm only through the external $Z_1$ kernels. We adopt Eq.~\eqref{eq:cbct} as our fiducial treatment and we give an intuitive explanation on the difference between these two approaches and a comparison of their performance for cosmological inference in Appendix~\ref{app:fog}.

\paragraph{Infrared Resummation.}
Large-scale bulk flows displace matter coherently, broadening the baryon acoustic oscillation feature \citep[e.g.,][]{Baldauf:2015xfa}. Since these displacements are not small compared to the width of the acoustic peak, their effect must be resummed rather than treated at fixed order in perturbation theory. We follow the infrared-resummation prescription derived in the time-sliced perturbation theory framework \citep{Blas:2015qsi,Blas:2016sfa,Ivanov:2018gjr}. At the order considered here, this amounts to replacing each occurrence of the linear power spectrum in the galaxy bispectrum and trispectrum theory by
\beq
P=P_L(k)\rightarrow P_{\rm nw}(k) + \exp\left[-k^2\Sigma_{\rm tot}^2(\mu)\right]P_{\rm w}(k),
\eeq
where $P_{\rm nw}$ and $P_{\rm w}$ denote the smooth (non-wiggly) and oscillatory (wiggly) components of the linear power spectrum, obtained from the wiggle/no-wiggle split of CLASS-PT, which isolates the two via a discrete sine transform \citep{Chudaykin:2020aoj}. In redshift space, the anisotropic damping scale is
\beq
\Sigma_{\rm tot}^2(\mu)
=
\Sigma^2\left[1+f\mu^2(2+f)\right]
+
\delta\Sigma^2 f^2\mu^2(\mu^2-1),
\eeq
such that the BAO component is suppressed by a factor dependent on scale and orientation. Here, $\mu$ is the cosine of the angle between the wavevector and the LoS, while $f$ is the logarithmic growth rate. As usual, the real-space result is recovered by setting $f=0$.
The damping scales are computed from the smooth power spectrum,
\beq
\Sigma^2
&=&
\frac{1}{6\pi^2}
\int_0^{k_S}dq\,P_{\rm nw}(q)
\left[
1-j_0(qr_{\rm BAO})+2j_2(qr_{\rm BAO})
\right],
\\
\delta\Sigma^2
&=&
\frac{1}{2\pi^2}
\int_0^{k_S}dq\,P_{\rm nw}(q)\,
j_2(qr_{\rm BAO}),
\eeq
where $r_{\rm BAO}$ is the acoustic scale and $k_S=0.05h{\rm Mpc}^{-1}$ separates the long-wavelength modes that are resummed from the short-wavelength modes treated perturbatively. 

\subsection{Estimator}\label{subsec: estimator}

We next describe how the trispectrum is measured from the galaxy density field. A direct evaluation of the four-point function is computationally expensive because the number of Fourier-space configurations grows rapidly with the size of the analysis grid. Efficient estimation therefore requires a separable form that can be evaluated using FFTs. For the power spectrum and bispectrum, this can be achieved while keeping all independent side lengths. The trispectrum introduces an additional complication, since a statistically isotropic tetrahedron is specified by six lengths, namely four external sides and two diagonals, whereas the separable estimator averages over one of the diagonals. The resulting observable is therefore five dimensional.

We use the same trispectrum estimator from \cite{Jung:2023kjh,Coulton_2024,Goldstein_2024}, which averages only over the $k_{14}$ diagonal. As illustrated in Fig.~\ref{fig:bird}, the retained diagonal joins the two triangles
$\{\boldsymbol{k}_1,\boldsymbol{k}_2,-\boldsymbol{k}_{12}\}$ and
$\{\boldsymbol{k}_3,\boldsymbol{k}_4,\boldsymbol{k}_{12}\}$.
The corresponding estimator is
\begin{equation}
\begin{aligned}
    \hat{T}_{\rm tot}(k_1,k_2,k_3,k_4,k_{12}) \propto \int_{\boldsymbol{p}_i \in k_i, \boldsymbol{p}_{12} \in k_{12}} & \Big[ (2\pi)^3 \delta_D(\boldsymbol{p}_1 + \boldsymbol{p}_2 - \boldsymbol{p}_{12})
    (2\pi)^3 \delta_D(\boldsymbol{p}_3 + \boldsymbol{p}_4 + \boldsymbol{p}_{12}) \\
    & \quad\delta_g(\boldsymbol{p}_1)\delta_g(\boldsymbol{p}_2)\delta_g(\boldsymbol{p}_3)\delta_g(\boldsymbol{p}_4) \Big],
    \label{eq:trispectrum_def}
\end{aligned}
\end{equation}
where we introduced the notation
\begin{equation}
    \int_{\boldsymbol{p}_i \in k_i} \equiv \int_{||\boldsymbol{p}_i|-k_i|<\Delta k/2} \frac{d\boldsymbol{p}_i}{(2\pi)^3},
\end{equation}
which indicates that all $\boldsymbol{p}_i$ modes within a $k$-shell of width $\Delta k$ centered on $k_i$ are integrated, for $i=\{1,2,3,4,12\}$. We can use FFTs to evaluate Eq.~\eqref{eq:trispectrum_def} if we replace the Dirac deltas by integrals. To see this, we can define 
\begin{equation}
    I_{k_i}(\boldsymbol{x}) = \int_{\boldsymbol{p}_i \in k_i} \delta_g(\boldsymbol{p}_i)e^{i\boldsymbol{x}\cdot\boldsymbol{p}_i}, \quad 
    J_{k_i}(\boldsymbol{x}) = \int_{\boldsymbol{p}_i \in k_i} e^{i\boldsymbol{x}\cdot\boldsymbol{p}_i},
\end{equation}
which represent the inverse Fourier transform over a $k$-shell of the galaxy overdensity field and a unit field, respectively, and
\begin{equation}
    \mathcal{I}_{k_i,k_j}(\boldsymbol{p}_{12}) = \int d\boldsymbol{x}\, e^{-i\boldsymbol{p}_{12}\cdot\boldsymbol{x}} \,I_{k_i}(\boldsymbol{x})I_{k_j}(\boldsymbol{x}), \quad 
    \mathcal{J}_{k_i,k_j}(\boldsymbol{p}_{12}) = \int d\boldsymbol{x}\, e^{-i\boldsymbol{p}_{12}\cdot\boldsymbol{x}} \,J_{k_i}(\boldsymbol{x})J_{k_j}(\boldsymbol{x}).
\end{equation}
Our estimator then takes the form
\begin{equation}
    \hat{T}_{\rm tot}(k_1,k_2,k_3,k_4,k_{12}) = N^{-1}\int_{\boldsymbol{p}_{12} \in k_{12}}  \mathcal{I}_{k_1,k_2}(\boldsymbol{p}_{12}) \mathcal{I}_{k_3,k_4}(-\boldsymbol{p}_{12}),
    \label{eq:trispectrum_estimator}
\end{equation}
where we defined the normalization
\begin{equation}
    N(k_1,k_2,k_3,k_4,k_{12})=\int_{\boldsymbol{p}_{12} \in k_{12}} \mathcal{J}_{k_1,k_2}(\boldsymbol{p}_{12}) \mathcal{J}_{k_3,k_4}(-\boldsymbol{p}_{12})
    \label{eq:trispectrum_normalization}
\end{equation}
that counts the number of tetrahedra in a bin. Eq.~\eqref{eq:trispectrum_estimator} can be efficiently implemented using FFTs. In order to select only the non-trivial components of the trispectrum and have a closed tetrahedron, we impose the conditions
\begin{equation}
    k_1 \leq k_2, \quad k_1 \leq k_3, \quad k_3 \leq k_4, \quad |k_1-k_2| \leq k_{12} \leq k_1 + k_2, \quad |k_3-k_4| \leq k_{12} \leq k_3 + k_4.
    \label{eq:tri_conds}
\end{equation}

\paragraph{Disconnected Subtraction.} The estimator of Eq.~\eqref{eq:trispectrum_estimator} measures the full
four-point function, $\langle\delta_g^4\rangle$, which contains both the
trispectrum $T$ and a Gaussian disconnected contribution built from
products of the power spectrum. Expanding the disconnected part into its Wick contractions gives
\begin{equation}
    \langle \delta_g(\boldsymbol{p}_1)\delta_g(\boldsymbol{p}_2)\delta_g(\boldsymbol{p}_3)\delta_g(\boldsymbol{p}_4)\rangle_{\rm disc}
    = \underbrace{\langle\delta_1\delta_2\rangle\langle\delta_3\delta_4\rangle}_{\boldsymbol{p}_{12}=0}
    + \underbrace{\langle\delta_1\delta_3\rangle\langle\delta_2\delta_4\rangle
    + \langle\delta_1\delta_4\rangle\langle\delta_2\delta_3\rangle}_{\text{disconnected bias}},
    \label{eq:wick}
\end{equation}
where $\delta_i\equiv\delta_g(\boldsymbol{p}_i)$. The first contraction requires $\boldsymbol{p}_1=-\boldsymbol{p}_2$ and therefore contributes only at $\boldsymbol{p}_{12}=\boldsymbol{0}$, which is not included in our measurement. The remaining two contractions require $\boldsymbol{p}_3=-\boldsymbol{p}_1$ and $\boldsymbol{p}_4=-\boldsymbol{p}_2$, or $\boldsymbol{p}_4=-\boldsymbol{p}_1$ and $\boldsymbol{p}_3=-\boldsymbol{p}_2$. Together with our ordering $k_1\leq k_2\leq k_3\leq k_4$ from Eq.~\eqref{eq:tri_conds}, either pairing implies $k_1=k_2=k_3=k_4\equiv k_a$. The disconnected bias therefore contributes only to equal-side bins. For all other bin combinations it vanishes identically, so that $\hat T_{\rm tot}=\hat T_{\rm conn}$ without any subtraction.

\paragraph{Optimal Estimator.}
We adopt the optimal, minimum-variance disconnected subtraction of
\cite{Goldstein_2024} (based on \citep[e.g.,][]{2017arXiv170903452S,2015arXiv150200635S}), which estimates the connected
trispectrum as
\begin{equation}
    \hat T = \hat T_{\rm tot} - \hat T_{\rm disc}^{\rm PD} + \hat T_{\rm disc}^{\rm PP},
    \label{eq:Tconn}
\end{equation}
corresponding to the schematic combination
\begin{equation}
    \hat T \leftrightarrow \delta^4 -6\delta^2\langle\delta^2\rangle + 3\langle\delta^2\rangle^2 ,
    \label{eq:opt}
\end{equation}
where $\langle\delta^2\rangle$ is evaluated using a chosen fiducial power spectrum, $P_{\rm fid}$. The term quadratic in the data (known as the ``realization-dependent bias piece'' in the CMB literature \citep{Namikawa:2012pe}),
$\hat T_{\rm disc}^{\rm PD}\!\leftrightarrow\!6\,\delta^2\langle\delta^2\rangle$,
subtracts twice the expected disconnected contribution when $P_{\rm fid}$ matches the true power spectrum. The data-independent term,
$\hat T_{\rm disc}^{\rm PP}\!\leftrightarrow\!3\,\langle\delta^2\rangle^2$,
then restores one of them, giving the combination
$\delta^4-6\delta^2\langle\delta^2\rangle+3\langle\delta^2\rangle^2$.
This construction cancels the disconnected bias exactly for the correct fiducial spectrum, while leaving only a contribution quadratic in any mismatch between $P_{\rm fid}$ and the true power spectrum. Compared with the two ``na\"ive'' prescriptions, $\delta^4-3\langle\delta^2\rangle^2$ and $\delta^4-3\langle\delta^2\rangle\delta^2$, the estimator in Eq.~\eqref{eq:opt} has lower bias and variance and is less sensitive to errors in $P_{\rm fid}$ \citep{Goldstein_2024,Namikawa:2012pe,2015arXiv150200635S}. More general data-driven methods for avoiding the disconnected Gaussian bias have been discussed in the literature \citep[e.g.,][]{Shen:2024vft}.

The two disconnected terms admit the same FFT-separable construction as $\hat T_{\rm tot}$. Restricting to the disconnected bins where they are nonzero, we define the windowed fiducial-power and data-power fields for shell $k_1=k_2=k_3=k_4\equiv k_a$,
\begin{equation}
    F^{P}_{k_a}(\boldsymbol{x}) = \int_{\boldsymbol{p}\in k_a} P_{\rm fid}(|\boldsymbol{p}|)\,e^{i\boldsymbol{x}\cdot\boldsymbol{p}},
    \qquad
    F^{D}_{k_a}(\boldsymbol{x}) = \int_{\boldsymbol{p}\in k_a} |\delta_g(\boldsymbol{p})|^2\,e^{i\boldsymbol{x}\cdot\boldsymbol{p}},
    \label{eq:Ffields}
\end{equation}
\textit{i.e.}\ the inverse Fourier transforms over the shell of the fiducial power and of the per-mode data power. Both disconnected-bias contractions in Eq.~\eqref{eq:wick} coincide on a disconnected bin, considering the $2$ and $4$ factors below. The disconnected pieces are then
\begin{align}
    \hat T_{\rm disc}^{\rm PP}(k_a,k_{12}) &= N^{-1}(k_a,k_{12})\int_{\boldsymbol{p}_{12}\in k_{12}} \int d\boldsymbol{x}\,e^{-i\boldsymbol{p}_{12}\cdot\boldsymbol{x}}\; 2\big[F^{P}_{k_a}(\boldsymbol{x})\big]^2 , \\[2pt]
    \hat T_{\rm disc}^{\rm PD}(k_a,k_{12}) &= N^{-1}(k_a,k_{12})\int_{\boldsymbol{p}_{12}\in k_{12}} \int d\boldsymbol{x}\,e^{-i\boldsymbol{p}_{12}\cdot\boldsymbol{x}} 4F^{P}_{k_a}(\boldsymbol{x})\,F^{D}_{k_a}(\boldsymbol{x}) ,
    \label{eq:PPPD}
\end{align}
where $N(k_a,k_{12})=\int_{\boldsymbol{p}_{12}\in k_{12}}\mathcal{J}_{k_a,k_a}(\boldsymbol{p}_{12})\mathcal{J}_{k_a,k_a}(-\boldsymbol{p}_{12})$
is the same tetrahedron count that normalizes Eq.~\eqref{eq:trispectrum_normalization}.
In practice we only ever need the $N_{\rm bin}$ diagonal fields
$F^{P}_{k_a},F^{D}_{k_a}$ (one per shell), rather than all pair products, so the disconnected subtraction adds negligible cost and memory on top of the $\hat T_{\rm tot}$ estimation. 

\paragraph{Fiducial Power Spectrum.}
Any mismatch between the fiducial and true power spectra enters the expectation value of $\hat T$ only at quadratic order, $\mathcal{O}\left([P_{\rm fid}-P_{\rm true}]^2\right)$, making the estimator insensitive to first-order errors in $P_{\rm fid}$. If $P_{\rm fid}$ equals the true ensemble power-spectrum, the subtraction is exact and the theory side then requires only the connected contribution. We therefore set $P_{\rm fid}$ to the measured power spectrum of the simulations themselves. When the data vector is an average over $N$ realizations, we use a leave-one-out fiducial for realization $i$,
\begin{equation}
    P_{\rm fid}^{(i)} = \frac{N\bar P - P^{(i)}}{N-1},
    \label{eq:leave-one-out}
\end{equation}
which avoids a spurious self-correlation in the covariance (the $P_{\rm fid}$ part of $\hat T_{\rm disc}^{\rm PD}$ would otherwise contain a $N^{-1}$ piece of the same realization) while keeping the mean unbiased.

The estimator of Eq.~\eqref{eq:trispectrum_estimator} can be used to measure the galaxy trispectrum in both real- and redshift-space, while in the latter it measures only the monopole (averaging over global rotations). For the disconnected subtraction in redshift space, $P_{\rm fid}(k)$ in Eq.~\eqref{eq:Ffields} is the full redshift-space power spectrum
\begin{equation}
    P_{\rm fid}(k,\mu) = P_0(k) + P_2(k)\,\mathcal{L}_2(\mu) + P_4(k)\,\mathcal{L}_4(\mu),
    \qquad \mu = \hat{\boldsymbol{k}}\cdot\hat{\boldsymbol{n}},
\end{equation}
where 
$\mathcal{L}_\ell$ are Legendre polynomials, and $P_{0,2,4}$ are the measured power-spectrum multipoles averaged over the simulations. For linear (Kaiser) case only $\ell=0,2,4$ are nonzero,  such that this truncation exactly captures the angular dependence of the Gaussian contribution in this limit. 

The trispectrum estimator used in this work is implemented in \textsc{LEFTfield}~\citep{Schmidt:2020ovm,Tucci:2023bag,Nguyen:2024yth}, which is written in \textsc{C++} with \textsc{python} bindings. We performed consistency checks in both real- and redshift-space on Gaussian random fields to test the disconnected subtraction, which is consistent with zero well below one percent. As an independent cross-check, the \textsc{LEFTfield} estimator agrees with the publicly available \textsc{PNGolin} implementation \citep{Goldstein_2024}, which uses the same disconnected subtraction procedure.

\subsection{Connecting the Trispectrum Model to Data}\label{subsec: model-data}

The estimator introduced in Eq.~\eqref{eq:trispectrum_def} does not measure the trispectrum at a single tetrahedron. Instead, it averages the second diagonal, $k_{14}$, over the finite shells defining the measured bin, and, in redshift space, over the orientation of the tetrahedron relative to the LoS. To compare the theoretical model with the data, we must therefore apply these same operations to our model. We first derive the diagonal average at fixed side lengths, then extend it to the redshift-space monopole and Alcock-Paczynski distortions, and finally account for binning effects.

The measured quantity is an average over all closed tetrahedra whose four sides lie in the shells centered on $\{k_1,k_2,k_3,k_4\}$ and whose retained diagonal lies in the $k_{12}$ shell, weighted by their phase-space density. The second diagonal, $k_{14}$, is summed over implicitly by the estimator. To make this dependence explicit, we insert
\begin{equation}
    1 = \int d\boldsymbol{p}_{14}\delta_D(\boldsymbol{p}_1+\boldsymbol{p}_4-\boldsymbol{p}_{14}),
    \qquad p_{14}\equiv|\boldsymbol{p}_{14}|=|\boldsymbol{p}_1+\boldsymbol{p}_4|=|\boldsymbol{p}_2+\boldsymbol{p}_3|,
    \label{eq:k14identity}
\end{equation}
into the numerator of Eq.~\eqref{eq:trispectrum_def}. Taking the ensemble average then gives
\begin{equation}
\begin{aligned}
    \langle\hat T\rangle \propto \int_{\boldsymbol{p}_i\in k_i,\boldsymbol{p}_{12}\in k_{12}}\int d\boldsymbol{p}_{14}\Big[
    &(2\pi)^3\delta_D(\boldsymbol{p}_1+\boldsymbol{p}_2-\boldsymbol{p}_{12})(2\pi)^3\delta_D(\boldsymbol{p}_3+\boldsymbol{p}_4+\boldsymbol{p}_{12})
    \\
    &\times \delta_D(\boldsymbol{p}_1+\boldsymbol{p}_4-\boldsymbol{p}_{14}) T(\boldsymbol{p}_1,\boldsymbol{p}_2,\boldsymbol{p}_3,\boldsymbol{p}_4)\Big] ,
\end{aligned}
\label{eq:Tbin_threedelta}
\end{equation}
where the three Dirac deltas close the tetrahedron. In real space, or after averaging the redshift-space trispectrum over its global orientation, the result depends only on the six lengths $\{k_1,k_2,k_3,k_4,k_{12},k_{14}\}$. 

Performing angular integrals by expanding the Dirac deltas in spherical Bessel functions and resumming the resulting Legendre series leaves an integral over the six magnitudes \citep{Fergusson_2010}, weighted by the phase-space density of tetrahedra,
\begin{equation}
    W = \frac{k_1k_2k_3k_4k_{12}k_{14}}{\sqrt{g_1}},
    \label{eq:Wmeasure}
\end{equation}
where $g_1$ is the Gram determinant of the three edge vectors that meet at one vertex,
\begin{equation}
\begin{aligned}
    g_1 = k_{12}^2k_{14}^2(\textstyle\sum_i k_i^2 - k_{12}^2 - k_{14}^2)
    - k_{12}^2\,\kappa_{23}\kappa_{14} + k_{14}^2\,\kappa_{12}\kappa_{34} - (k_1^2 k_3^2 - k_2^2 k_4^2)(\kappa_{12}+\kappa_{34}),
\end{aligned}
\label{eq:g1}
\end{equation}
for $\kappa_{ij}\equiv k_i^2 - k_j^2$.
Equivalently, the determinant is related to the Euclidean volume of the tetrahedron through $\sqrt{g_1}=12V_{\rm tet}$, with $V_{\rm tet}$ obtained, for example, from the Cayley-Menger determinant \cite{Liberti:2014distance}. At fixed values of the four sides and $k_{12}$, the observable is therefore the $W$-weighted average over the remaining diagonal,
\begin{equation}
    T(k_1,k_2,k_3,k_4,k_{12}) = \mathcal{N}^{-1}
    \int dk_{14}\; W\, T(k_1,k_2,k_3,k_4,k_{12},k_{14}),
    \label{eq:Tbin}
\end{equation}
where the normalization is $\mathcal{N}\equiv\int dk_{14}\,W$. 

In real space, Eq.~\eqref{eq:Tbin} gives the diagonal-averaged trispectrum at fixed side lengths. In redshift-space, however, the trispectrum acquires a dependence on the orientation of the tetrahedron relative to the LoS $\hat{\boldsymbol{n}}$, and we have two extra integrals, \textit{i.e.}
\begin{equation}
    T_0(k_1,k_2,k_3,k_4,k_{12})
    =\mathcal{N}^{-1}\int dk_{14}\,W \,\int_0^{2\pi}\frac{d\varphi}{4\pi}\int_{-1}^{1}d\mu\,T(k_1, k_2,k_3,k_4,k_{12},k_{14},\mu,\varphi).
    \label{eq:Tmonopole}
\end{equation}

Now we turn to a few geometric relations which will be useful in the following. First, we fix the orientation by the polar angle of the binned diagonal $\boldsymbol{k}_{12}$ and the azimuth about it,
\begin{equation}
    \mu \equiv \mu_{12} = \hat{\boldsymbol{k}}_{12}\cdot\hat{\boldsymbol{n}}, \qquad \varphi\in[0,2\pi), \qquad
    \hat{\boldsymbol{n}}=\big(\sqrt{1-\mu^2}\cos\varphi,\;\sqrt{1-\mu^2}\sin\varphi,\;\mu\big),
    \label{eq:los}
\end{equation}
working in the frame where $\boldsymbol{k}_{12}=k_{12}\hat{\boldsymbol{z}}$. The tetrahedron is two triangles sharing $\boldsymbol{k}_{12}$, namely $\{\boldsymbol{k}_1,\boldsymbol{k}_2,-\boldsymbol{k}_{12}\}$ and $\{\boldsymbol{k}_3,\boldsymbol{k}_4,\boldsymbol{k}_{12}\}$, with their relative fold fixed by the second diagonal $k_{14}$. Placing the first triangle in the $xz$ plane, we can identify the cosine of the angles of the tetrahedron,
\begin{equation}
    \cos\theta_1=\frac{k_1^2+k_{12}^2-k_2^2}{2k_1k_{12}}, \quad
    \cos\theta_2=\frac{k_4^2+k_{12}^2-k_3^2}{2k_4k_{12}}, \quad
    \cos\theta_3=\frac{k_4^2-k_{12}^2-k_3^2}{2k_3k_{12}}.
    \label{eq:thetas}
\end{equation}
The angles $\theta_i$ fix the orientation of the sides within the two triangles shown in Fig.~\ref{fig:bird}: $\theta_1$ is the angle between $\boldsymbol{k}_1$ and $\boldsymbol{k}_{12}$, $\theta_2$ the angle between $\boldsymbol{k}_4$ and $-\boldsymbol{k}_{12}$, and $\theta_3$ the angle between $\boldsymbol{k}_3$ and $\boldsymbol{k}_{12}$. The cosines of the four sides and the second diagonal with the line of sight are then \citep{Gualdi:2020mattertrispectrum,Paul_2024}
\be
\begin{split}
    \mu_1 &= \cos\theta_1\,\mu + \sin\theta_1\sqrt{1-\mu^2}\,\cos\varphi, \quad\quad\quad\,\,\,
    \mu_2 = (k_{12}\mu - k_1\mu_1)/k_2, \\
    \mu_3 &= \cos\theta_3\,\mu + \sin\theta_3\sqrt{1-\mu^2}\,\cos(\varphi+\psi), \quad
    \mu_4 = -(k_{12}\mu + k_3\mu_3)/k_4, \\
    \mu_{14} &= (k_1\mu_1 + k_4\mu_4)/k_{14}, \label{eq:mu14}
\end{split}
\ee
where the angle $\psi$, shown in Fig.~\ref{fig:bird}, is the folding angle between the planes of the two triangles sharing $\boldsymbol{k}_{12}$. At fixed side lengths, $\psi$ is set by $k_{14}$ and satisfies
\begin{equation}
    \cos\psi=\frac{k_1^2\sin^2\theta_1 + k_3^2\sin^2\theta_3 + (k_{12}+k_3\cos\theta_3-k_1\cos\theta_1)^2 - k_{14}^2}
                  {2\,k_1 k_3\sin\theta_1\sin\theta_3}.
    \label{eq:psi}
\end{equation}
The remaining cosines follow directly from closure: $k_1\mu_1+k_2\mu_2=k_{12}\mu$, $k_3\mu_3+k_4\mu_4=-k_{12}\mu$, and $k_1\mu_1+k_4\mu_4=k_{14}\mu_{14}$. A non-degenerate tetrahedron has $g_1>0$ and $|\cos\psi|<1$, while $g_1=0$ corresponds to the planar configurations with $|\cos\psi\,|=1$.

The $k_{14}$ integral in Eq.~\eqref{eq:Tbin} includes only values for which the tetrahedron closes in three dimensions, corresponding to $g_1>0$. Its endpoints, $k_{14}^{\min}$ and $k_{14}^{\max}$, are the two planar configurations for which the folding angle around $k_{12}$ is $0$ or $\pi$. The two limits can be written as
\begin{equation}
    k_{14}^{\max/\min}
    =
    [k_1^2+k_4^2-2k_1k_4\cos(\theta_1\pm\theta_2)]^{1/2} .
    \label{eq:dihedral}
\end{equation}
As illustrated by the wing motion in Fig.~\ref{fig:bird}, varying $k_{14}$ is equivalent to changing the folding angle $\psi$ between the two triangles sharing $k_{12}$. Since $W$ diverges as $g_1^{-1/2}$ at the two endpoints, we make the change of variables
\begin{equation}
    k_{14}
    =
    k_{14}^{c}
    +\Delta k_{14}\sin\frac{\pi t}{2},
    \qquad
    k_{14}^{c}=\frac{k_{14}^{\max}+k_{14}^{\min}}{2},
    \qquad
    \Delta k_{14}=\frac{k_{14}^{\max}-k_{14}^{\min}}{2},
    \label{eq:sinetransform}
\end{equation}
with $t\in[-1,1]$. After this redefinition, the $k_{14}$ integral is straightforward to evaluate with Gauss--Legendre quadrature as described below. Alternatively, one could use a uniform $k_{14}$ grid restricted to $g_1>0$; this gives consistent results in practice.

\paragraph{AP effect.}
When converting the observed galaxy redshifts and angular positions into comoving coordinates, one must assume a fiducial cosmology. If this differs from the true cosmology, radial and transverse distances are rescaled differently, producing the Alcock-Paczynski effect \cite{Alcock:1979mp,Ballinger:1996cd}. Following \citep{Ivanov:2021kcd}, the true wavenumbers and angles $(q,\nu)$ are related to those defined in the fiducial cosmology, $(k,\mu)$, by
\begin{equation}
    q^2=k^2\big[\alpha_\parallel^{-2}\mu^2+\alpha_\perp^{-2}(1-\mu^2)\big],
    \qquad
    \nu^2=\frac{\alpha_\parallel^{-2}\mu^2}{\alpha_\parallel^{-2}\mu^2+\alpha_\perp^{-2}(1-\mu^2)},
    \label{eq:apmap}
\end{equation}
\begin{equation}
    \alpha_\parallel=\frac{H_{\rm fid}(z)}{H_{\rm true}(z)}\frac{H_{0,\rm true}}{H_{0,\rm fid}},
    \qquad
    \alpha_\perp=\frac{D_{A,\rm true}(z)}{D_{A,\rm fid}(z)}\frac{H_{0,\rm true}}{H_{0,\rm fid}}.
\end{equation}
The factors of $H_{0,\rm true}/H_{0,\rm fid}$ arise because we express wavenumbers in units of $h{\rm Mpc}^{-1}$. 

The AP transformation acts linearly on the Cartesian components of every wavevector, with the transverse components rescaled by $\alpha_\perp^{-1}$ and the line-of-sight component by $\alpha_\parallel^{-1}$. It therefore preserves momentum conservation, so a quadrilateral that closes in the fiducial cosmology also closes after the AP transformation. We apply the same mapping to each of the four sides and to the two diagonals,
\begin{equation}
\begin{aligned}
    &q_i=q[k_i,\mu_i],\quad \nu_i=\nu[\mu_i]\quad (i=1,2,3,4),\\
    &q_{12}=q[k_{12},\mu],\quad q_{14}=q[k_{14},\mu_{14}],\quad
     \nu_{12}=\nu[\mu],\quad \nu_{14}=\nu[\mu_{14}].
\end{aligned}
\end{equation}

There is also an overall volume factor. Writing
$J=\alpha_\perp^2\alpha_\parallel$, the $N$-point function transforms as $J^{-(N-1)}$, giving the usual factor of $J^{-1}$ for the power spectrum, $J^{-2}$ for the bispectrum \citep{Song:2015gca,Ivanov:2021kcd}, and therefore $J^{-3}=\alpha_\perp^{-6}\alpha_\parallel^{-3}$ for the trispectrum. The AP-distorted redshift-space monopole is then
\begin{equation}
    T_0(k_1,k_2,k_3,k_4,k_{12})
    =\alpha_\perp^{-6}\alpha_\parallel^{-3}\mathcal{N}^{-1}
    \int dk_{14}\,W\,\int_0^{2\pi}\frac{d\varphi}{4\pi}\int_{-1}^{1}d\mu\,T\big(q_i,q_{12},q_{14};\,\nu_i,\nu_{12},\nu_{14}\big).
    \label{eq:Tbin_monopole}
\end{equation}

\paragraph{Binning Effects.}

The measured trispectrum is averaged over finite bins, so we average the theory over the same bins in $(k_1,k_2,k_3,k_4,k_{12})$.
We follow the same general approach used for the bispectrum
\citep{Ivanov:2021kcd,Philcox:2022b1loop,Bakx:2025pop}. If $k_i$ denotes
the center of a bin, the continuous bin average is
\beq
\label{eq:Tk-binning}
\begin{split}
\hat T_{0,\rm int}
=
\alpha_\perp^{-6}\alpha_\parallel^{-3}\mathcal N_{\rm bin}^{-1}
&
\prod_{i\in\{1,2,3,4,12\}}
\int_{k_i-\Delta k/2}^{k_i+\Delta k/2}
dp_i\,p_i
\int_{p_{14}^{\rm min}}^{p_{14}^{\rm max}}
\frac{dp_{14}\,p_{14}}{\sqrt{g_1}}
\\
&\times
\int_0^{2\pi}\frac{d\varphi}{4\pi}
\int_{-1}^{1}d\mu\,
T\big(q_i,q_{12},q_{14};
\nu_i,\nu_{12},\nu_{14}\big).
\end{split}
\eeq
Here, $p_i$ are the fiducial wavenumbers integrated within the bins. At each integration point, the corresponding $(q,\nu)$ are obtained from the AP mapping of Eq.~\eqref{eq:apmap}. The limits $k_{14}^{\rm min}$ and $k_{14}^{\rm max}$ are given by Eq.~\eqref{eq:dihedral}. The normalization $\mathcal{N}_{\rm bin}$ is the similarly binned approximation to $\mathcal{N}$,
\beq
    \mathcal{N}_{\rm bin} = \Bigg[\prod_{i\in\{1,2,3,4,12\}}\int_{k_i-\Delta k/2}^{k_i+\Delta k/2}\,dp_i\, p_i\,\int_{p_{14}^{\rm min}}^{p_{14}^{\rm max}}\, \frac{1}{\sqrt{g_1}} \, dp_{14}\,p_{14} \Bigg]\,,
    \label{eq:bin_den}
\eeq
which gives the number count of tetrahedra in the bin and is the analytic counterpart of the mode count of the estimator. At each point in the integral, we keep only closed tetrahedra satisfying $g_1>0$. This is analogous to imposing the closed triangle condition in the bispectrum bin integral of \cite{Bakx:2025pop}.\footnote{This differs from the original treatment of \cite{Ivanov:2021kcd}, where closure was not imposed point by point inside the bin, leading to large discreteness effects. The prescription was later rectified in
\cite{Bakx:2025pop}.}

We evaluate Eqs.~\eqref{eq:Tk-binning} and \eqref{eq:bin_den} using Gauss-Legendre (GL) quadrature. Similar numerical integrations are used for bispectrum binning and angular averages in \citep{Ivanov:2021kcd,Philcox:2022b1loop,Bakx:2025pop}. We use $n_{k_{\rm bin}}$ nodes for each of the five binned momenta $\{k_1,k_2,k_3,k_4,k_{12}\}$, $n_{k_{14}}$ nodes for $k_{14}$ (first applying the sine transformation of Eq.~\eqref{eq:sinetransform}), and $n_{\rm ang}$ nodes for each of the two orientation angles $(\mu,\varphi)$.

To assess the convergence of the GL integration, we compare a high-resolution (HR) evaluation of the theory vector with a lower-resolution (LR) one, obtained by reducing one or more of the GL node counts. The HR reference is chosen as the coarsest node configuration whose $\chi^2$ no longer changes after further refinement. Since the power spectrum and bispectrum do not change in the two cases, the difference $\Delta\boldsymbol{x}=\boldsymbol{x}^{\rm LR}-\boldsymbol{x}^{\rm HR}$ is nonzero only for the trispectrum.

Rather than comparing the theory vectors directly, we quantify if this numerical difference is large enough to affect the inferred parameters. Using the standard Fisher-bias approximation, we estimate the parameter shift induced by $\Delta\boldsymbol{x}$ as
\be\label{eq:gl-shift}
\Delta\boldsymbol{\theta} = (\boldsymbol{F}+\boldsymbol{P})^{-1}\boldsymbol{D}^{T}\boldsymbol{C}^{-1}\,\Delta\boldsymbol{x},
\qquad
\boldsymbol{F} = \boldsymbol{D}^{T}\boldsymbol{C}^{-1}\boldsymbol{D},
\ee
where $\boldsymbol{D}=\partial\boldsymbol{x}/\partial\boldsymbol{\theta}$ contains the model derivatives with respect to the cosmological and leading bias parameters, $\boldsymbol{C}$ is the covariance, and $\boldsymbol{P}$ is the prior Fisher matrix including the parameters that are marginalized analytically. We consider the GL integration converged when
\be
\frac{|\Delta\theta_i|}{\sigma(\theta_i)}
<
\epsilon_{\rm GL}
\ee
for every cosmological parameter, where $\sigma(\theta_i)$ is the marginalized uncertainty and $\epsilon_{\rm GL}$ is our adopted convergence tolerance. We use $\epsilon_{\rm GL}=0.3$ throughout and report the node counts used for inference in Sec.~\ref{sec: data}.

\paragraph{Discreteness Weights.}

The binned model in Eq.~\eqref{eq:Tk-binning} replaces the sum over Fourier modes by a continuous integral. In a finite box this is only an approximation, and the difference is largest at low $k$, where each bin holds only a few modes. Following \citep{Bakx:2025pop}, this can be corrected with a discreteness weight. We compute this weight for both the bispectrum and the trispectrum in Appendix~\ref{app:discreteness}. The weight remains within $2\%$ of unity for all but one bispectrum bin and for $98\%$ of the trispectrum bins, so we do not include it in the analysis.

\subsection{Covariance}
\label{sec:cov}
A joint analysis of the power spectrum, bispectrum and trispectrum needs a covariance for the full data vector. For the power spectrum and bispectrum we use the standard analytic Gaussian covariance, which is diagonal apart from the cross terms between the power-spectrum multipoles \cite{Ivanov:2021kcd}. We treat the trispectrum in the same spirit. We neglect non-Gaussian contributions to the trispectrum covariance in our baseline model, showing in Appendix~\ref{app:cov} that these are small, and in Appendix~\ref{app:fisher} that the trispectrum can be treated as effectively independent of the power spectrum and bispectrum.

\begin{figure}
\centering
\includegraphics[width=\textwidth]{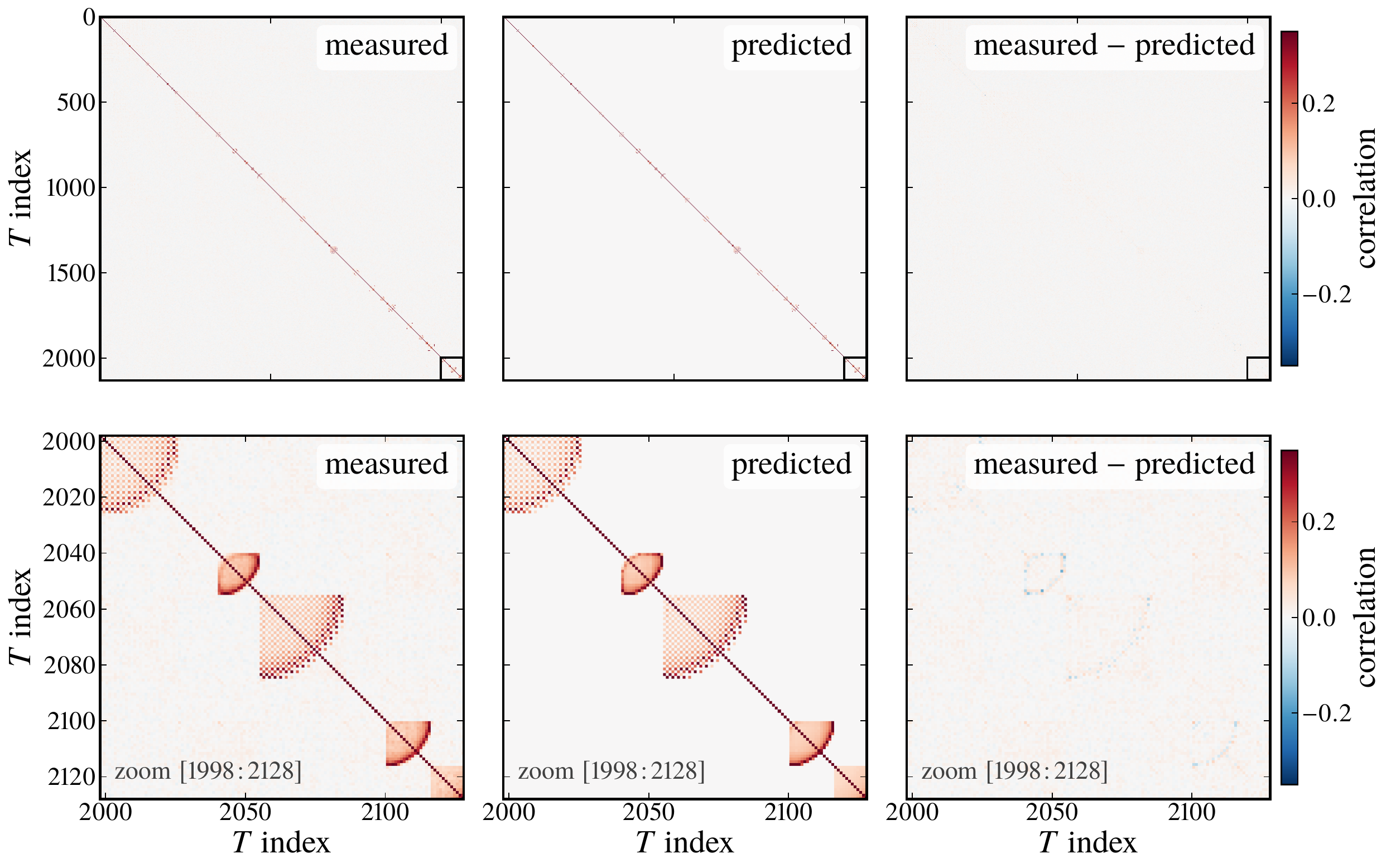}
\caption{Correlation matrix of the redshift-space trispectrum, measured from the \textsc{Quijote} simulations at $z=0.5$). From left to right: measured, our Gaussian prediction, and the residual. The top row shows the full $2132$-bin matrix for $k_{\rm max}=0.1\hMpc$, while the bottom row zooms in on the densest sibling bins, namely bins with the same four sides but a different diagonal. The Gaussian covariance model reproduces both the diagonal variance and the arc-like sibling correlations, leaving a residual consistent with noise.}
\label{fig:Tcov}
\end{figure}

A key difference between the trispectrum and lower-order statistics is that our estimator keeps only one of the two independent diagonals, which introduces an off-diagonal structure to the Gaussian covariance. We label a trispectrum bin by
\begin{equation}
    I=(k_1,k_2,k_3,k_4,k_{12}^I),
\end{equation}
where $k_{12}^I$ is the centre of the retained-diagonal bin (see Fig.~\ref{fig:bird}). Consider first two ``sibling'' bins,
\begin{equation}
    I=(k_1,k_2,k_3,k_4,k_{12}^I),
    \quad\text{and}\quad
    J=(k_1,k_2,k_3,k_4,k_{12}^J).
\end{equation}
These have the same four external side bins, but are allowed to have different retained diagonals $k_{12}$. These bins are correlated already in the Gaussian covariance, because the estimator averages over the second diagonal $k_{14}$. Equivalently, a Wick contraction can match the retained diagonal of bin $J$ not only to $k_{12}$ of bin $I$, but also to the other pair diagonals $k_{14}$ and $k_{13}$.

The Gaussian covariance $C^T_{IJ}\equiv\text{Cov}[T_0(k_1,k_2,k_3,k_4,k_{12}^I), T_0(k_1,k_2,k_3,k_4,k_{12}^J)]$ is given by
\begin{equation}
    C^T_{IJ}
    =
    V^2\, P(k_1)P(k_2)P(k_3)P(k_4)\,
    \frac{
    c_{12}N_{12}(I,J)
    +c_{14}N_{14}(I,J)
    +c_{13}N_{13}(I,J)
    }{
    N_I N_J
    } 
    \label{eq:Tcov_exact}
\end{equation}
(see Appendix~\ref{app:cov} for a derivation). Here the integers $c_{12}$, $c_{14}$ and $c_{13}$ are the corresponding symmetry factors. They are fixed by the side-bin degeneracies and are discussed in Appendix~\ref{app:cov}. $N_I$ and $N_J$ are the numbers of tetrahedra in each bin defined as Eq.~\eqref{eq:trispectrum_normalization}, e.g., $N_I=N(k_1,k_2,k_3,k_4,k_{12}^I)$. Moreover, the quantities $N_{12}$, $N_{14}$ and $N_{13}$ count the overlap in the three possible diagonals. For example,
\begin{equation}
    N_{14}(I,J) =
    \sum_{\boldsymbol p_i\in k_i}
    \sum_{\boldsymbol p_{12}\in k_{12}^I}
    \sum_{\boldsymbol p_{14}\in k_{12}^J}
    \delta^K_{\boldsymbol p_1+\boldsymbol p_2,\boldsymbol p_{12}}\,
    \delta^K_{\boldsymbol p_{1}+\boldsymbol p_{4},\boldsymbol p_{14}}
    \delta^K_{\boldsymbol p_1+\boldsymbol p_2+\boldsymbol p_3+\boldsymbol p_4,\boldsymbol 0},
\label{eq:NX_discrete}
\end{equation}
for $i\in\{1,2,3,4\}$, where $\delta^K$ is the Kronecker delta, and equivalently for $N_{13}(I,J)$ replacing $\boldsymbol{p}_{14}$ by $\boldsymbol{p}_{13}$ and $\boldsymbol{p}_4$ by $\boldsymbol{p}_3$. Note that $N_{12}(I,I)=N_I$. For sibling bins the $k_{12}$ piece vanishes, so the covariance comes only from the diagonals $k_{14}$ and $k_{13}$, giving the arc-like structure seen in Fig.~\ref{fig:Tcov}. 

The diagonal variance is the special case $I=J$ (with $k_{12}^I=k_{12}^J$). If some of the four side bins are equal, this expression is only multiplied by the retained diagonal symmetry factor $c_{12}$ (defined in Appendix~\ref{app:cov}). Thus the trispectrum diagonal covariance is
\begin{equation}
    C^T_{II}
    =
    \frac{V^2}{N_I}
    c_{12}
    P(k_1)P(k_2)P(k_3)P(k_4).
    \label{eq:Tcov_exact_diag_full}
\end{equation}

We test the covariance model of Eq.~\eqref{eq:Tcov_exact} against the sample redshift-space trispectrum covariance from $15{,}000$ \textsc{Quijote} simulations (Sec.~\ref{sec: data}). Fig.~\ref{fig:Tcov} compares the measured and predicted correlation matrices. The model captures both the diagonal variance and the off-diagonal correlations between sibling bins, and the residual is consistent with noise. We show the ratio of the measured and predicted diagonal covariance in Appendix~\ref{app:cov}, along with the cross-covariance of the trispectrum with the power spectrum and bispectrum and an enumeration of the non-Gaussian terms for the diagonal trispectrum covariance.

\subsection{Free Parameters}\label{subsec: free-params}
Combining our model for the one-loop galaxy power spectrum, tree-level galaxy bispectrum and tree-level galaxy trispectrum results in 22 free parameters:
\beq \label{eq:free_params}
    &&\,\,\,\,\,\{\textcolor{red}{b_1},\textcolor{red}{b_2},
    \textcolor{red}{b_{\mathcal{G}_2}},\textcolor{purple}{b_{\Gamma_3}},b_3,b_{\mathcal{G}_3},b_{\delta\mathcal{G}_2}\}
    \\\nonumber
    &&\times\{\textcolor{blue}{c_0},\textcolor{blue}{c_2},\textcolor{blue}{c_4},\textcolor{blue}{\tilde c},\textcolor{magenta}{c_B},c_T
    \}\\\nonumber
    &&\times\{\textcolor{red}{P_{\rm shot}}, \textcolor{blue}{a_0}, 
    \textcolor{blue}{a_2},
    \textcolor{orange}{A_{\rm shot}},
    \textcolor{orange}{B_{\rm shot}},C_{\rm shot},D_{\rm shot},E_{\rm shot},F_{\rm shot}\} ,
\eeq
which splits into parameters appearing only in the power spectrum (shown in \textcolor{blue}{blue}), only in the bispectrum (\textcolor{magenta}{magenta}), only in the trispectrum (black), in both power-spectrum and trispectrum (\textcolor{purple}{purple}), in both bispectrum and trispectrum (\textcolor{orange}{orange}), and in all three statistics (\textcolor{red}{red}). The three lines give bias parameters, UV counterterms, and stochasticity parameters respectively. We only sample $\{b_1,b_2,b_{\mathcal{G}_2}\}$, while all other parameters are analytically marginalized, following \citep{Philcox:2020zyp}, reducing considerably the computational cost required for convergence of the likelihood analyses.
In Eq.~\eqref{eq:stoch_pieces}, there is one term in $T_{\rm stoch}^{(n^{-2})}$ that is quadratic in $P_{\rm shot}$ and $B_{\rm shot}$. To facilitate efficient computation, we linearize it as
\beq
\begin{aligned}    
\bigl[b_1 B_{\rm shot} + f(1+P_{\rm shot})\mu_{ij}^{2}\bigr]^{2}
&\approx
\bigl[b_1 B_{\rm fid} + f(1+P_{\rm fid})\mu_{ij}^{2}\bigr]^{2}
\\
&+
2\bigl[b_1 B_{\rm fid} + f(1+P_{\rm fid})\mu_{ij}^{2}\bigr]
b_1\bigl(B_{\rm shot}-B_{\rm fid}\bigr)
\\
&+
2\bigl[b_1 B_{\rm fid} + f(1+P_{\rm fid})\mu_{ij}^{2}\bigr]
f\mu_{ij}^{2}\bigl(P_{\rm shot}-P_{\rm fid}\bigr),
\end{aligned}
\eeq
where the model is evaluated at the first line and the derivatives for analytical marginalization are obtained from second and third lines.

We sample the cosmological parameters $\{h,\,\omega_{\rm cdm},\,\ln(10^{10}A_s)\}$ (fixing $n_s$ and $\omega_b$ to their fiducial values), and we adopt the following
priors displayed in Tab.~\ref{tab:priors} (rescaling parameters by factors of $A\equiv \sigma_8/\sigma_{8,\rm ref}$ to minimize prior projection effects \citep[e.g.,][]{Maus:2024sbb,Chudaykin:2025aux,Tsedrik:2025hmj}). We center the Gaussian prior of $b_3$ on its polynomial fit from \cite{Lazeyras:2015lgp},
\beq\label{eq:b3b1}
b_3(b_1) = -1.028 + 7.646 b_1 - 6.227 b_1^2 + 0.912 b_1^3,
\eeq
while the other three third-order bias parameters are centered on the local-in-Lagrangian coevolution relations of \citep{Eggemeier:2020umu,Lazeyras:2017hxw},
\beq\label{eq:coev}
    b_{\Gamma_3} = \frac{23}{42}(b_1-1), \quad
    b_{\cG_3} = -\frac{22}{63}(b_1-1), \quad
    b_{\delta\cG_2} = -\frac{2}{7}b_2.
\eeq
Note that $b_{\Gamma_3}$ is usually also centered at this value in EFTofLSS analysis where the one-loop power-spectrum is included (see \cite[e.g.,][]{Philcox:2022b1loop}). The shot-noise amplitudes are dimensionless and centered on their Poisson values, as the data are not shot-noise subtracted.  

\begin{table}\label{tab:priors}
    \centering
{\small
    \begin{tabular}{llll}
        \hline
        & parameter & prior & units \\
        \hline

        \textbf{cosmology}
        & $h$
        & $\mathcal{U}[0.4,1.0]$
        & --- \\

        & $\omega_{\rm cdm}$
        & $\mathcal{U}[0.001,0.99]$
        & --- \\

        & $\ln(10^{10}A_s)$
        & $\mathcal{U}[1.61,3.91]$
        & --- \\

        \hline

        \textbf{nuisance}
        & $A\,b_1$
        & $\mathcal{U}[0,3]$
        & --- \\

        (sampled)

        & $A^2 b_2$
        & $\mathcal{N}(0,5^2)$
        & --- \\

        & $A^2 b_{\mathcal{G}_2}$
        & $\mathcal{N}(0,5^2)$
        & --- \\

        \hline

        \textbf{nuisance}
        
        & $A^3 b_{\Gamma_3}$
        & $\mathcal{N}\!\left(
            23(b_1-1)/42,1^2
          \right)$
        & --- \\

        (analytically

        & $A^3 b_3$
        & $\mathcal{N}(b_3(b_1),1^2)$
        & --- \\

        marginalized)

        & $A^3 b_{\mathcal{G}_3}$
        & $\mathcal{N}\!\left(
            -22(b_1-1)/63,1^2
          \right)$
        & --- \\

        & $A^3 b_{\delta\mathcal{G}_2}$
        & $\mathcal{N}\!\left(
            -2b_2/7,1^2
          \right)$
        & --- \\

        & $A^2c_0$
        & $\mathcal{N}(0,30^2)$
        & $[\mathrm{Mpc}/h]^2$ \\

        & $A^2c_2$
        & $\mathcal{N}(30,30^2)$
        & $[\mathrm{Mpc}/h]^2$ \\

        & $A^2c_4$
        & $\mathcal{N}(0,30^2)$
        & $[\mathrm{Mpc}/h]^2$ \\

        & $A^4\tilde c$
        & $\mathcal{N}(400,400^2)$
        & $[\mathrm{Mpc}/h]^4$ \\

        & $c_B$
        & $\mathcal{N}(0,5^2)$
        & --- \\

        & $c_T$
        & $\mathcal{N}(0,5^2)$
        & --- \\

        & $P_{\rm shot}$
        & $\mathcal{N}(0,1^2)$
        & --- \\
        
        & $a_0$
        & $\mathcal{N}(0,1^2)$
        & --- \\

        & $a_2$
        & $\mathcal{N}(0,1^2)$
        & --- \\

        & $A_{\rm shot}$
        & $\mathcal{N}(1,1^2)$
        & --- \\

        & $B_{\rm shot}$
        & $\mathcal{N}(1,1^2)$
        & --- \\

        & $C_{\rm shot}$
        & $\mathcal{N}(1,1^2)$
        & --- \\

        & $D_{\rm shot}$
        & $\mathcal{N}(1,1^2)$
        & --- \\

        & $E_{\rm shot}$
        & $\mathcal{N}(1,1^2)$
        & --- \\

        & $F_{\rm shot}$
        & $\mathcal{N}(1,1^2)$
        & --- \\

        \hline
    \end{tabular}
    }
    \caption{
    Model parameters and priors used in the analysis. Here, $\mathcal{U}$ denotes a uniform prior and $\mathcal{N}(\mu,\sigma^2)$ a Gaussian prior with mean $\mu$ and variance $\sigma^2$. The bias parameters are rescaled by the appropriate power of the amplitude $A\equiv\sigma_8(z)/\sigma_{8,\text{ref}}(z)$. The third-order bias parameters are assigned Gaussian priors centered on their coevolution predictions, while $b_3$ is centered in the $b_3(b_1)$ fit from Eq.~\eqref{eq:b3b1} \cite{Lazeyras:2015lgp}.
    }
    \label{tab:priors}
\end{table}

\subsection{Numerical Results}

\begin{figure}
    \centering
    \includegraphics[width=0.85\linewidth]{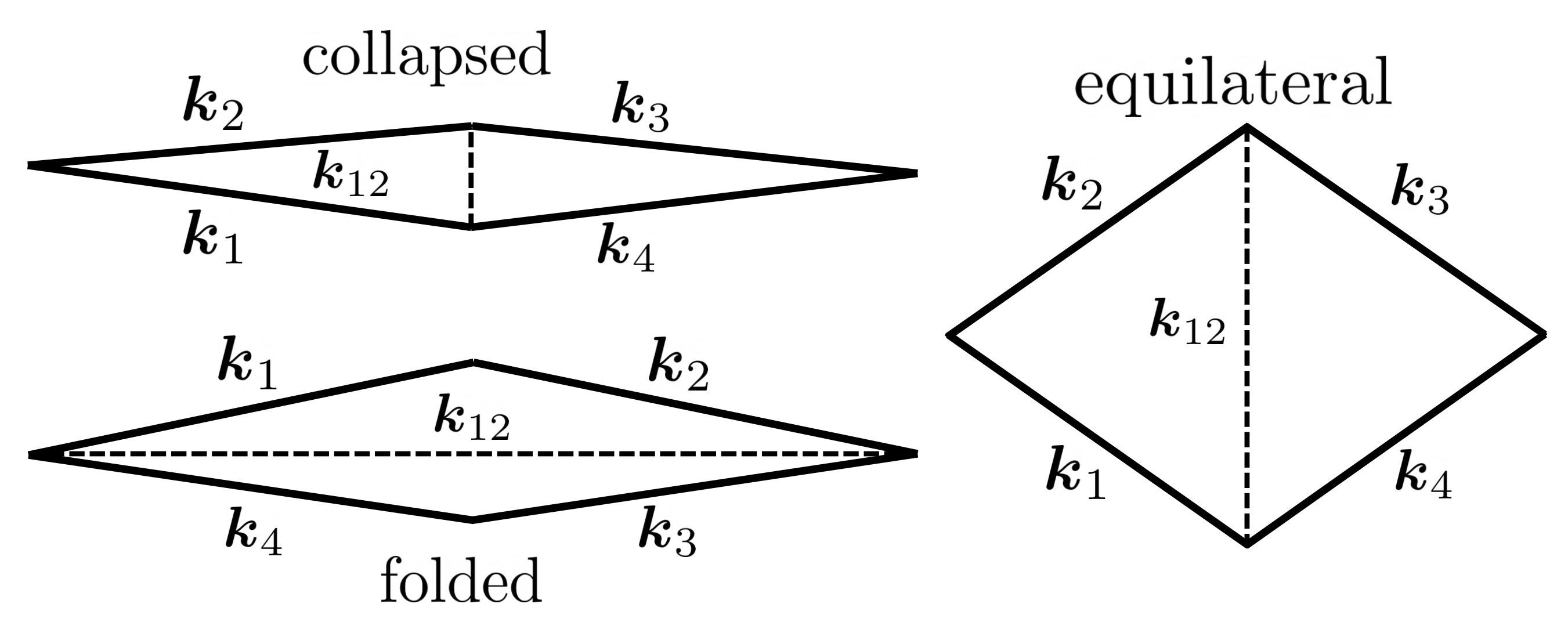}
    \caption{Schematic illustration of the collapsed, folded, and equilateral shapes of the trispectrum, with the binned diagonal $k_{12}$ shown as a dashed line. In the collapsed configuration, $k_{12}$ is short compared to the external sides, while in the folded configuration it approaches the sum of the lengths of two sides, giving a nearly collinear shape. In this representation, the external sides have equal side lengths.}
    \label{fig:tri_shapes}
\end{figure}

Before proceeding to inference, we first examine the form of the model. The trispectrum monopole depends on five length parameters: $\{k_1,k_2,k_3,k_4,\kd\}$ (after averaging the LoS); for visualization, we fix all four sides to be equal, $k_1=k_2=k_3=k_4=k$, and consider three diagonal configurations (illustrated in Fig.~\ref{fig:tri_shapes}):
\emph{equilateral} (a regular tetrahedron, $\kd=\sqrt{4/3}k$), \emph{folded} ($\kd=1.9k$, approaching the colinear limit $\kd\to k_1+k_2$), and \emph{collapsed} ($\kd=0.35\,k$, a soft binned diagonal). As usual, we integrate over the second diagonal, $k_{14}$, via Eq.~\eqref{eq:Tk-binning} and the redshift-space monopole by averaging over the LoS angles $(\mu,\varphi)$.

Taking the linear bias $b_1=1.7472$ (the best fit from our redshift-space $P_\ell+B_0+T_0$ analysis, cf.\,Sec.~\ref{sec: results-rsd}), we fix the bias parameters using the $N$-body fit of \citep{Lazeyras:2015lgp} for the quadratic bias, and the local-in-Lagrangian coevolution relations of \citep{Eggemeier:2020umu,Lazeyras:2017hxw} for the cubic and tidal biases as described above, and we use $b_3=-3b_2$ instead of the fit from \cite{Lazeyras:2015lgp}. We adopt the \textsc{AbacusSummit} cosmology at $z=1.1$ (see Sec.~\ref{sec: data}, with $\sigma_8(z{=}1.1)=0.470$, $f=0.888$, and the IR-resummed power spectrum), set $\alpha_\parallel=\alpha_\perp=1$, and use a number density of $\bar n\simeq3.2\times10^{-3}\,h^3{\rm Mpc}^{-3}$, which is the same obtained for the dark-matter halos used for inference at this redshift in the following. For the stochastic contribution, all shot-noise amplitudes are set to their Poisson expectations.

\begin{figure}
    \centering
    \includegraphics[width=\textwidth]{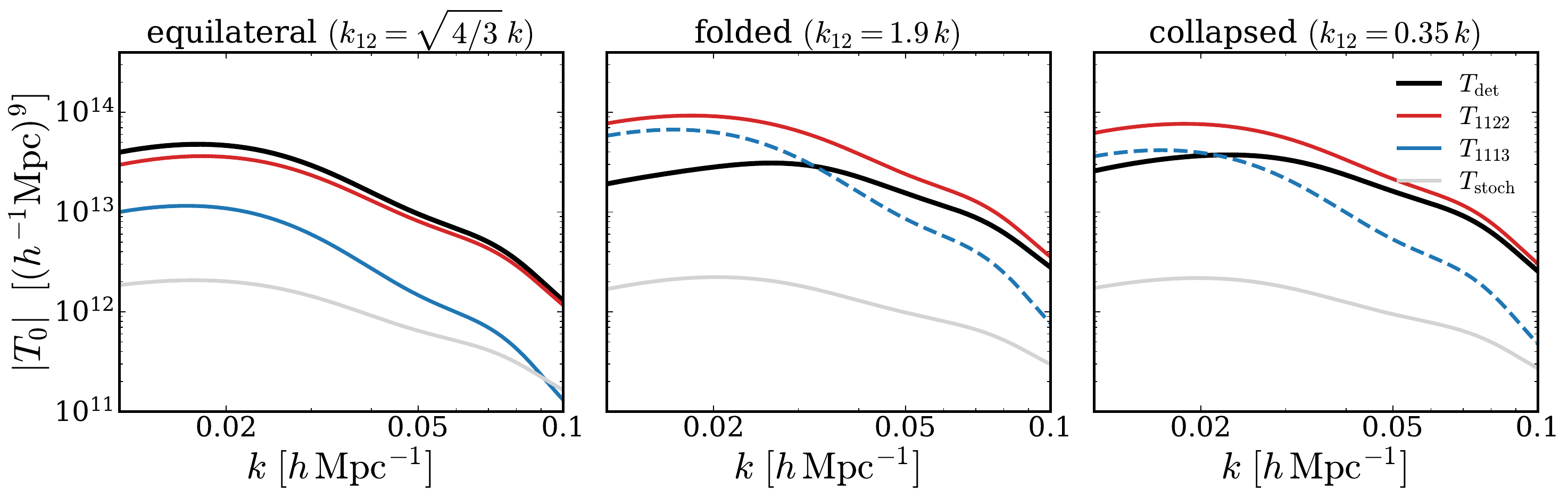}
    \caption{Contributions to the tree-level galaxy trispectrum monopole in redshift-space, for equilateral, folded, and collapsed tetrahedra (left to right) as a function of the overall scale $k$. We show the two deterministic blocks $T_{1122}$ (red) and $T_{1113}$ (blue), their sum $T_{\rm det}$ (black), and the stochastic contribution $T_{\rm stoch}$ (grey), with solid and dashed lines denoting positive and negative values respectively.}
    \label{fig:T-plot}
\end{figure}

Fig.~\ref{fig:T-plot} shows the resulting trispectrum, with the deterministic terms $T_{1122}$ and $T_{1113}$ of Eq.~\eqref{eq: Tk-det}, their sum $T_{\rm det}=T_{1122}+T_{1113}$, and the stochastic piece $T_{\rm stoch}$ of Eq.~\eqref{eq:stoch_pieces} plotted separately for each shape. The $T_{1122}$ piece dominates the deterministic signal and is positive on all scales. For the equilateral tetrahedron, $T_{1113}$ is positive and reaches $10$–$35\%$ of $T_{1122}$, while for the folded and collapsed configurations it is negative, reaching up to $\sim75\%$ and $\sim60\%$ of $T_{1122}$ in magnitude respectively. The stochastic term is positive and comparatively flat, dominated on these scales by the $T_{\rm stoch}^{(n^{-1})}$ contribution in Eq.~\eqref{eq:stoch_pieces}, as shown in Appendix~\ref{app:stoch}. Due to the high density of this sample, it lies roughly an order of magnitude below $T_{\rm det}$ but grows in relative importance toward smaller scales.

\section{Results}\label{sec: results}

\subsection{Data and Analysis Details}\label{sec: data}

Our analysis uses two suites of $N$-body simulations: \textsc{AbacusSummit}~\citep{Garrison:2021lfa,Maksimova:2021distinct}, which we use to run cosmological inference, and \textsc{Quijote}~\citep{Villaescusa-Navarro:2019fdia}, which is used to test the covariance model. From \textsc{AbacusSummit} we take $25$ independent realizations of the base $L=2\,h^{-1}{\rm Gpc}$ boxes at $z=1.1$ and $z=0.5$, run in the Planck-2018 baseline cosmology, with $\Omega_m=0.3152$, $h=0.6736$, $n_s=0.9649$, and a single massive neutrino, giving $\sigma_8(z=1.1)=0.470$ and $\sigma_8(z=0.5)=0.622$. We apply a mass cut of $12.0\le\log_{10}(M/h^{-1}M_\odot)\le13.5$ to \textsc{CompaSO} halos \citep{Hadzhiyska:2021zbd}, giving a number density $\bar n\simeq3.23\times10^{-3}h^3{\rm Mpc}^{-3}$ at $z=1.1$ and $\bar n\simeq3.95\times10^{-3}h^3{\rm Mpc}^{-3}$ at $z=0.5$. These should roughly resemble bias and number densities of Emission Line Galaxies (ELGs) and Luminous Red Galaxies (LRGs) from DESI. We place the halos in redshift space in the plane-parallel approximation, taking the LoS along the box $z$-axis and displacing each halo by its peculiar velocity, $\boldsymbol{s}=\boldsymbol{x}+[(1+z)v_z]\,\hat{\boldsymbol{z}}/{H(z)}$,
with $H(z)$ the Hubble rate, re-imposing periodic boundaries. All galaxy density fields are assigned to a $256^3$ mesh with a non-uniform FFT~\citep{Barnett_2019}.

The power-spectrum multipoles $P_\ell(k)$ ($\ell=0,2,4$) and the bispectrum monopole $B_0$ are measured with the estimators of \textsc{PolyBin3D}~\citep{Philcox:2024rqr}, while the trispectrum monopole $T_0$ is measured with \textsc{LEFTfield}~\citep{Schmidt:2020ovm,Tucci:2023bag,Nguyen:2024yth}, subtracting the disconnected part as discussed above. We use linear bins of width $\Delta k=0.01\,h\,{\rm Mpc}^{-1}$ with $k_{\rm min}=0.01\,h\,{\rm Mpc}^{-1}$, discarding the fundamental bin as it is dominated by discreteness effects and typically affected by large-scale systematics in real surveys~\citep{Alam:2016hwk}.

First, we perform cosmological inference on an effective volume of $200 h^{-3}{\rm Gpc}^3$ both in real-space (Sec.~\ref{sec: results-real}) and redshift space (Sec.~\ref{sec: results-rsd}), in each case taking the mean of the data vector over the $25$ realizations at $z=1.1$. In these cases, the residual statistical error lies well below the survey level errors of interest and the model is tested to very high precision. Lastly, in Sec.~\ref{sec: desi} we rescale our covariance to $50 h^{-3}{\rm Gpc}^3$ to study the scale cuts and information content of the galaxy trispectrum in a DESI-like volume at both $z=1.1$ and $z=0.5$.

Our covariance model discussed in Sec.~\ref{sec:cov} is validated against the sample covariance of $15{,}000$ fiducial \textsc{Quijote} realizations at $z=0.5$. These which have $L=1\,h^{-1}{\rm Gpc}$, and processed through the identical redshift-space procedure, estimators and binning, but with halos identified with the Friends-of-Friends (FoF) algorithm \citep{Davis:1985rj} instead of \textsc{CompaSO}.

\paragraph{Likelihood.} We assume a Gaussian likelihood for the joint $(P,B,T)$ data vector $\boldsymbol{D}$,
\be
-2\ln\mathcal{L}=\big[\boldsymbol{D}-\boldsymbol{T}(\boldsymbol{\theta})\big]^{\rm T}\mathbf{C}^{-1}\big[\boldsymbol{D}-\boldsymbol{T}(\boldsymbol{\theta})\big],
\ee
where $\boldsymbol{T}(\boldsymbol{\theta})$ is the binned theory $(P_{\rm 1loop},B_{\rm tree},T_{\rm tree})$ of Sec.~\ref{sec: model}, and $\mathbf{C}$ is the analytical Gaussian covariance model of Sec.~\ref{sec:cov}. The analytical marginalization discussed in Sec.~\ref{subsec: free-params} replaces $\mathbf{C}^{-1}$ by the prior-corrected covariance using the Woodbury identity and adds its associated term to the likelihood \citep[e.g.,][]{Philcox:2020zyp}. The priors are listed in Sec.~\ref{subsec: free-params}. The remaining parameters (the cosmological set $\{h,\omega_{\rm cdm},\ln(10^{10}A_s)\}$ together with $\{b_1,b_2,b_{\mathcal{G}_2}\}$) are sampled with \textsc{MontePython}~\citep{Audren:2012vy,Brinckmann:2018cvx}, and we quote the derived clustering amplitude $\sigma_8$ and the matter fraction $\Omega_m$. The convergence of the MCMC chains is verified via Gelman--Rubin diagnostic, discarding the first $30\%$ of each chain as burn-in and requiring the potential scale reduction factor to satisfy $R-1<0.01$ for every sampled parameter.

To remove the dominant $\sigma_8$--bias degeneracy we sample the rescaled combinations $A^{n}b_n$ (the quantities to which $P,B,T$ are directly sensitive, see Tab.~\ref{tab:priors}), where $A\equiv\sigma_8(z_{\rm eff})/\sigma_{8,\rm ref}$ and $\sigma_{8,\rm ref}$ is the fiducial amplitude at the effective redshift, and convert to physical biases \textit{a posteriori} by dividing out the appropriate power of $A$. 

\paragraph{GL Node Choices.} 
As previously mentioned, we perform bin integration by means of Gauss-Legendre (GL) quadrature. In real space the trispectrum monopole is isotropic, so a single angular node is sufficient (since nothing depends on $\mu,\phi$) and we set $(n_{k_{\rm bin}},n_{k_{14}},n_{\rm ang})=(2,8,1)$. In redshift space we use $(2,6,6)$ at all three scale cuts. With these choices the shift in the cosmological parameters (from Eq.~\eqref{eq:gl-shift}) stays well within the statistical error at the lower $k_{\rm max}$ cuts, the largest shift on any cosmological parameter being $0.04\sigma$ at $\kmax^T=0.08\hMpc$ and $0.24\sigma$ at $0.10\hMpc$, with $\sigma_8$ moving by less than $0.06\sigma$ in both cases. At $\kmax^T=0.12\hMpc$ the linearized diagnostic suggests that our GL precision will result in a slightly low $\sigma_8$; increasing the precision further is computationally prohibitive at this $\kmax$ however.

The most demanding analysis is the joint $P_\ell{+}B_0{+}T_0$ fit at the highest cut $\kmax^B=\kmax^T=0.12\hMpc$, which we run as $20$ parallel chains on four CPUs each, for $80$ CPUs in total, reaching $R-1<0.01$ after about four days of wall-clock time and roughly $7{,}700$ CPU-hours. This is dominated by the per-step tree-level trispectrum evaluation, and could be improved by precomputing the bin-integrals, including through a factorization scheme such as \textsc{cobra} \citep{Bakx:2024zgu}.

\subsection{Real Space}\label{sec: results-real}

We first present cosmological constraints from \textsc{AbacusSummit} in real space, setting $f=0$ in the redshift-space model of Sec.~\ref{sec: model} and dropping the LoS-dependent EFT operators (\textit{i.e.}\ all the parameters in the second line of Eq.~\eqref{eq:free_params} with except of $c_0$). This gives a $(P_{\rm 1loop},B_{\rm tree},T_{\rm tree})$ analysis pipeline that is structurally identical to the redshift-space one of Sec.~\ref{sec: results-rsd}. Recovering the clustering amplitude and matter fraction in this controlled setting validates both the theoretical modelling and the trispectrum measurement before we deal with the additional complications of redshift space. 

We use the same $25$ \textsc{AbacusSummit} base cosmology simulations at $z=1.1$ and the same halo mass selection described in Sec.~\ref{sec: data}, but without the redshift-space displacement. In real-space, we keep not only $n_s$ and $\omega_b$ fixed to the \textsc{AbacusSummit} fiducial values but also $h$ (due to the increased parameter degeneracies found therein), so that the sampled cosmological parameters are $\{\omega_{\rm cdm},\ln(10^{10}A_s)\}$. For this test, we use the scale cuts $\kmax^P=0.20\hMpc$ and $\kmax^B=\kmax^T=0.10\hMpc$. This yields data vectors of size $N_{\rm bins}=19$ ($P$ only), $134$ ($P{+}B$), and $2712$ ($P{+}B{+}T$).

\begin{figure}
\centering
\includegraphics[width=\textwidth]{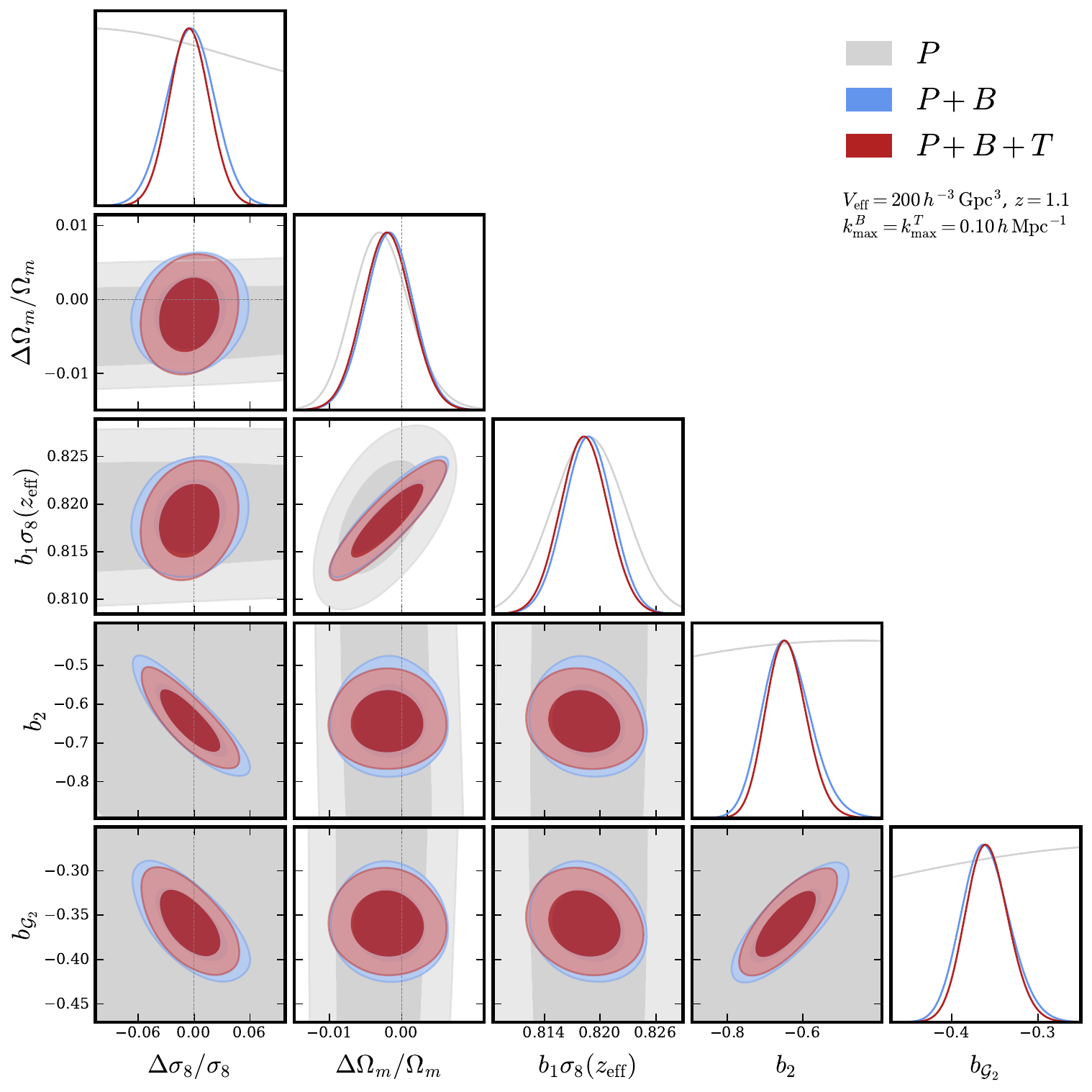}
\caption{Marginalized posterior contours ($68\%$ and $95\%$) from the real-space \textsc{AbacusSummit} $z=1.1$ data, comparing the three data vectors $P$ (grey), $P{+}B$ (blue) and $P{+}B{+}T$ (red), with scale cuts $\kmax^P=0.20\hMpc$ and $\kmax^B=\kmax^T=0.10\hMpc$. The $P$-alone posterior does not constrain $\sigma_8$ in real space and is shown for illustration only.}
\label{fig:reals}
\end{figure}

Fig.~\ref{fig:reals} compares the marginalised posteriors from $P$ alone, $P{+}B$ and $P{+}B{+}T$, with one-dimensional constraints shown in Tab.~\ref{tab:reals}. The two cosmological parameters behave very differently, and thus will be discussed separately below.

The clustering amplitude is carried entirely by the higher-order statistics. At tree level the real-space power spectrum monopole measures only the combination $b_1^2\sigma_8^2$, so $P$ on its own cannot separate $b_1$ from $\sigma_8$. The amplitude posterior spans most of its prior range and the value in Tab.~\ref{tab:reals} is set by the prior rather than by the data. Adding the bispectrum breaks the degeneracy, because $B$ scales as $b_1^3\sigma_8^4$ while $P$ scales as $b_1^2\sigma_8^2$, and this alone turns an unconstrained direction into a $2.5\%$ measurement of $\sigma_8$. Adding the trispectrum tightens $\sigma_8$ by $16\%$, from $\sigma(\sigma_8)/\sigma_8=0.025$ to $0.021$. This is broadly consistent with the Fisher analysis found by \cite{Spezzati_2025}, despite the differences in the tracer biases and number density. The bispectrum also improves the higher-order bias parameters constrains by more than an order of magnitude, from $b_2=0.25\pm1.54$ and $b_{\mathcal{G}_2}=-0.37\pm0.71$ with $P$ alone to $b_2=-0.642\pm0.063$ and $b_{\mathcal{G}_2}=-0.361\pm0.027$, and the trispectrum tightens these by a further $15\%$ and $11\%$ respectively.

The matter fraction behaves in the opposite way. $\Omega_m$ is fixed here by the shape of the power spectrum, which is already measured to $0.4\%$ by the $19$ monopole bins out to $\kmax^P=0.20\hMpc$. The bispectrum improves this only slightly, to $0.33\%$, and the trispectrum adds nothing further at this scale cut. The information the trispectrum adds in real space is therefore specific to the amplitude, not to the shape.

Both $P{+}B$ and $P{+}B{+}T$ recover the true cosmological parameters. The amplitude is consistent with the true value to within $0.4\%$ and $0.5\%$ respectively (which is $0.2\sigma$ in both cases), and $\Omega_m$ is consistent to $0.2\%$ (about $0.5\sigma$). The fits are good, with $\chi^2/\nu=1.25$ for $P{+}B$ and $1.01$ for $P{+}B{+}T$ on $2712$ data points. We conclude that at $\kmax^B=\kmax^T=0.10\hMpc$ the tree-level trispectrum model is unbiased in real space, which is the prerequisite for the redshift-space analysis that follows.

\begin{table}[!htb]
\centering
{\small
\begin{tabular}{l|c|c|c}
\hline\hline
\multicolumn{4}{c}{$\kmax^P=0.20$, $\kmax^B=\kmax^T=0.10~\hMpc$}\\ \hline
Parameter & $P$ & $P{+}B$ & $P{+}B{+}T$ \\ \hline
$\Delta\sigma_8/\sigma_8$              & unconstr. & $-0.004\pm0.025$ & $-0.005\pm0.021$ \\
$\Delta\Omega_m/\Omega_m$              & $-0.003\pm0.004$ & $-0.002\pm0.003$ & $-0.002\pm0.003$ \\
$b_1\sigma_8(z_{\rm eff})$             & $0.819\pm0.004$ & $0.819\pm0.003$ & $0.818\pm0.003$ \\
$b_2$                                  & $0.25\pm1.54$ & $-0.642\pm0.063$ & $-0.644\pm0.053$ \\
$b_{\mathcal{G}_2}$                    & $-0.37\pm0.71$ & $-0.361\pm0.027$ & $-0.359\pm0.024$ \\
\hline\hline
\end{tabular}
}
\caption{One-dimensional marginalized constraints ($68\%$) from the real-space \textsc{AbacusSummit} $z=1.1$ data using the baseline scale cuts $\kmax^P=0.20\hMpc$ and $\kmax^B=\kmax^T=0.10\hMpc$, for the three data vectors $P$, $P{+}B$ and $P{+}B{+}T$. The clustering amplitude is normalised to the fiducial $\sigma_8(z{=}0)=0.8077$, and the bias combination $b_1\sigma_8(z_{\rm eff})$ is reported because it is the well-measured quantity at tree level, while the remaining bias parameters $b_2$ and $b_{\mathcal{G}_2}$ are quoted in their physical values, $b_n=b_n^{\rm samp}/\sigma_8(z_{\rm eff})^n$ with $\sigma_8(z_{\rm eff})=\sigma_8(z{=}0)\,D(z_{\rm eff})/D(0)$ and $z_{\rm eff}=1.1$. The power spectrum alone does not constrain $\sigma_8$ in real space, where only the combination $b_1\sigma_8$ is measured at tree level, so its amplitude entry is marked as unconstrained. The corresponding $b_2$ and $b_{\mathcal{G}_2}$ values are prior-dominated. $\Omega_m$ is derived from the sampled $\omega_{\rm cdm}$ at fixed $h$, and unlike $\sigma_8$ it is already well measured by the power spectrum alone.}
\label{tab:reals}
\end{table}

The ordering of the gains also matches what is seen in redshift space (Fig.~\ref{fig:rsds}) and in the existing real-space literature~\cite{Ivanov:2021kcd}, with the bispectrum responsible for the bulk of the improvement on the amplitude and the bias parameters, and the trispectrum contributing a smaller additional gain. One difference is worth noting is that in redshift space the power spectrum multipoles already constrain $\sigma_8$ through the Kaiser term, so the higher-order statistics improve an existing measurement, whereas in real space they provide the only source of information (beyond loop corrections), bolstering their relative utility.

\subsection{Redshift Space}\label{sec: results-rsd}

\begin{figure}
\centering
\includegraphics[width=\textwidth]{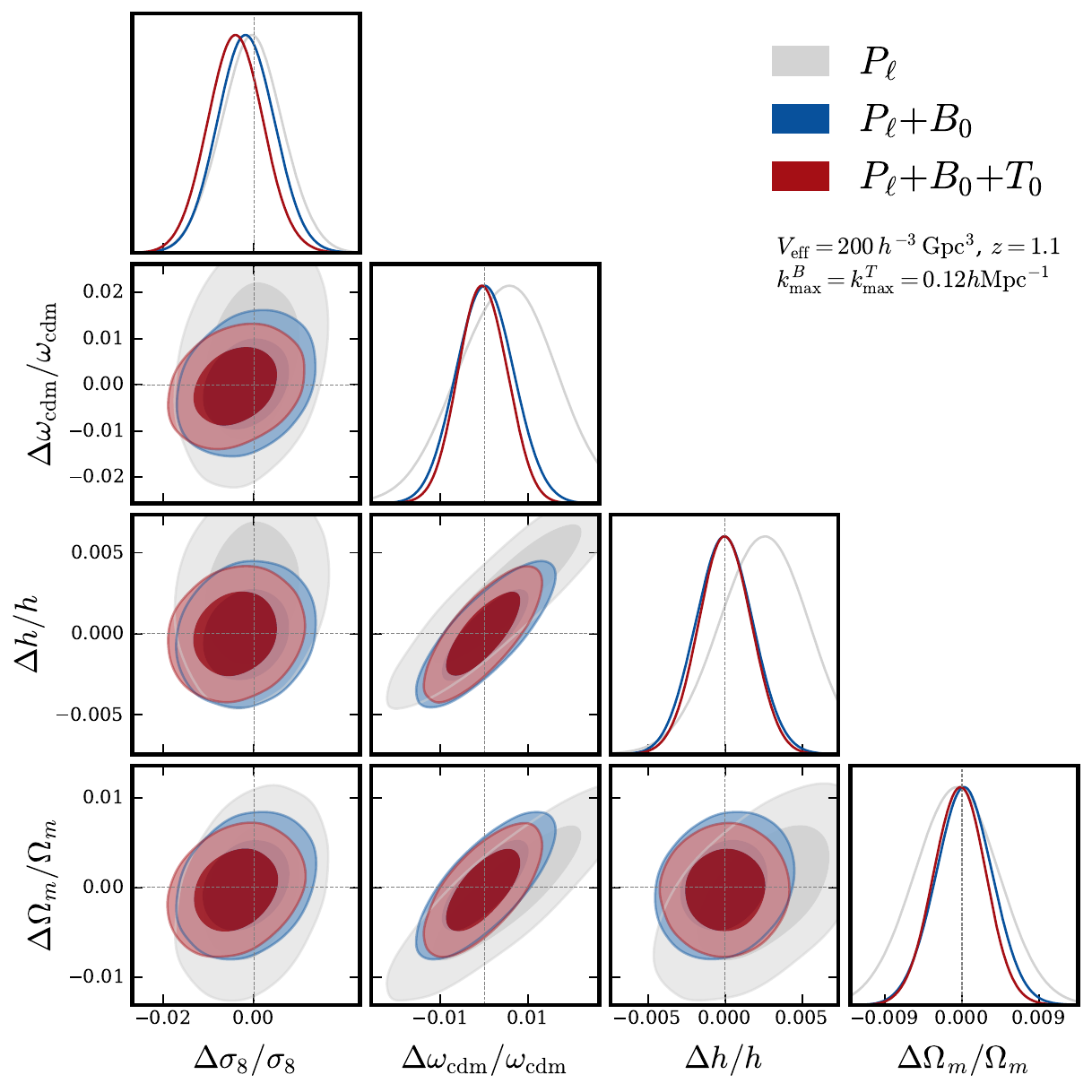}
\caption{
Marginalized cosmological constraints ($68\%$ and $95\%$) from the redshift-space \textsc{AbacusSummit} $z=1.1$ data at $V_{\rm eff}=200\,h^{-3}{\rm Gpc}^3$, shown as fractional deviations $\Delta\boldsymbol{\theta}/\boldsymbol{\theta}$ of $\{\sigma_8,\omega_{\rm cdm},h,\Omega_m\}$ from their true values (dashed lines at zero). We overlay the power spectrum alone ($P_\ell$, grey), $P_\ell{+}B_0$ (blue) and $P_\ell{+}B_0{+}T_0$ (red) at the highest cut $\kmax^B=\kmax^T=0.12\hMpc$, with $\kmax^P=0.18\hMpc$. One-dimensional constraints for all three scale cuts are listed in Tab.~\ref{tab:rsds}.
}
\label{fig:rsds}
\end{figure}

In this section, we analyse the full redshift-space data vector, which includes the power spectrum multipoles, the bispectrum monopole $B_0$ and the trispectrum monopole $T_0$. This matches the configuration of a real galaxy survey such as BOSS~\cite{Alam:2016hwk} or DESI~\cite{Aghamousa:2016zmz}. Our aim is to quantify the cosmological information that the trispectrum monopole adds in redshift space within the EFTofLSS, and to find the largest wavenumber at which the tree-level model stays unbiased at the \textsc{AbacusSummit} effective volume of $200 h^{-3}{\rm Gpc}^3$ .

Throughout this section, we fix $\kmax^P=0.18\hMpc$, which gives $17$ bins per multipole and $51$ power spectrum data points. This is the largest $\kmax^P$ at which the one-loop model returns unbiased cosmology from $P_\ell$ alone, which we note to be significantly larger than the one used for the PT challenge simulations \cite{Nishimichi:2020tvu} (whose effective volume is around three times larger than that of this work). 

We compare three data vectors, the multipoles $P_\ell$ on their own, the pair $P_\ell{+}B_0$, and the full $P_\ell{+}B_0{+}T_0$. The bispectrum and trispectrum share a common scale cut $\kmax^B=\kmax^T$ that we vary over $\{0.08,0.10,0.12\}\hMpc$ while keeping $\kmax^P=0.18\hMpc$ fixed. The bispectrum binning keeps every closed triangle, giving $62$, $115$ and $191$ configurations at the three cuts. The trispectrum bins the four sides together with the first diagonal $k_{12}$ and keeps every closed tetrahedron, giving $951$, $2578$ and $5885$ bins. In total, the $P_\ell{+}B_0{+}T_0$ data vector has $1064$, $2744$ and $6127$ bins also at the three cuts.
\begin{table}[!htb]
\centering
{\small
\begin{tabular}{l|c|c|c}
\hline\hline
\multicolumn{4}{c}{$\kmax^B=\kmax^T=0.08\,\hMpc$}\\ \hline
Parameter & $P_\ell$ & $P_\ell{+}B_0$ & $P_\ell{+}B_0{+}T_0$ \\ \hline
$\Delta\sigma_8/\sigma_8$                & $-0.0006\pm0.0069$ & $+0.0018\pm0.0065$ & $+0.0008\pm0.0065$ \\
$\Delta\omega_{\rm cdm}/\omega_{\rm cdm}$ & $+0.0051\pm0.0113$ & $+0.0052\pm0.0097$ & $+0.0036\pm0.0095$ \\
$\Delta h/h$                             & $+0.0025\pm0.0029$ & $+0.0017\pm0.0025$ & $+0.0014\pm0.0024$ \\
$\Delta\Omega_m/\Omega_m$                & $-0.0007\pm0.0049$ & $+0.0011\pm0.0044$ & $+0.0002\pm0.0043$ \\ \hline
$b_1$                                    & $1.763\pm0.020$ & $1.747\pm0.016$ & $1.747\pm0.016$ \\
$b_2$                                    & $0.09\pm0.63$ & $-0.826\pm0.089$ & $-0.731\pm0.086$ \\
$b_{\mathcal{G}_2}$                      & $-0.45\pm0.36$ & $-0.504\pm0.057$ & $-0.499\pm0.055$ \\
\hline\hline
\end{tabular}

\vspace{0.30cm}

\begin{tabular}{l|c|c|c}
\hline\hline
\multicolumn{4}{c}{$\kmax^B=\kmax^T=0.10\,\hMpc$}\\ \hline
Parameter & $P_\ell$ & $P_\ell{+}B_0$ & $P_\ell{+}B_0{+}T_0$ \\ \hline
$\Delta\sigma_8/\sigma_8$                & $-0.0006\pm0.0069$ & $+0.0018\pm0.0065$ & $+0.0014\pm0.0065$ \\
$\Delta\omega_{\rm cdm}/\omega_{\rm cdm}$ & $+0.0051\pm0.0113$ & $+0.0068\pm0.0082$ & $+0.0049\pm0.0075$ \\
$\Delta h/h$                             & $+0.0025\pm0.0029$ & $+0.0018\pm0.0022$ & $+0.0014\pm0.0021$ \\
$\Delta\Omega_m/\Omega_m$                & $-0.0007\pm0.0049$ & $+0.0020\pm0.0040$ & $+0.0012\pm0.0037$ \\ \hline
$b_1$                                    & $1.763\pm0.020$ & $1.748\pm0.015$ & $1.746\pm0.015$ \\
$b_2$                                    & $0.09\pm0.63$ & $-0.827\pm0.053$ & $-0.803\pm0.048$ \\
$b_{\mathcal{G}_2}$                      & $-0.45\pm0.36$ & $-0.445\pm0.036$ & $-0.436\pm0.033$ \\
\hline\hline
\end{tabular}
}

\vspace{0.30cm}
{\small
\begin{tabular}{l|c|c|c}
\hline\hline
\multicolumn{4}{c}{$\kmax^B=\kmax^T=0.12\,\hMpc$}\\ \hline
Parameter & $P_\ell$ & $P_\ell{+}B_0$ & $P_\ell{+}B_0{+}T_0$ \\ \hline
$\Delta\sigma_8/\sigma_8$                & $-0.0006\pm0.0069$ & $-0.0017\pm0.0063$ & $-0.0040\pm0.0061$ \\
$\Delta\omega_{\rm cdm}/\omega_{\rm cdm}$ & $+0.0051\pm0.0113$ & $+0.0002\pm0.0065$ & $-0.0003\pm0.0055$ \\
$\Delta h/h$                             & $+0.0025\pm0.0029$ & $-0.0000\pm0.0018$ & $-0.0000\pm0.0017$ \\
$\Delta\Omega_m/\Omega_m$                & $-0.0007\pm0.0049$ & $+0.0003\pm0.0033$ & $-0.0003\pm0.0030$ \\ \hline
$b_1$                                    & $1.763\pm0.020$ & $1.746\pm0.014$ & $1.748\pm0.013$ \\
$b_2$                                    & $0.09\pm0.63$ & $-0.773\pm0.041$ & $-0.754\pm0.036$ \\
$b_{\mathcal{G}_2}$                      & $-0.45\pm0.36$ & $-0.389\pm0.025$ & $-0.400\pm0.022$ \\
\hline\hline
\end{tabular}
}
\caption{One-dimensional marginalized constraints ($68\%$) from the redshift-space \textsc{AbacusSummit} $z=1.1$ data ($V_{\rm eff}=200\,h^{-3}{\rm Gpc}^3$), using
$\kmax^P=0.18\hMpc$ throughout, at the three scale cuts $\kmax^B=\kmax^T=\{0.08,0.10,0.12\}\hMpc$. Each panel shows the power spectrum alone ($P_\ell$, identical across panels), and its combination with the bispectrum monopole ($P_\ell{+}B_0$) and additionally the trispectrum monopole ($P_\ell{+}B_0{+}T_0$). Cosmological parameters are normalized to their true values. The leading bias parameters $\{b_1,b_2,b_{\mathcal{G}_2}\}$ are quoted directly, as their true values are unknown.}
\label{tab:rsds}
\end{table}

The results are shown in Tab.~\ref{tab:rsds} and we plot the $\kmax^B=\kmax^T=0.12\hMpc$ contours in Fig.~\ref{fig:rsds}. At the conservative cut $\kmax^B=\kmax^T=0.08\hMpc$ every parameter is recovered unbiased. The power spectrum alone already fixes the amplitude to $\Delta\sigma_8/\sigma_8=-0.1\%\pm0.7\%$, and adding the bispectrum and then the trispectrum leaves the central value on truth while tightening the error a little. The other cosmological parameters $\omega_{\rm cdm}$, $h$ and $\Omega_m$ agree with their true values well below $1\sigma$. The same holds at $\kmax^B=\kmax^T=0.10\hMpc$, where the amplitude stays unbiased for both $P_\ell{+}B_0$ ($+0.18\%$) and $P_\ell{+}B_0{+}T_0$ ($+0.14\%$). At the highest cut $\kmax^B=\kmax^T=0.12\hMpc$, $P_\ell{+}B_0$ is still accurate at $-0.17\%$ and $P_\ell{+}B_0{+}T_0$ at $-0.40\%$, about $0.7\sigma$. We read this as still unbiased but at the edge of the tree-level description, since the $P_\ell{+}B_0{+}T_0$ errors stay at or below the $P_\ell{+}B_0$ errors on every parameter.

\begin{figure}
\centering
\includegraphics[width=0.99\textwidth]{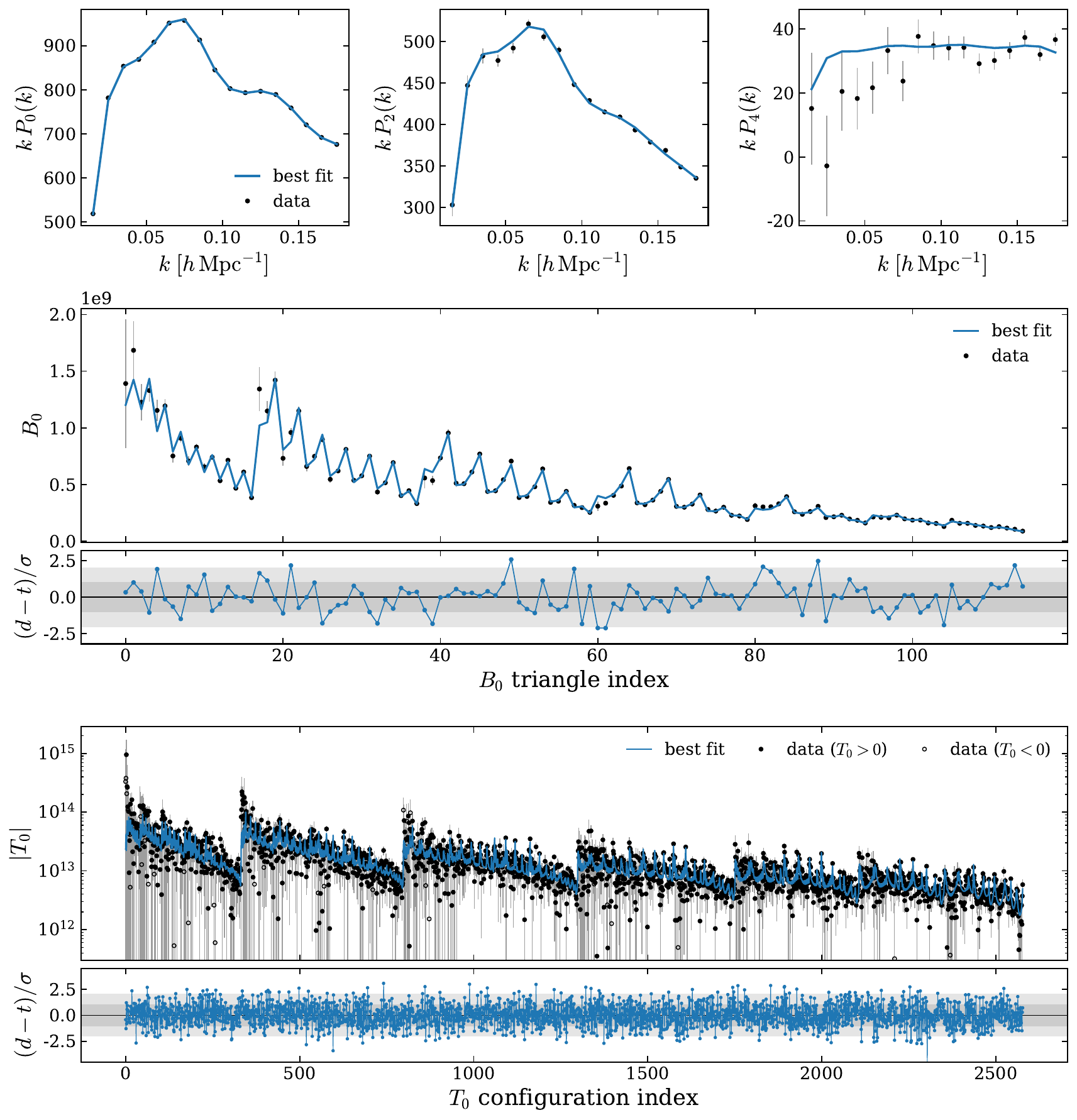}
\caption{
Measured power spectrum multipoles $P_0,P_2,P_4$ (top), bispectrum monopole $B_0$ (middle) and connected
trispectrum monopole $T_0$ (bottom) of the redshift-space \textsc{AbacusSummit} $z=1.1$ data (black, with
$1\sigma$ error bars in grey), compared with the best-fit $P_\ell{+}B_0{+}T_0$ tree-level model (blue) at
the baseline cut $\kmax^B=\kmax^T=0.10\hMpc$. The small panels below $B_0$ and $T_0$ show the per-bin residuals
$(d-t)/\sigma$, with grey bands at $\pm1\sigma$ and $\pm2\sigma$. The trispectrum panel is plotted as $|T_0|$ on a logarithmic
axis, with filled (open) markers denoting $T_0>0$ ($T_0<0$). The fit is good, with
$\chi^2/{\rm dof}=3092/2719=1.14$, and the trispectrum is reproduced across roughly four orders of magnitude in $|T_0|$. Its
large per-configuration error bars ($\sigma_T\!\gtrsim\!|T_0|$ for many bins) reflect the low
signal-to-noise that limits its constraining power.
}
\label{fig:rsd_dvt}
\end{figure}
The clearest gain from the higher-order statistics is on the galaxy bias. From the power spectrum alone $b_2$ and $b_{\mathcal{G}_2}$ are prior-dominated, with $b_2=0.1\pm0.6$ and $b_{\mathcal{G}_2}=-0.45\pm0.36$. The bispectrum improves both by about an order of magnitude, to $b_2\simeq-0.83\pm0.09$ and $b_{\mathcal{G}_2}\simeq-0.50\pm0.06$ at the conservative cut. The linear bias $b_1$ stays well measured throughout at $b_1\simeq1.75\pm0.02$. This is similar to the real-space findings of Sec.~\ref{sec: results-real} and of \cite{Ivanov:2021kcd}.

The trispectrum adds a smaller cosmological gain, but one that grows with the scale cut. Its extra information is marginal at $\kmax^T=0.08\hMpc$, where the $\sigma_8$ error is unchanged, because the signal comes mostly from relatively higher $k$. It starts to help by $\kmax^T=0.10\hMpc$, with larger gains found for $\kmax^T=0.12\hMpc$. Writing the ratio of $P_\ell{+}B_0{+}T_0$ to $P_\ell{+}B_0$ marginal errors on $(\omega_{\rm cdm},h,\Omega_m,\sigma_8)$, we find $(0.98,0.98,0.99,1.00)$ at $0.08\hMpc$, $(0.92,0.95,0.94,1.01)$ at $0.10\hMpc$ and $(0.85,0.94,0.91,0.97)$ at $0.12\hMpc$. The trispectrum tightens the $\omega_{\rm cdm}$ errors the most, by up to about $15\%$, with a smaller effect on $\sigma_8$. This trend is similar to the results showed in \cite{Philcox:2022b1loop,Bakx:2025pop}, where the information gain of the one-loop bispectrum over the tree-level bispectrum was more significant for $\omega_{\rm cdm}$ than for $\sigma_8$.

Notably, our tree-level bispectrum analyses are unbiased to much higher $\kmax^B$ than in previous works \citep[e.g.,][]{Ivanov:2021kcd}. This is due to our improved treatment of the FoG effect, via the FoG counterterms $c_B$ and $c_T$ (Sec.~\ref{sec: fog}), as demonstrated in Appendix~\ref{app:fog}. These parameters have best-fit amplitudes of order unity, and marginalizing over them keeps $\sigma_8$ unbiased through $0.12\hMpc$. We note, however, that the halo catalogs used in \citep[e.g.,][]{Ivanov:2021kcd} were constructed with the \textsc{Rockstar} halo finder, whereas our analysis uses a different halo definition. Since we have not tested our FoG counterterms on those catalogs, it remains possible that their range of validity, and hence $\kmax^B$, would be lower for that sample.

Because the higher-order bias and counterterm parameters are marginalized analytically, we can still recover their posteriors after sampling, finding well-constrained FoG amplitudes $c_B=0.63\pm0.07$ and $c_T=0.73\pm0.07$ at $\kmax^{B}=\kmax^T=0.12\hMpc$, together with the cubic bias $b_{\Gamma_3}$ that the power spectrum leaves at its prior ($0.94\pm1.02$) but the bispectrum tightens more than five times ($0.78\pm0.19$) and the trispectrum by a further five percent ($0.71\pm0.18$).

For the same scale cuts, at $z=1.1$ we recover $b_{\Gamma_3}=0.73\pm0.17$, $b_3=0.20\pm0.30$, $b_{\mathcal{G}_3}=-0.61\pm0.15$, and $b_{\delta\mathcal{G}_2}=0.12\pm0.09$. The bias $b_{\delta\mathcal{G}_2}$ is fully consistent with its coevolution value, while $b_{\Gamma_3}$ lies $1.9\sigma$ above and $b_{\mathcal{G}_3}$ $2.4\sigma$ below their respective coevolution predictions. The largest deviation is seen for the cubic bias $b_3$, which lies about $6\sigma$ above the $b_3(b_1)$ fit relation. We explicitly checked that this insensitive to the choice of prior center, this shift reflects a genuine preference of the data rather than a prior effect. Such departures are not unexpected, since the coevolution and fit relations provide approximate predictions for the higher-order bias parameters, while the large effective volume considered here allows the data to resolve deviations from these relations.

Previous measurements of third-order halo bias parameters have been obtained using simulation-based techniques. In particular, $b_3$ has been accurately calibrated with separate-universe simulations \citep{Lazeyras:2015lgp}, while \citep{Lazeyras:2017hxw} and \citep{Abidi:2018eyd} measured the complete cubic-bias sector using compressed halo-matter-matter-matter trispectrum estimators. These methods use the knowledge of the initial density field and substantial cosmic-variance cancellation, and thus differ from an inference based only on late-time tracer clustering. Among analyses based on the latter, the one-loop power spectrum and bispectrum study of \citep{Eggemeier:2021cam} provided some of the strongest previous simultaneous constraints, although $b_3$, $b_{\mathcal G_3}$, and $b_{\delta\mathcal G_2}$ enter there only through loop corrections and remained weakly determined. In our $(P_{\text{1-loop}},B_{\text{tree}})$ analysis these three parameters do not enter the model, while the inclusion of the tree-level trispectrum gives significantly more precise measurements.

At $\kmax^B=0.14\hMpc$ the tree-level bispectrum breaks down, where the $P_\ell{+}B_0$ constrains on $\sigma_8$ are low by $1.2\sigma$ and on $\omega_{\rm cdm}$ by $1.5\sigma$, so we do not extend the trispectrum analysis to this scale cut. Extending the model to one-loop order and higher bispectrum and trispectrum multipoles is a natural next step that we leave to future work. 

Fig.~\ref{fig:rsd_dvt} compares the best-fit $P_\ell{+}B_0{+}T_0$ model with the measured data at $\kmax^B=\kmax^T=0.1\hMpc$, which we choose instead of $\kmax^B=\kmax^T=0.12\hMpc$ for visualization purposes. The model provides an accurate match to the data across all three statistics, with $\chi^2/{\rm dof}=3092/2719=1.14$. We discuss the detection significance of $B_0$ and $T_0$ in Appendix~\ref{app:detection}.

\begin{figure}
\centering
\includegraphics[width=\textwidth]{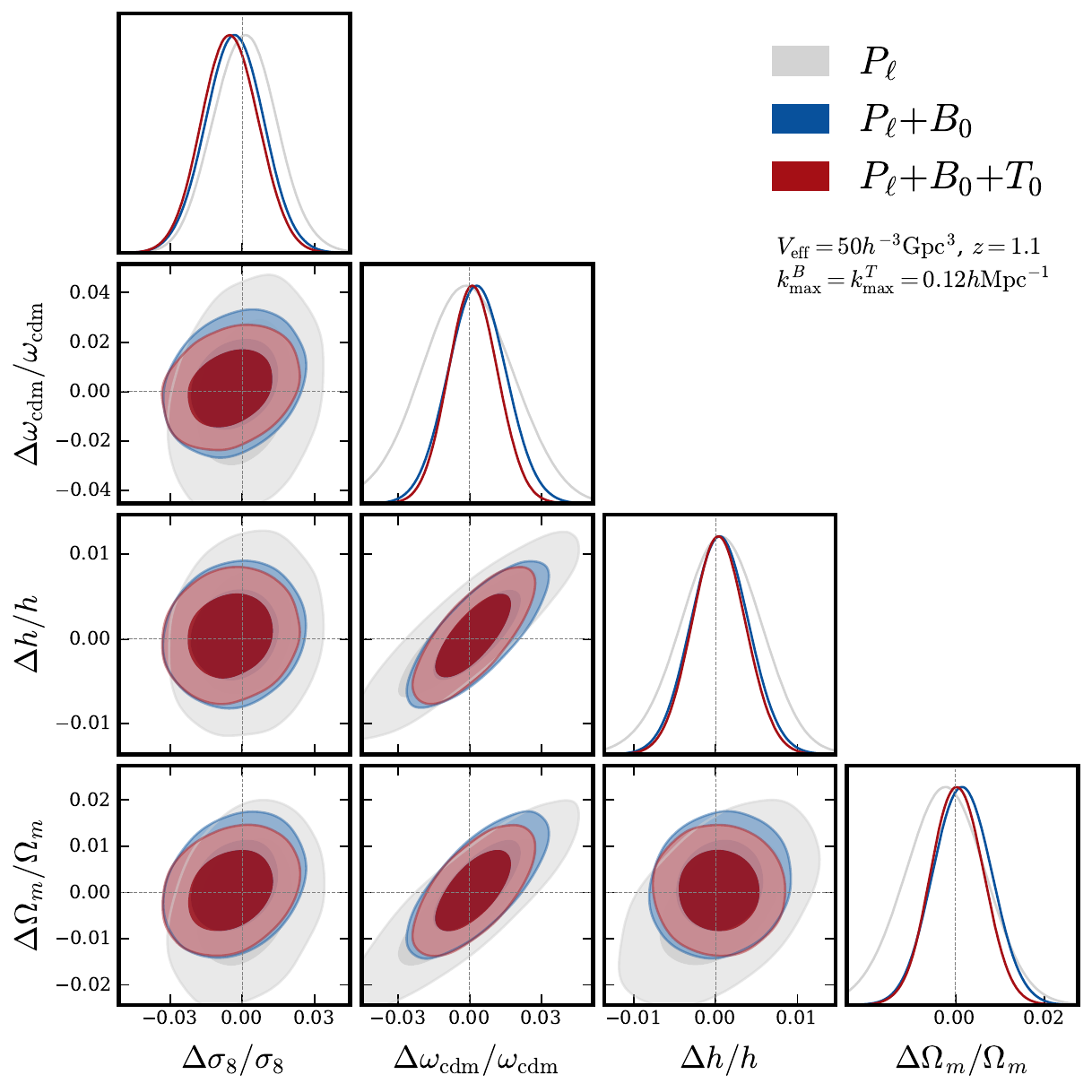}
\caption{
Marginalized cosmological constraints ($68\%$ and $95\%$) at the DESI-like volume $V_{\rm eff}=50\,h^{-3}{\rm Gpc}^3$ for the redshift-space \textsc{AbacusSummit} $z=1.1$ data at $\kmax^B=\kmax^T=0.12\hMpc$ and $\kmax^P=0.18\hMpc$ with the $N_A$-in-$B$ covariance. Parameters are shown as fractional deviations $\Delta\boldsymbol{\theta}/\boldsymbol{\theta}$ from their true values, with $P_\ell$ (grey), $P_\ell{+}B_0$ (blue) and $P_\ell{+}B_0{+}T_0$ (red). One-dimensional constraints are given in Tab.~\ref{tab:rsd_v50_z11}.
}
\label{fig:rsd_v50}
\end{figure}

\begin{figure}
\centering
\includegraphics[width=\textwidth]{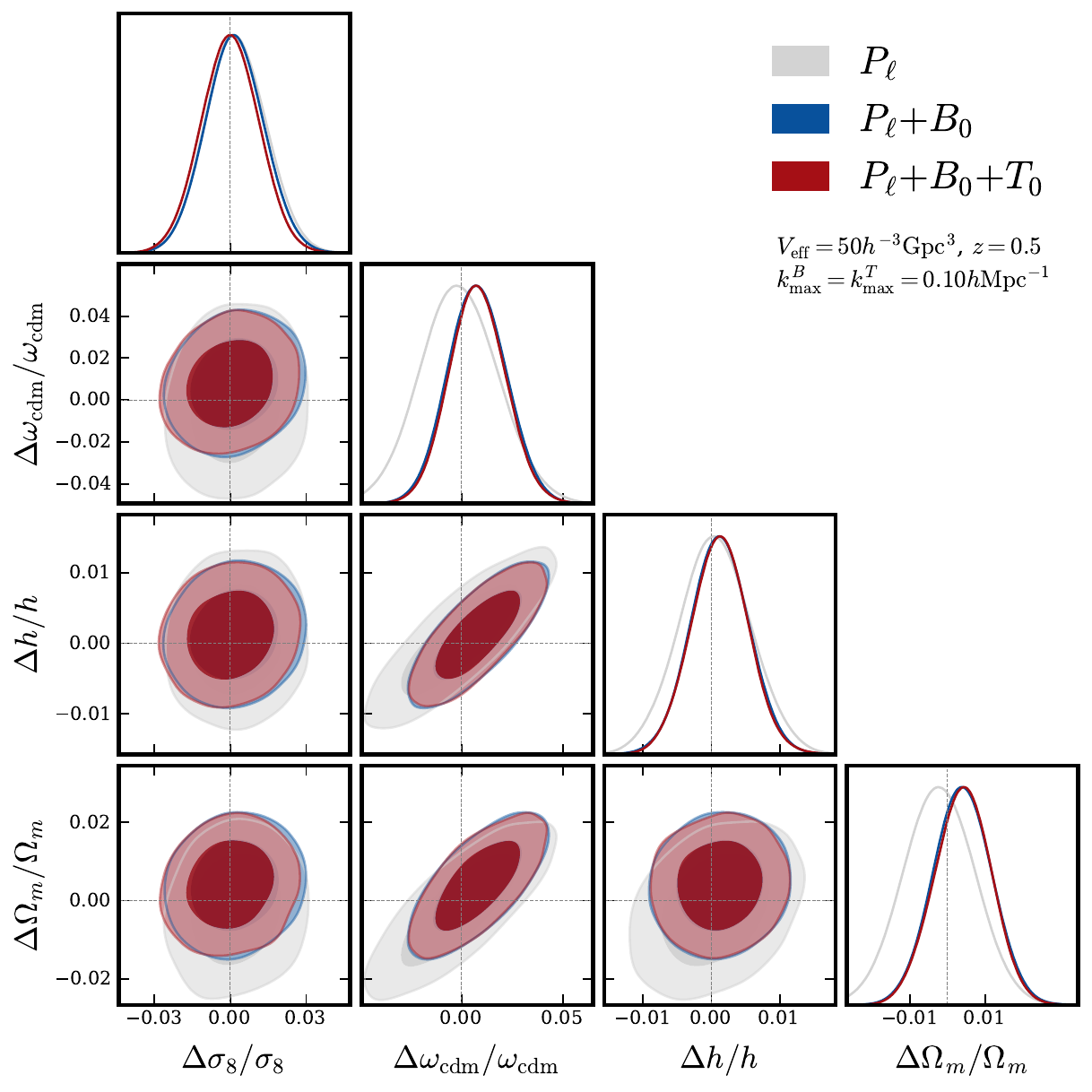}
\caption{
As Fig.~\ref{fig:rsd_v50} but for the \textsc{AbacusSummit} $z=0.5$ data at $\kmax^B=\kmax^T=0.10\hMpc$. One-dimensional constraints are given in Tab.~\ref{tab:rsd_v50_z05}.
}
\label{fig:rsd_v50_z05}
\end{figure}

\subsection{Implications for the DESI Survey}\label{sec: desi}

In this section, we rescale the covariance to an effective volume $V_{\rm eff}=50\,h^{-3}{\rm Gpc}^3$, close to that of a survey like DESI \cite{Aghamousa:2016zmz}, and repeat the analysis at both redshifts of the \textsc{AbacusSummit} sample. The data and the model are unchanged, so this isolates the effect of the survey-level errors. The detection significance of these samples is given in Appendix~\ref{app:detection}. Tabs.~\ref{tab:rsd_v50_z11}~\&~\ref{tab:rsd_v50_z05}, together with Figs.~\ref{fig:rsd_v50}~\&~\ref{fig:rsd_v50_z05}, show that the cosmological parameters are recovered unbiased up to $\kmax^B=\kmax^T=0.12\hMpc$ at $z=1.1$ and $0.10\hMpc$ at $z=0.5$. For $P_\ell{+}B_0{+}T_0$ the shift in the amplitude is $\Delta\sigma_8/\sigma_8=-0.50\%$, about $0.4\sigma$, at $z=1.1$, and at $-0.01\%$ at $z=0.5$, within $0.1\sigma$, with every other cosmological parameter within about $0.6\sigma$ of its true value.

\begin{table}[!htb]
\centering
{\small
\begin{tabular}{l|c|c|c}
\hline\hline
\multicolumn{4}{c}{$\kmax^B=\kmax^T=0.10\,\hMpc$}\\ \hline
Parameter & $P_\ell$ & $P_\ell{+}B_0$ & $P_\ell{+}B_0{+}T_0$ \\ \hline
$\Delta\sigma_8/\sigma_8$                & $+0.0012\pm0.0131$ & $+0.0010\pm0.0123$ & $+0.0001\pm0.0122$ \\
$\Delta\omega_{\rm cdm}/\omega_{\rm cdm}$ & $-0.0013\pm0.0194$ & $+0.0088\pm0.0143$ & $+0.0076\pm0.0135$ \\
$\Delta h/h$                             & $+0.0006\pm0.0049$ & $+0.0021\pm0.0039$ & $+0.0019\pm0.0038$ \\
$\Delta\Omega_m/\Omega_m$                & $-0.0024\pm0.0091$ & $+0.0032\pm0.0073$ & $+0.0026\pm0.0070$ \\ \hline
$b_1$                                    & $1.745\pm0.035$ & $1.751\pm0.028$ & $1.750\pm0.027$ \\
$b_2$                                    & $-0.15\pm0.84$ & $-0.82\pm0.10$ & $-0.783\pm0.091$ \\
$b_{\mathcal{G}_2}$                      & $-0.33\pm0.39$ & $-0.471\pm0.066$ & $-0.456\pm0.060$ \\
\hline\hline
\end{tabular}
}
\vspace{0.30cm}
{\small
\begin{tabular}{l|c|c|c}
\hline\hline
\multicolumn{4}{c}{$\kmax^B=\kmax^T=0.12\,\hMpc$}\\ \hline
Parameter & $P_\ell$ & $P_\ell{+}B_0$ & $P_\ell{+}B_0{+}T_0$ \\ \hline
$\Delta\sigma_8/\sigma_8$                & $+0.0012\pm0.0131$ & $-0.0032\pm0.0120$ & $-0.0050\pm0.0117$ \\
$\Delta\omega_{\rm cdm}/\omega_{\rm cdm}$ & $-0.0013\pm0.0194$ & $+0.0031\pm0.0121$ & $+0.0015\pm0.0104$ \\
$\Delta h/h$                             & $+0.0006\pm0.0049$ & $+0.0005\pm0.0035$ & $+0.0004\pm0.0033$ \\
$\Delta\Omega_m/\Omega_m$                & $-0.0024\pm0.0091$ & $+0.0016\pm0.0065$ & $+0.0005\pm0.0058$ \\ \hline
$b_1$                                    & $1.745\pm0.035$ & $1.752\pm0.027$ & $1.751\pm0.025$ \\
$b_2$                                    & $-0.15\pm0.84$ & $-0.767\pm0.079$ & $-0.737\pm0.069$ \\
$b_{\mathcal{G}_2}$                      & $-0.33\pm0.39$ & $-0.403\pm0.048$ & $-0.408\pm0.041$ \\
\hline\hline
\end{tabular}
}
\caption{One-dimensional marginalized constraints ($68\%$) from the redshift-space \textsc{AbacusSummit} $z=1.1$ data at the smaller effective volume $V_{\rm eff}=50\,h^{-3}{\rm Gpc}^3$, representative of a DESI-like survey, using $\kmax^P=0.18\hMpc$, at the two scale cuts $\kmax^B=\kmax^T=\{0.10,0.12\}\hMpc$. The columns and their normalization follow Tab.~\ref{tab:rsds}.}
\label{tab:rsd_v50_z11}
\end{table}

The two redshifts reach different scales. At $z=1.1$ the tree-level model stays unbiased to $0.12\hMpc$, while at $z=0.5$ we find a slight drift in parameters, with $P_\ell{+}B_0{+}T_0$ pulling $\sigma_8$ low by $-1.2\%$, about $1.2\sigma$, so we adopt $0.10\hMpc$ for that sample. This follows from the lower redshift of the $z=0.5$ box, where the FoG are stronger, so at a fixed wavenumber the tree-level description breaks down sooner.

\begin{table}[!htb]
\centering
{\small
\begin{tabular}{l|c|c|c}
\hline\hline
\multicolumn{4}{c}{$\kmax^B=\kmax^T=0.10\,\hMpc$}\\ \hline
Parameter & $P_\ell$ & $P_\ell{+}B_0$ & $P_\ell{+}B_0{+}T_0$ \\ \hline
$\Delta\sigma_8/\sigma_8$                & $+0.0022\pm0.0118$ & $+0.0017\pm0.0114$ & $-0.0000\pm0.0112$ \\
$\Delta\omega_{\rm cdm}/\omega_{\rm cdm}$ & $-0.0013\pm0.0192$ & $+0.0077\pm0.0143$ & $+0.0075\pm0.0135$ \\
$\Delta h/h$                             & $+0.0006\pm0.0052$ & $+0.0012\pm0.0042$ & $+0.0011\pm0.0041$ \\
$\Delta\Omega_m/\Omega_m$                & $-0.0022\pm0.0094$ & $+0.0040\pm0.0077$ & $+0.0040\pm0.0074$ \\ \hline
$b_1$                                    & $1.255\pm0.026$ & $1.253\pm0.021$ & $1.255\pm0.020$ \\
$b_2$                                    & $-0.28\pm0.47$ & $-0.835\pm0.059$ & $-0.809\pm0.053$ \\
$b_{\mathcal{G}_2}$                      & $-0.11\pm0.29$ & $-0.235\pm0.040$ & $-0.219\pm0.038$ \\
\hline\hline
\end{tabular}
}
\vspace{0.30cm}
{\small
\begin{tabular}{l|c|c|c}
\hline\hline
\multicolumn{4}{c}{$\kmax^B=\kmax^T=0.12\,\hMpc$}\\ \hline
Parameter & $P_\ell$ & $P_\ell{+}B_0$ & $P_\ell{+}B_0{+}T_0$ \\ \hline
$\Delta\sigma_8/\sigma_8$                & $+0.0022\pm0.0118$ & $-0.0049\pm0.0110$ & $-0.0122\pm0.0105$ \\
$\Delta\omega_{\rm cdm}/\omega_{\rm cdm}$ & $-0.0013\pm0.0192$ & $+0.0009\pm0.0119$ & $-0.0006\pm0.0103$ \\
$\Delta h/h$                             & $+0.0006\pm0.0052$ & $-0.0014\pm0.0038$ & $-0.0015\pm0.0035$ \\
$\Delta\Omega_m/\Omega_m$                & $-0.0022\pm0.0094$ & $+0.0035\pm0.0069$ & $+0.0025\pm0.0063$ \\ \hline
$b_1$                                    & $1.255\pm0.026$ & $1.257\pm0.019$ & $1.264\pm0.018$ \\
$b_2$                                    & $-0.28\pm0.47$ & $-0.793\pm0.045$ & $-0.753\pm0.039$ \\
$b_{\mathcal{G}_2}$                      & $-0.11\pm0.29$ & $-0.180\pm0.030$ & $-0.168\pm0.026$ \\
\hline\hline
\end{tabular}
}
\caption{As Tab.~\ref{tab:rsd_v50_z11} but for the \textsc{AbacusSummit} $z=0.5$ data ($V_{\rm eff}=50\,h^{-3}{\rm Gpc}^3$), at the two scale cuts $\kmax^B=\kmax^T=\{0.10,0.12\}\hMpc$.}
\label{tab:rsd_v50_z05}
\end{table}

At these cuts the ratio of $P_\ell{+}B_0{+}T_0$ to $P_\ell{+}B_0$ marginal errors on $(\omega_{\rm cdm},h,\Omega_m,\sigma_8)$ is $(0.86,0.94,0.89,0.98)$ for $z=1.1$ and $(0.97,1.00,0.97,0.99)$ for $z=0.5$. As at the full volume, the trispectrum tightens $\omega_{\rm cdm}$ the most, by up to about $14\%$ at $z=1.1$, and changes $\sigma_8$ only slightly.

After recovering the parameters that are marginalized analytically, we find that at the highest scale cut for each redshift $c_B=0.61\pm0.13$ and $c_T=0.70\pm0.14$ at $z=1.1$ ($\kmax^B=\kmax^T=0.12\hMpc$), and $c_B=1.18\pm0.22$ and $c_T=1.56\pm0.27$ at $z=0.5$ ($\kmax^B=\kmax^T=0.10\hMpc$). Both counterterms come out of order unity and are well inside their prior widths, and their recovered errors are more than an order of magnitude below the prior width. The recovered values are consistent with the fact that, at lower redshifts, we expect the FoG effects to become more important. 
For the same scale cuts, at $z=1.1$ we find $b_{\Gamma_3}=0.75\pm0.28$, $b_3=-0.24\pm0.49$, $b_{\mathcal{G}_3}=-0.54\pm0.28$ and $b_{\delta\mathcal{G}_2}=0.03\pm0.16$, while at $z=0.5$ we find $b_{\Gamma_3}=0.47\pm0.14$, $b_3=1.09\pm0.33$, $b_{\mathcal{G}_3}=-0.14\pm0.21$ and $b_{\delta\mathcal{G}_2}=-0.06\pm0.10$. The tidal biases $b_{\mathcal{G}_3}$ and $b_{\delta\mathcal{G}_2}$ are within about one standard deviation of their coevolution values at both redshifts, and $b_{\Gamma_3}$ is slightly above its coevolution prediction, by $1.2\sigma$ at $z=1.1$ and $2.3\sigma$ at $z=0.5$. The cubic bias $b_3$ shows the largest departure from its fit relation \cite{Lazeyras:2017hxw}, pulled up by about $3\sigma$ at $z=1.1$ and $2\sigma$ at $z=0.5$. It is also the least constrained of the four and retains the most prior information, and if we replace the prior center by the $-3b_2$ coevolution value the recovered $b_3$ is unchanged.

\section{Conclusions and Discussion}\label{sec: discussion}

In this work, we have built and validated the first complete tree-level EFTofLSS model of the galaxy trispectrum monopole, in both real and redshift space, including galaxy bias, stochasticity, infrared resummation, Fingers of God counterterms, Alcock-Paczynski distortions, binning, and covariance. We have run the joint $P_\ell{+}B_0{+}T_0$ likelihood on the \textsc{AbacusSummit} dark-matter halos, using 25 realizations to test our model against an effective volume of $200\,h^{-3}{\rm Gpc}^3$. We have also validated our covariance model using $15{,}000$ \textsc{Quijote} simulations. 

In real space, the bispectrum and trispectrum together recover $\sigma_8$ unbiased to a precision of $2\%$ up to $\kmax^B=\kmax^T=0.10\hMpc$ with $\kmax^P=0.20\hMpc$. In redshift-space, at the $200\,h^{-3}{\rm Gpc}^3$ volume we recover unbiased cosmology up to $\kmax^B=\kmax^T=0.12\hMpc$ for $z=1.1$, and at the DESI-like volume up to $0.12\hMpc$ ($0.1\hMpc$) for the $z=1.1$ ($z=0.5$) sample. At these volumes the trispectrum adds a modest amount of $\Lambda$CDM information beyond the power spectrum and bispectrum, tightening $\omega_{\rm cdm}$ by up to about $15\%$ and leaving $\sigma_8$ almost unchanged. At the large effective volume the statistical errors are tiny, so a sub-percent error in any single ingredient, whether the modeling, the binning, or the covariance, would have shown up as a bias in the recovered cosmology; getting all aspects of the pipeline to agree at once was the main challenge of this work.

Several extensions can follow naturally. The trispectrum should be measured with a window-free estimator, in the spirit of the quadratic and cubic estimators already built for the power spectrum and bispectrum \cite{Philcox:2020vbm,Philcox:2021ukg,Philcox:2024rqr}, to avoid the need to model the survey window function. A real-data analysis will also need to control observational systematics such as fiber collisions, along the lines of recent reanalyses of DESI data \cite{Chudaykin:2025aux}. The large dimensionality of the joint $P+B+T$ data vector will also make covariance estimation from a limited number of realistic mocks challenging. A complete likelihood analysis should furthermore systematically assess non-Gaussian covariance contributions, as discussed in Appendix~\ref{app:fisher}. This could be addressed either through suitable data compression or, if the covariance remains close to Gaussian, through large ensembles of inexpensive Gaussian surrogate fields with realistic survey geometry \cite{Hotinli:2025tul}. On the model side, the reach in $\kmax$ can be extended by moving from the tree-level bispectrum and trispectrum to a one-loop bispectrum \cite{Philcox:2022b1loop,DAmico:2022ukl,Bakx:2025pop} and a one-loop trispectrum, and we plan to add higher bispectrum and trispectrum multipoles into the analysis. An interesting comparison would also be with field-level inference and alternative forward-modeling approaches to higher-order statistics, such as \textsc{LEFTfield}, along the lines of \cite{Beyond-2pt:2024mqz,Nguyen:2024yth,Stadler:2026ufu}.

The fact that the tree-level trispectrum returns unbiased cosmology on realistic halo samples encourages us to apply it to real galaxy data, further compelled by the high detection signifcance forecasted in Appendix~\ref{app:detection}. We are particularly interested in using it to search for primordial non-Gaussianity of the $\tau_{\rm NL}$ and $g_{\rm NL}$ type and for cosmological collider signatures, which leave distinguishable imprints on the galaxy four-point function. The model developed in this work is therefore the first step toward turning the galaxy trispectrum into a competitive probe of inflation.

\paragraph{Acknowledgments} 
We would like to thank Thomas Bakx, Sam Goldstein, Misha Ivanov, Nick Kokron, Noah Sailer and Fabian Schmidt for valuable discussions, and the CMBxLSS group at Stanford for suggesting the bird analogy to our trispectrum estimator.
OHEP is supported in part by the National Science Foundation under grant
number PHY-2609396.
Some of the computing for this project was performed on the Sherlock cluster. We would like to thank Stanford University and Stanford Research Computing for providing computational resources and support that contributed to these research results.
\appendix 

\section{Perturbation Theory Kernels}\label{app:kernels}

\paragraph{Real Space.}\label{app:kernels-real}
The real-space perturbation theory kernels at third order in Eq.~\eqref{eq: delta-n-exp} are given by \cite{Desjacques:2016bnm,Chudaykin:2020aoj}
\beq\label{eq: real-kernels}
    K_1(\vq_1) &=& b_1 , \\\nonumber
    K_2(\vq_1,\vq_2) &=& b_1 F_2(\vq_1,\vq_2) + \frac{b_2}{2} + b_{\mathcal{G}_2}\kappa(\vq_1,\vq_2),
    \\\nonumber
    K_3(\vq_1,\vq_2,\vq_3)&=& b_1 F_3(\vq_1,\vq_2,\vq_3) + b_2F_2(\vq_1,\vq_2) + 2 b_{\mathcal{G}_2} \kappa(\vq_{12},\vq_3) F_2(\vq_1,\vq_2) 
    \\\nonumber
    &&\,+ \frac{b_3}{6}-\frac{4b_{\Gamma_3}}{7}\kappa(\vq_1,\vq_2)\kappa(\vq_{12},\vq_3) -\frac{b_{\mathcal{G}_3}}{2} L(\vq_1,\vq_2,\vq_3)  + b_{\delta\mathcal{G}_2}\kappa(\vq_1,\vq_2),
\eeq
where $F_n$ and $G_n$ are the standard Eulerian density and velocity kernels \cite{Bernardeau:2001qr}, $\vq_{i\cdots j}\equiv \vq_i+\cdots+\vq_j$ and we define the angles
\beq\label{eq: ang-def}
    \kappa(\vq_1,\vq_2) &=& (\hq_1\cdot\hq_2)^2-1 , \\\nonumber
    L(\vq_1,\vq_2,\vq_3) &=& 2(\hq_1\cdot\hq_2)(\hq_2\cdot\hq_3)(\hq_3\cdot\hq_1)-(\hq_1\cdot\hq_2)^2-(\hq_2\cdot\hq_3)^2-(\hq_3\cdot\hq_1)^2+1.
\eeq
The kernels have to be further symmetrized over their arguments. The same kernels appear in the one-loop galaxy power spectrum, though the dependence on $b_3$, $b_{\mathcal{G}_3}$ and $b_{\delta\mathcal{G}_2}$ is there removed by renormalization \cite{Assassi:2014fva}.

\paragraph{Redshift Space.} \label{app:kernels-rsd} After expanding the RSD mapping in Eq.~\eqref{eq: rsd-mapping}, we obtain the third-order redshift-space kernels \citep[cf.][]{Bernardeau:2001qr,Desjacques:2016bnm,Chudaykin:2020aoj}:
\be
\begin{aligned}\label{eq: RSD-kernels}
Z_1(\mathbf{q}_1) &= K_1 + f\mu_1^2, \\[6pt]
Z_2(\mathbf{q}_1,\mathbf{q}_2) &= K_2(\mathbf{q}_1,\mathbf{q}_2) + f\mu_{12}^2 G_2(\mathbf{q}_1,\mathbf{q}_2)
+ \frac{f\mu_{12}q_{12}}{2} K_1 \left[\frac{\mu_1}{q_1} + \frac{\mu_2}{q_2}\right]
+ \frac{(f\mu_{12}q_{12})^2}{2}\frac{\mu_1}{q_1}\frac{\mu_2}{q_2}, \\[6pt]
Z_3(\mathbf{q}_1,\mathbf{q}_2,\mathbf{q}_3) &= K_3(\mathbf{q}_1,\mathbf{q}_2,\mathbf{q}_3) + f\mu_{123}^2 G_3(\mathbf{q}_1,\mathbf{q}_2,\mathbf{q}_3) \notag\\
&\quad + (f\mu_{123}q_{123})
\left[
\frac{\mu_{12}}{q_{12}} K_1 G_2(\mathbf{q}_1,\mathbf{q}_2)
+ \frac{\mu_3}{q_3} K_2(\mathbf{q}_1,\mathbf{q}_2)
\right] \notag\\
&\quad + \frac{(f\mu_{123}q_{123})^2}{2}
\left[
2\frac{\mu_{12}}{q_{12}}\frac{\mu_3}{q_3} G_2(\mathbf{q}_1,\mathbf{q}_2)
+ \frac{\mu_1}{q_1}\frac{\mu_2}{q_2} K_1
\right] \notag\\
&\quad + \frac{(f\mu_{123}q_{123})^3}{6}
\frac{\mu_1}{q_1}\frac{\mu_2}{q_2}\frac{\mu_3}{q_3},
\end{aligned}
\ee
where $\mu_{i\cdots j}\equiv \mu_{\vq_i+\cdots+\vq_j}$.

\section{Stochastic Contributions}
\label{app:stoch}

In this appendix we derive the stochastic trispectrum quoted in
Eq.~\eqref{eq:stoch_pieces}. We work at leading order in derivatives of the
stochastic fields and remove the momentum-conserving Dirac delta from all
correlators, as denoted by a prime. To simplify the notation, we define
\begin{equation}
\delta_{\rm det}^{(n)}(\mathbf{k})\equiv
\delta_{g,\rm det}^{(s,n)}(\mathbf{k}),
\qquad
\delta_{\rm stoch}^{(n)}(\mathbf{k})\equiv
\delta_{g,\rm stoch}^{(s,n)}(\mathbf{k}),
\label{eq:stoch_short_notation}
\end{equation}
\textit{i.e.}\ omitting the $(s)$ and $g$ labels from the redshift-space galaxy densities. When the perturbative-order superscript is omitted, $\delta_{\rm det}$ and
$\delta_{\rm stoch}$ denote the full deterministic and stochastic fields.

The deterministic redshift-space density needed below is
\begin{equation}
\delta_{\rm det}^{(1)}(\mathbf{k})
=Z_1(\mathbf{k})\delta^{(1)}(\mathbf{k}),\quad
\delta_{\rm det}^{(2)}(\mathbf{k})
=\int_{\mathbf q}
Z_2(\mathbf q,\mathbf k-\mathbf q)
\delta^{(1)}(\mathbf q)\delta^{(1)}(\mathbf k-\mathbf q),
\label{eq:stoch_det_expansion}
\end{equation}
where $\int_{\mathbf q}\equiv\int d^3q/(2\pi)^3$. The stochastic density is
split according to its order in the linear density \cite{Senatore:2014eva,Desjacques:2016bnm},
\begin{equation}
\delta_{g,\rm stoch}^{(s)}
=\delta_{\rm stoch}^{(0)}
+\delta_{\rm stoch}^{(1)}
+\delta_{\rm stoch}^{(2)},
\label{eq:stoch_order_expansion}
\end{equation}
with
\begin{equation}
\begin{aligned}
\delta_{\rm stoch}^{(0)}(\mathbf{k})={}&\epsilon(\mathbf{k}),\\
\delta_{\rm stoch}^{(1)}(\mathbf{k})={}&
[\epsilon_\delta\delta^{(1)}]_{\mathbf{k}}
-ifk_z[\epsilon u_z]_{\mathbf{k}},\\
\delta_{\rm stoch}^{(2)}(\mathbf{k})={}&
[\epsilon_\delta\delta^{(2)}]_{\mathbf{k}}
-ifk_z[\epsilon u_z^{(2)}]_{\mathbf{k}}
+\frac{1}{2}[\epsilon_{\delta^2}(\delta^{(1)})^2]_{\mathbf{k}}
+[\epsilon_{\mathcal G_2}\mathcal G_2]_{\mathbf{k}}
\\
&-ifk_z[\epsilon_\delta\delta^{(1)}u_z]_{\mathbf{k}}
-\frac{1}{2}f^2k_z^2[\epsilon u_z^2]_{\mathbf{k}}.
\end{aligned}
\label{eq:stoch_field_app}
\end{equation}
Here $u_z(\mathbf q)=i\mu_{\mathbf q}\delta^{(1)}(\mathbf q)/q$ at linear
order and the stochastic fields are uncorrelated with $\delta^{(1)}$ by definition.

We define the white-noise stochastic correlators directly in terms of the amplitudes
used in the likelihood,
\begin{equation}
\begin{aligned}
\langle\epsilon\epsilon\rangle'
&=\frac{1+P_{\rm shot}}{\bar n},
&
\langle\epsilon\epsilon_\delta\rangle'
&=\frac{b_1B_{\rm shot}}{2\bar n},
&
\langle\epsilon\epsilon\epsilon\rangle'_c
&=\frac{A_{\rm shot}}{\bar n^2},
\\
\langle\epsilon\epsilon_{\delta^2}\rangle'
+\langle\epsilon_\delta\epsilon_\delta\rangle'
&=\frac{b_2C_{\rm shot}}{2\bar n},
&
\langle\epsilon\epsilon_{\mathcal G_2}\rangle'
&=\frac{b_{\mathcal G_2}D_{\rm shot}}{2\bar n},
&
\langle\epsilon\epsilon\epsilon_\delta\rangle'_c
&=\frac{b_1E_{\rm shot}}{3\bar n^2},
\\
&&
\langle\epsilon\epsilon\epsilon\epsilon\rangle'_c
&=\frac{F_{\rm shot}}{\bar n^3}.
\end{aligned}
\label{eq:shot_correlators_app}
\end{equation}
With this convention the Poisson values are $A_{\rm shot}=\cdots=F_{\rm shot}=1$ and $P_{\rm shot}=0$.

The stochastic trispectrum can be organized by the number of stochastic
external legs,
\begin{equation}
\begin{aligned}
T_{\rm stoch}={}&
\langle
\delta_{g,\rm det}^{(s)}
\delta_{g,\rm det}^{(s)}
\delta_{g,\rm stoch}^{(s)}
\delta_{g,\rm stoch}^{(s)}
\rangle'
+\text{5 perm.}
\\
&+\langle
\delta_{g,\rm det}^{(s)}
\delta_{g,\rm stoch}^{(s)}
\delta_{g,\rm stoch}^{(s)}
\delta_{g,\rm stoch}^{(s)}
\rangle'
+\text{3 perm.}
\\
&+\langle
\delta_{g,\rm stoch}^{(s)}
\delta_{g,\rm stoch}^{(s)}
\delta_{g,\rm stoch}^{(s)}
\delta_{g,\rm stoch}^{(s)}
\rangle'
\\
\equiv{}&T_{\rm stoch}^{(\bar n^{-1})}
+T_{\rm stoch}^{(\bar n^{-2})}
+T_{\rm stoch}^{(\bar n^{-3})},
\end{aligned}
\label{eq:stoch_decomposition_app}
\end{equation}
and we derive each of these terms in the following.

\paragraph{Two Stochastic Legs.}
Consider the first line in Eq.~\eqref{eq:stoch_decomposition_app}, where the first-order deterministic legs depend on $\mathbf{k}_1,\mathbf{k}_2$ and the first-order stochastic legs depend on $\mathbf{k}_3,\mathbf{k}_4$. At tree level,
\begin{equation}
\begin{aligned}
&\langle
\delta_{\rm det}(\mathbf{k}_1)
\delta_{\rm det}(\mathbf{k}_2)
\delta_{\rm stoch}(\mathbf{k}_3)
\delta_{\rm stoch}(\mathbf{k}_4)
\rangle'
\\
&={}
\langle
\delta_{\rm det}^{(1)}(\mathbf{k}_1)
\delta_{\rm det}^{(1)}(\mathbf{k}_2)
\delta_{\rm stoch}^{(1)}(\mathbf{k}_3)
\delta_{\rm stoch}^{(1)}(\mathbf{k}_4)
\rangle'
\\
&+\langle
\delta_{\rm det}^{(1)}(\mathbf{k}_1)
\delta_{\rm det}^{(2)}(\mathbf{k}_2)
\delta_{\rm stoch}^{(0)}(\mathbf{k}_3)
\delta_{\rm stoch}^{(1)}(\mathbf{k}_4)
\rangle'
+(1\leftrightarrow2)+(3\leftrightarrow4)
\\
&+\langle
\delta_{\rm det}^{(1)}(\mathbf{k}_1)
\delta_{\rm det}^{(1)}(\mathbf{k}_2)
\delta_{\rm stoch}^{(0)}(\mathbf{k}_3)
\delta_{\rm stoch}^{(2)}(\mathbf{k}_4)
\rangle'
+(3\leftrightarrow4)
,
\end{aligned}
\label{eq:two_stoch_tree_split}
\end{equation}
where the $(i\leftrightarrow j)$ symbols correspond to permutating the arguments ($\mathbf{k}_i\leftrightarrow\mathbf{k}_j$). Starting with the second line of Eq.~\eqref{eq:two_stoch_tree_split}, we have
\begin{equation}
\begin{aligned}
&\langle
\delta_{\rm det}^{(1)}(\mathbf{k}_1)
\delta_{\rm det}^{(1)}(\mathbf{k}_2)
\delta_{\rm stoch}^{(1)}(\mathbf{k}_3)
\delta_{\rm stoch}^{(1)}(\mathbf{k}_4)
\rangle'
\\
&=\langle
Z_1(\mathbf{k}_1)\delta^{(1)}(\mathbf{k}_1)
Z_1(\mathbf{k}_2)\delta^{(1)}(\mathbf{k}_2)
([
\epsilon_\delta\delta^{(1)}
]_{\mathbf{k}_3}-ifk_{3z}[\epsilon u_z]_{\mathbf{k}_3})
([
\epsilon_\delta\delta^{(1)}
]_{\mathbf{k}_4}-ifk_{4z}[\epsilon u_z]_{\mathbf{k}_4})
\rangle'
\\
&=
\frac{Z_1(\mathbf{k}_1)Z_1(\mathbf{k}_2)P(k_1)P(k_2)}{\bar n}
\Big[
2\langle\epsilon_\delta\epsilon_\delta\rangle'
+\frac{b_1B_{\rm shot}}{2}fR_{12}
+2f^2(1+P_{\rm shot})
\frac{k_3k_4}{k_1k_2}\mu_1\mu_2\mu_3\mu_4
\Big],
\end{aligned}
\label{eq:two_stoch_first_first_piece}
\end{equation}
where we defined
\begin{equation}
R_{12}\equiv
\mu_1^2+\mu_2^2+
\left(\frac{k_1}{k_2}+\frac{k_2}{k_1}\right)\mu_1\mu_2.
\label{eq:R12_app}
\end{equation}
Next, we compute the third line of Eq.~\eqref{eq:two_stoch_tree_split}:
\begin{equation}
\begin{aligned}
&\langle
\delta_{\rm det}^{(1)}(\mathbf{k}_1)
\delta_{\rm det}^{(2)}(\mathbf{k}_2)
\delta_{\rm stoch}^{(0)}(\mathbf{k}_3)
\delta_{\rm stoch}^{(1)}(\mathbf{k}_4)
\rangle'
+(1\leftrightarrow2)+(3\leftrightarrow4)
\\
&=\langle
Z_1(\mathbf{k}_1)\delta^{(1)}(\mathbf{k}_1)
\delta_{\rm det}^{(2)}(\mathbf{k}_2)
\epsilon(\mathbf{k}_3)
([\epsilon_\delta\delta^{(1)}]_{\mathbf{k}_4}
-ifk_{4z}[\epsilon u_z]_{\mathbf{k}_4})
\rangle'
+(1\leftrightarrow2)+(3\leftrightarrow4)
\\
&=
\frac{P(k_{12})}{\bar n}
[2b_1B_{\rm shot}+2f(1+P_{\rm shot})\mu_{12}^2]
[
Z_1(\mathbf{k}_1)Z_2(\mathbf{k}_1,-\mathbf{k}_{12})P(k_1)
\\
&\hspace{6.5cm}
+Z_1(\mathbf{k}_2)Z_2(\mathbf{k}_2,-\mathbf{k}_{12})P(k_2)
].
\end{aligned}
\label{eq:two_stoch_Z2_piece}
\end{equation}
Finally, we decompose the last line of Eq.~\eqref{eq:two_stoch_tree_split} into the $\delta_{\rm stoch}^{(2)}$ contributions given in Eq.~\eqref{eq:stoch_field_app}. Taking first into account the $[\epsilon\delta^{(2)}]$ and $[\epsilon u_z^{(2)}]$ pieces, we have
\begin{equation}
\begin{aligned}
&\langle
\delta_{\rm det}^{(1)}(\mathbf{k}_1)
\delta_{\rm det}^{(1)}(\mathbf{k}_2)
\delta_{\rm stoch}^{(0)}(\mathbf{k}_3)
\delta_{\rm stoch}^{(2)}(\mathbf{k}_4)
\rangle'_{\delta^{(2)},u_z^{(2)}}
+(3\leftrightarrow4)
\\
&=\langle
Z_1(\mathbf{k}_1)\delta^{(1)}(\mathbf{k}_1)
Z_1(\mathbf{k}_2)\delta^{(1)}(\mathbf{k}_2)
\epsilon(\mathbf{k}_3)
([\epsilon_\delta\delta^{(2)}]_{\mathbf{k}_4}
-ifk_{4z}[\epsilon u_z^{(2)}]_{\mathbf{k}_4}
)
\rangle'
+(3\leftrightarrow4)
\\
&=
\frac{Z_1(\mathbf{k}_1)Z_1(\mathbf{k}_2)P(k_1)P(k_2)}{\bar n}
[
2b_1B_{\rm shot}F_2(\mathbf{k}_1,\mathbf{k}_2)
+2f(1+P_{\rm shot})\mu_{12}^2G_2(\mathbf{k}_1,\mathbf{k}_2)
].
\end{aligned}
\label{eq:two_stoch_delta2_piece}
\end{equation}
The $[\epsilon_{\delta^2}\delta^2]$ and $[\epsilon_{\mathcal G_2}\mathcal G_2]$ contributions give
\begin{equation}
\begin{aligned}
&\langle
\delta_{\rm det}^{(1)}(\mathbf{k}_1)
\delta_{\rm det}^{(1)}(\mathbf{k}_2)
\delta_{\rm stoch}^{(0)}(\mathbf{k}_3)
\delta_{\rm stoch}^{(2)}(\mathbf{k}_4)
\rangle'_{\delta^2,\mathcal G_2}
+(3\leftrightarrow4)
\\
&=\langle
Z_1(\mathbf{k}_1)\delta^{(1)}(\mathbf{k}_1)
Z_1(\mathbf{k}_2)\delta^{(1)}(\mathbf{k}_2)
\epsilon(\mathbf{k}_3)
[
\tfrac12[\epsilon_{\delta^2}(\delta^{(1)})^2]_{\mathbf{k}_4}
+[\epsilon_{\mathcal G_2}\mathcal G_2]_{\mathbf{k}_4}
]
\rangle'
+(3\leftrightarrow4)
\\
&=
\frac{Z_1(\mathbf{k}_1)Z_1(\mathbf{k}_2)P(k_1)P(k_2)}{\bar n}
[
2\langle\epsilon\epsilon_{\delta^2}\rangle'
+2b_{\mathcal G_2}D_{\rm shot}\,
\kappa(\mathbf{k}_1,\mathbf{k}_2)
].
\end{aligned}
\label{eq:two_stoch_local_piece}
\end{equation}
The remaining contribution from $\delta_{\rm stoch}^{(2)}$ follows from $[\epsilon_\delta\delta u_z]$ and
$[\epsilon u_z^2]$,
\begin{equation}
\begin{aligned}
&\langle
\delta_{\rm det}^{(1)}(\mathbf{k}_1)
\delta_{\rm det}^{(1)}(\mathbf{k}_2)
\delta_{\rm stoch}^{(0)}(\mathbf{k}_3)
\delta_{\rm stoch}^{(2)}(\mathbf{k}_4)
\rangle'_{\delta u_z,u_z^2}
+(3\leftrightarrow4)
\\
&=\langle
Z_1(\mathbf{k}_1)\delta^{(1)}(\mathbf{k}_1)
Z_1(\mathbf{k}_2)\delta^{(1)}(\mathbf{k}_2)
\epsilon(\mathbf{k}_3)
[
-ifk_{4z}[\epsilon_\delta\delta^{(1)}u_z]_{\mathbf{k}_4}
-\tfrac12f^2k_{4z}^2[\epsilon u_z^2]_{\mathbf{k}_4}
]
\rangle'
+(3\leftrightarrow4)
\\
&=
\frac{Z_1(\mathbf{k}_1)Z_1(\mathbf{k}_2)P(k_1)P(k_2)}{\bar n}
\Big[
\frac{b_1B_{\rm shot}}{2}fR_{12}
+f^2(1+P_{\rm shot})
\frac{\mu_1\mu_2}{k_1k_2}
(k_3^2\mu_3^2+k_4^2\mu_4^2)
\Big].
\end{aligned}
\label{eq:two_stoch_mapping_second_piece}
\end{equation}
Using momentum conservation, $k_3\mu_3+k_4\mu_4=-k_{12}\mu_{12}$, thus the two $f^2$ terms in
Eqs.~\eqref{eq:two_stoch_mapping_second_piece} and
\eqref{eq:two_stoch_first_first_piece} combine as
\begin{equation}
\frac{\mu_1\mu_2}{k_1k_2}
(k_3^2\mu_3^2+k_4^2\mu_4^2+2k_3k_4\mu_3\mu_4)
=
\frac{k_{12}^2}{k_1k_2}\mu_1\mu_2\mu_{12}^2.
\label{eq:stoch_velocity_identity_app}
\end{equation}
Similarly, the two terms proportional to $R_{12}$ sum to
$b_1B_{\rm shot}fR_{12}$. The constant pieces in Eq.~\eqref{eq:two_stoch_first_first_piece} and Eq.~\eqref{eq:two_stoch_local_piece} combine according to
Eq.~\eqref{eq:shot_correlators_app},
\begin{equation}
2[
\langle\epsilon_\delta\epsilon_\delta\rangle'+
\langle\epsilon\epsilon_{\delta^2}\rangle'
]=\frac{b_2C_{\rm shot}}{\bar n}.
\end{equation}
Collecting all contributions gives
\begin{equation}
\begin{aligned}
&\langle
\delta_{\rm det}(\mathbf{k}_1)
\delta_{\rm det}(\mathbf{k}_2)
\delta_{\rm stoch}(\mathbf{k}_3)
\delta_{\rm stoch}(\mathbf{k}_4)
\rangle'
\\
&=
\frac{Z_1(\mathbf{k}_1)Z_1(\mathbf{k}_2)P(k_1)P(k_2)}{\bar n}
\Big[
b_1B_{\rm shot}
(2F_2(\mathbf{k}_1,\mathbf{k}_2)+fR_{12})
+b_2C_{\rm shot}
+2b_{\mathcal G_2}D_{\rm shot}\,
\kappa(\mathbf{k}_1,\mathbf{k}_2)
\\
&\hspace{5cm}+(1+P_{\rm shot})
(2f\mu_{12}^2G_2(\mathbf{k}_1,\mathbf{k}_2)
+f^2\frac{k_{12}^2}{k_1k_2}\mu_1\mu_2\mu_{12}^2)
\Big]
\\
&+\frac{P(k_{12})}{\bar n}
[2b_1B_{\rm shot}+2f(1+P_{\rm shot})\mu_{12}^2]
\\
&\hspace{1.8cm}\times[
Z_1(\mathbf{k}_1)Z_2(\mathbf{k}_1,-\mathbf{k}_{12})P(k_1)
+Z_1(\mathbf{k}_2)Z_2(\mathbf{k}_2,-\mathbf{k}_{12})P(k_2)
].
\end{aligned}
\label{eq:two_stoch_channel_final}
\end{equation}
The full $\bar n^{-1}$ contribution is obtained by adding the other five
permutations from Eq.~\eqref{eq:stoch_decomposition_app}.

\paragraph{Three Stochastic Legs.}
For a deterministic external leg $\mathbf{k}_1$, only the first-order
stochastic density contributes at tree-level,
\begin{equation}
\begin{aligned}
&\langle
\delta_{\rm det}^{(1)}(\mathbf{k}_1)
\delta_{\rm stoch}(\mathbf{k}_2)
\delta_{\rm stoch}(\mathbf{k}_3)
\delta_{\rm stoch}(\mathbf{k}_4)
\rangle'
\\
&=
\langle
\delta_{\rm det}^{(1)}(\mathbf{k}_1)
\delta_{\rm stoch}^{(0)}(\mathbf{k}_2)
\delta_{\rm stoch}^{(0)}(\mathbf{k}_3)
\delta_{\rm stoch}^{(1)}(\mathbf{k}_4)
\rangle'
+(2\leftrightarrow4)+(3\leftrightarrow4)
\\
&=Z_1(\mathbf{k}_1)
\langle
\delta^{(1)}(\mathbf{k}_1)
\epsilon(\mathbf{k}_2)
\epsilon(\mathbf{k}_3)
([
\epsilon_\delta\delta^{(1)}
]_{\mathbf{k}_4}-ifk_{4z}[\epsilon u_z]_{\mathbf{k}_4})
\rangle'
+(2\leftrightarrow4)+(3\leftrightarrow4)
\\
&=
\frac{Z_1(\mathbf{k}_1)P(k_1)}{\bar n^2}
[b_1E_{\rm shot}+f\mu_1^2A_{\rm shot}].
\end{aligned}
\label{eq:three_stoch_explicit}
\end{equation}
Adding the four permutations from Eq.~\eqref{eq:stoch_decomposition_app} gives the first $\bar n^{-2}$ term.

\paragraph{Four Stochastic Legs.}
The connected four-point function of the zeroth-order stochastic density gives
immediately
\begin{equation}
\begin{aligned}
&\langle
\delta_{\rm stoch}^{(0)}(\mathbf{k}_1)
\delta_{\rm stoch}^{(0)}(\mathbf{k}_2)
\delta_{\rm stoch}^{(0)}(\mathbf{k}_3)
\delta_{\rm stoch}^{(0)}(\mathbf{k}_4)
\rangle'_c
=\langle
\epsilon(\mathbf{k}_1)\epsilon(\mathbf{k}_2)
\epsilon(\mathbf{k}_3)\epsilon(\mathbf{k}_4)
\rangle'_c
=\frac{F_{\rm shot}}{\bar n^3}.
\end{aligned}
\label{eq:four_stoch_white_piece}
\end{equation}
At order $\bar n^{-2}P$, 
\begin{equation}
\begin{aligned}
&\langle
\delta_{\rm stoch}^{(0)}(\mathbf{k}_1)
\delta_{\rm stoch}^{(0)}(\mathbf{k}_2)
\delta_{\rm stoch}^{(1)}(\mathbf{k}_3)
\delta_{\rm stoch}^{(1)}(\mathbf{k}_4)
\rangle'
+\text{5 perm.}
\\
&=\langle
\epsilon(\mathbf{k}_1)\epsilon(\mathbf{k}_2)
([
\epsilon_\delta\delta^{(1)}
]_{\mathbf{k}_3}-ifk_{3z}[\epsilon u_z]_{\mathbf{k}_3})
([
\epsilon_\delta\delta^{(1)}
]_{\mathbf{k}_4}-ifk_{4z}[\epsilon u_z]_{\mathbf{k}_4})
\rangle'
+\text{5 perm.}
\end{aligned}
\label{eq:four_stoch_P_explicit}
\end{equation}
This contains the two-point contractions
$\langle\epsilon\epsilon_\delta\rangle^2$,
$\langle\epsilon\epsilon_\delta\rangle\langle\epsilon\epsilon\rangle$,
and $\langle\epsilon\epsilon\rangle^2$. For a given diagonal
$\mathbf{k}_{ij}=\mathbf{k}_i+\mathbf{k}_j$, these terms are
\begin{equation}
\frac{1}{\bar n^2}
\sum_{ij\in\{12,13,14\}}
P(k_{ij})
[b_1B_{\rm shot}+f(1+P_{\rm shot})\mu_{ij}^2]^2.
\label{eq:four_stoch_P_final}
\end{equation}

\paragraph{Final Result.}
Combining the two-, three-, and four-stochastic pieces contributions, the stochastic trispectrum is given in Eq.~\eqref{eq:stoch_pieces}. The real-space expression follows by setting $f=0$ and replacing $Z_n\rightarrow K_n$.

\begin{figure}
\centering
\includegraphics[width=\textwidth]{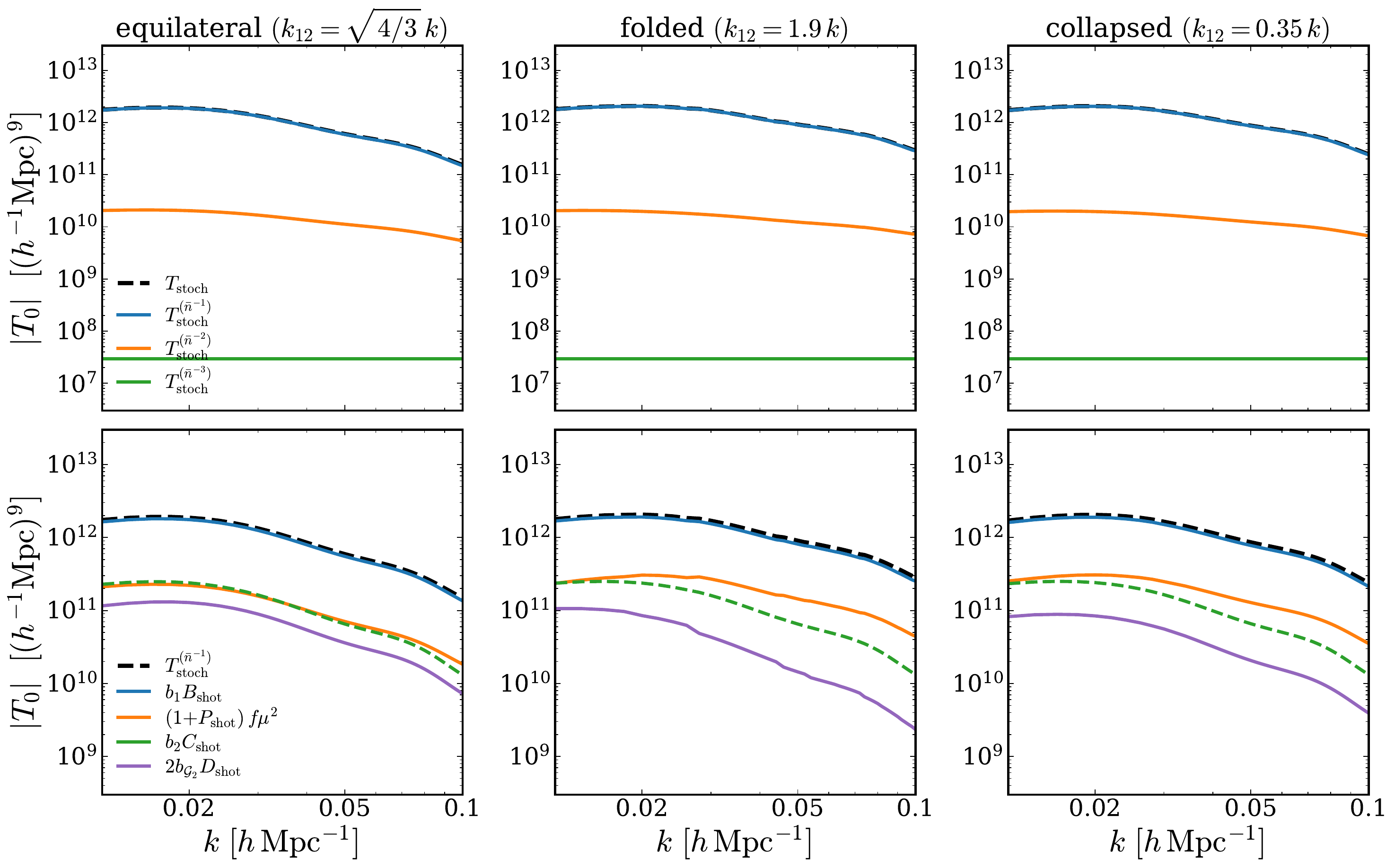}
\caption{Decomposition of the tree-level stochastic trispectrum monopole
$T_{\rm stoch}$ for the equilateral, folded and collapsed shapes (left to
right), at the Poisson shot-noise fiducial and the cosmology and biases of
Fig.~\ref{fig:T-plot}. \emph{Top:} split by power of inverse number density
(Eq.~\eqref{eq:stoch_pieces}), showing the total $T_{\rm stoch}$ (dashed) and the
$\bar n^{-1}$ (blue), $\bar n^{-2}$ (orange) and $\bar n^{-3}$ (green) terms.
\emph{Bottom:} the dominant $\bar n^{-1}$ block (dashed) split into its contributions: $b_1 B_{\rm shot}$ (blue), $(1+P_{\rm shot})f\mu^2$
(orange), $b_2 C_{\rm shot}$ (green) and $2b_{\mathcal{G}_2}D_{\rm shot}$ (purple).
Solid and dashed coloured lines denote positive and negative values respectively.}
\label{fig:stoch-decomp}
\end{figure}

\paragraph{Numerical Evaluation.}
Fig.~\ref{fig:stoch-decomp} shows each order of stochastic contribution to the redshift-space monopole for the equilateral, folded and collapsed shape configuations of Fig.~\ref{fig:T-plot}, evaluated at the same fiducial cosmology and biases with all shot-noise amplitudes set to their Poisson values ($A_{\rm shot}=\ldots=F_{\rm shot}=1$, $P_{\rm shot}=0$) and $\bar n\simeq3.2\times10^{-3}\,h^3{\rm Mpc}^{-3}$.

The top panels split $T_{\rm stoch}$ by power of $\bar n^{-1}$. The hierarchy is set by the factor $\bar n P(k)\gg1$ on these scales: the $\bar n^{-1}$ term ($\propto P(k)^2/\bar n$) dominates and is essentially indistinguishable from the total, the $\bar n^{-2}$ term ($\propto P(k)/\bar n^2$) lies roughly two orders of magnitude lower, and the $\bar n^{-3}$ term is the scale-independent shot floor $F_{\rm shot}/\bar n^3$. The relative importance of the subleading terms grows mildly towards small scales as $P(k)$ falls, and would differ for a less dense galaxy sample.

The bottom panels split the dominant $\bar n^{-1}$ block into the four operator contributions of Eq.~\eqref{eq:stoch_pieces}: the linear-bias term $b_1 B_{\rm shot}$, the redshift-space velocity term $\propto(1+P_{\rm shot})f\mu^2$, the quadratic term $b_2 C_{\rm shot}$, and the tidal term $2b_{\mathcal{G}_2}D_{\rm shot}$. The $b_1 B_{\rm shot}$ piece dominates and traces the full $\bar n^{-1}$ piece, with the remaining three each at the $\sim\!10\%$ level; the $b_2 C_{\rm shot}$ contribution is negative (dashed) since $b_2<0$ for our sample, partially canceling the others. As for the deterministic blocks, the shape dependence is weak since the stochastic contributions are controlled by the power spectra and the shot noise amplitudes rather than by the precise tetrahedra configuration.

\section{FoG Counterterms}
\label{app:fog}

\begin{figure}
\centering
\includegraphics[width=\textwidth]{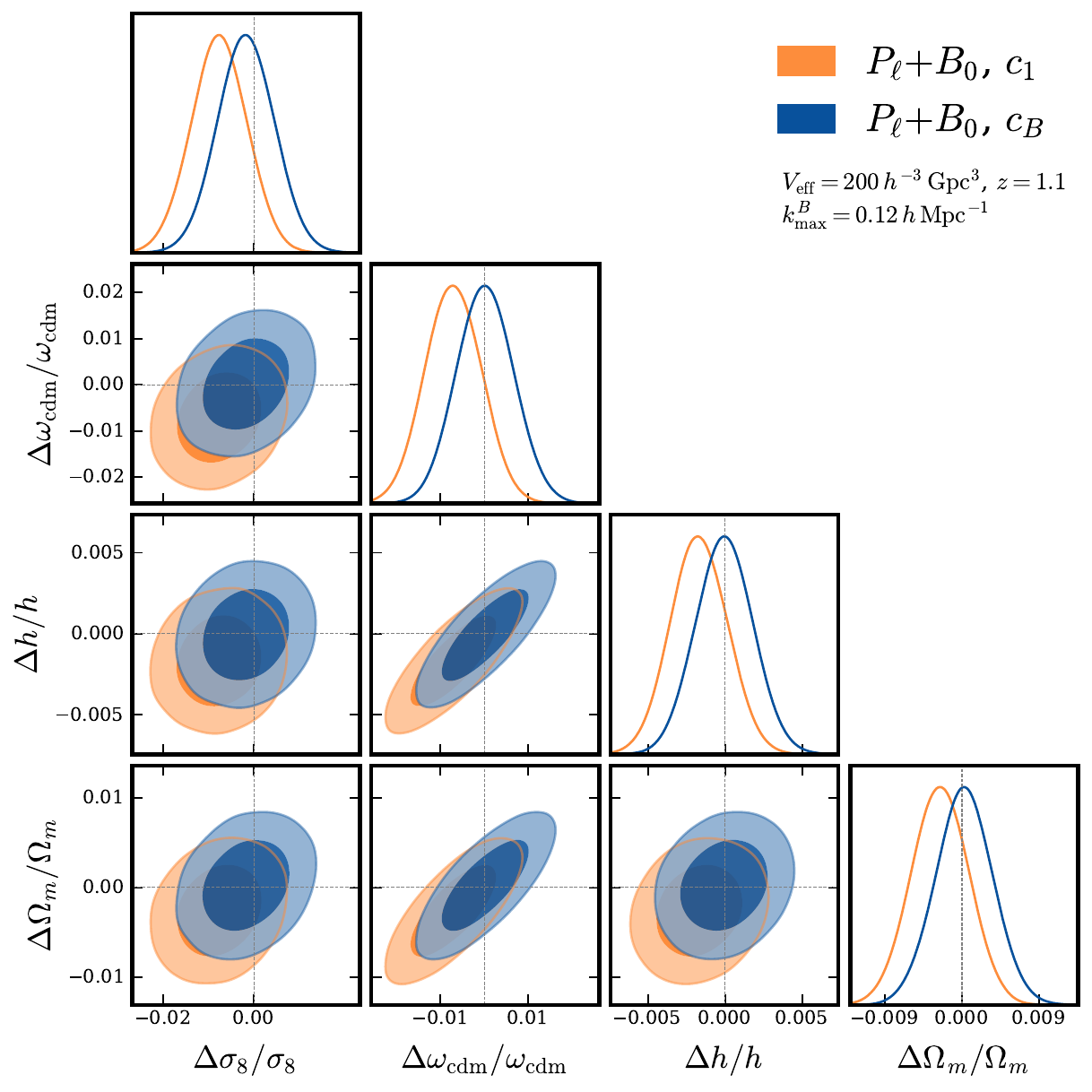}
\caption{
Marginalized cosmological constraints ($68\%$ and $95\%$) from the redshift-space \textsc{AbacusSummit} $z=1.1$ $P_\ell{+}B_0$ data at $\kmax^B=0.12\hMpc$, with $\kmax^P=0.18\hMpc$ and $V_{\rm eff}=200\,h^{-3}{\rm Gpc}^3$, comparing the two bispectrum FoG counterterm choices. Orange is the $c_1$ model commonly used in the literature and blue is the coherent Gaussian $c_B$ model of this work. Parameters are shown as fractional deviations $\Delta\boldsymbol{\theta}/\boldsymbol{\theta}$ from their true values, with dashed lines at zero. The $c_1$ counterterm pulls the whole cosmology low, by $1.3\sigma$ in $\sigma_8$ and $1.1\sigma$ in $\omega_{\rm cdm}$, while $c_B$ keeps it within $0.3\sigma$ of the truth. The $c_B$ contours are identical to the $P_\ell{+}B_0$ result of Sec.~\ref{sec: results-rsd} at this cut.
}
\label{fig:c1_vs_cB}
\end{figure}

In this appendix, we describe in more detail the FoG counterterm model of Sec.~\ref{sec: fog}, which differs from the method adopted for modeling the tree-level galaxy bispectrum monopole in some of the EFTofLSS literature (see \citep{Ivanov:2021kcd} and subsequent works). For the redshift-space density field, the leading real- and redshift-space higher-derivative corrections are usually written schematically as
\be
\label{eq:ctrs}
\delta^{\rm ctr.}(\vk)
=
-\left[
c_0\left(\frac{k}{k_{\rm NL}}\right)^2
+
\left(c_1\mu^2+c_2\mu^4\right)
\left(\frac{k}{k^r_{\rm NL}}\right)^2
\right]\delta^{(1)}(\vk),
\ee
where $\mu=\hat{\vk}\cdot\hat{\vn}$, $k_{\rm NL}$ is the nonlinear scale for real-space, and $k^r_{\rm NL}=0.3\hMpc$ (as defined in Sec.~\ref{sec: fog}). The isotropic coefficient $c_0$ is the usual real-space higher-derivative counterterm (from $\nabla^2\delta$ terms in $\delta_{g,\text{det}}$), while the anisotropic coefficients $c_1$ and $c_2$ absorb the leading order FoG effects ~\citep{Perko:2016puo,Ivanov:2021kcd}. For the monopole, the $\mu^4$ shape associated with $c_2$ is strongly degenerate with the $\mu^2$ shape associated with $c_1$, hence $c_1$ is the only free term when using only the bispectrum monopole. After expanding to first order in the counterterms, the $Z_1$ kernel appearing in the bispectrum model is typically redefined to absorb the $c_1$ dependence, as shown in Eq.~\eqref{eq:Z1fog}. 

Let us comment on the distinction between the commonly-used $c_1$ model and the coherent FoG damping used in this work, parametrized by $c_B$. Consider one cyclic contribution to the bispectrum,
\be
\label{eq:Bij_channel}
B_{\rm det}
\supset
2Z_1(\vk_1)Z_1(\vk_2)Z_2(\vk_1,\vk_2)P(k_1)P(k_2).
\ee
The model adopted in this work coherently damps this term by
$-\left(
k_1^2\mu_1^2+
k_2^2\mu_2^2+
k_3^2\mu_3^2
\right)$ with the factor $c_B/(k^r_{\rm NL})^2$. By contrast, the $c_1$ model would correct only the two explicit $Z_1$ kernels in this cyclic term, which expanded to first order gives
\be
\label{eq:Bc1_external}
B_{\rm ct}
\supset
-2c_1\left[\mu_2^2
\left(\frac{k_2}{k^r_{\rm NL}}\right)^2 Z_1(\vk_1) + \mu_1^2\left(\frac{k_1}{k^r_{\rm NL}}\right)^2 Z_1(\vk_2)\right]Z_2(\vk_1,\vk_2)P(k_1)P(k_2).
\ee
The difference is that the $c_B$ model also damps the external momentum $\vk_3$ produced by the quadratic kernel $Z_2$, while the $c_1$ prescription does not. 

The same logic applies to the trispectrum. For example, in $T_{1122}$ and $T_{1113}$, the $c_T$ template introduced in Sec.~\ref{sec: fog} damps all four external momenta. On the other hand, if one were to adopt a FoG countertem for the trispectrum inspired in the $c_1$ bispectrum convention, a $c_{1,T}$ parameter could be added to correct the two explicit $Z_1$ kernels appearing in $T_{1122}$ and $T_{1113}$. Again, we emphasize that this prescription is not intended to replace a full one-loop EFT counterterm model. Rather, it should effectively capture the redshift-space velocity corrections in the tree-level bispectrum and trispectrum monopoles using only one parameter for each. 

In practice, we find that the coherent FoG term adopted here is more effective than the $c_1$ model at absorbing the FoG effects. In the joint $P_\ell+B_0+T_0$ analysis of the large \textsc{AbacusSummit} volume, $c_B$ and $c_T$ are detected at high significance and remove the low-$\sigma_8$ bias that the tree-level model would otherwise show at higher scale cuts.  Notably, a bias in $\sigma_8$ is observed in a $P_\ell+B_0$ only analysis when using $c_1$ instead of $c_B$, as shown in Fig.~\ref{fig:c1_vs_cB} at $\kmax^B=0.12\hMpc$. The two runs are identical and differ only in the counterterm, $c_1$ in place of $c_B$. The $c_1$ model pulls the cosmology low, with $\Delta\sigma_8/\sigma_8=-0.8\%$ ($1.3\sigma$) and $\Delta\omega_{\rm cdm}/\omega_{\rm cdm}=-0.7\%$ ($1.1\sigma$), while $c_B$ stays within $0.3\sigma$ of the truth on every parameter. The blue $c_B$ contours are the same $P_\ell{+}B_0$ constraints shown at this cut in the redshift-space analysis of Sec.~\ref{sec: results-rsd}. Marginalizing over $c_B$ and $c_T$ therefore extends the reach of our model to higher $\kmax$ than the $c_1$ counterterms allows. We therefore adopt Eq.~\eqref{eq:cbct} as our fiducial counterterm model throughout this work.

\section{Discreteness Weights}
\label{app:discreteness}

\begin{figure}
\centering
\includegraphics[width=\textwidth]{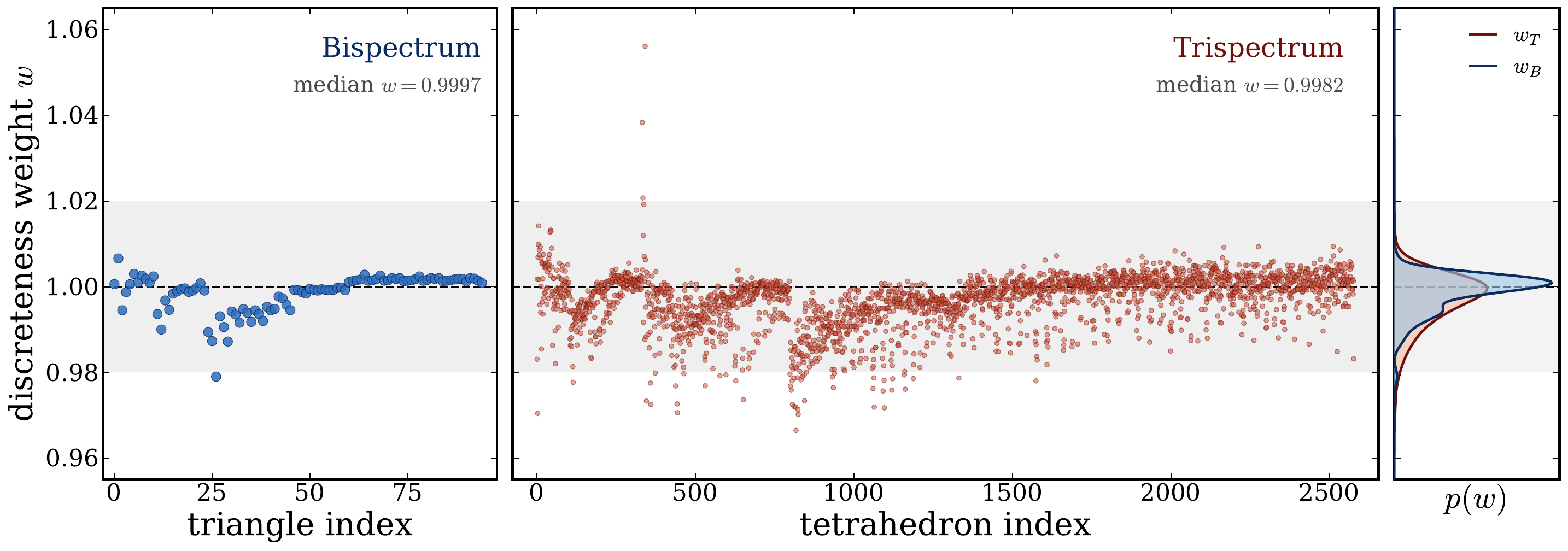}
\caption{Discreteness weight $w=\hat X_{0,\rm disc}/\hat X_{0,\rm int}$ for the bispectrum (left, $X=B$) and the trispectrum (middle, $X=T$), in real space on the \textsc{AbacusSummit} binning. Each point is one triangle or tetrahedron, ordered by bin index. The right panel shows the distribution of $w$ for both statistics, and the grey band marks $\pm2\%$. The largest departures come from bins with short sides, whose shells contain the fewest lattice modes. For the trispectrum these are the regular tetrahedra with all four sides at $k=0.025\hMpc$, with $w_T=1.056$ for $k_{12}=0.045\hMpc$ and $w_T=1.038$ for $k_{12}=0.015\hMpc$, followed by tetrahedra with three sides at $0.035\hMpc$ and a much longer fourth side between $0.075$ and $0.095\hMpc$, which sit about $3\%$ below unity. For the bispectrum the largest departure is the triangle with sides $(0.035,0.035,0.055)\hMpc$, with $w_B=0.979$.}
\label{fig:discreteness}
\end{figure}

The binned model of Eq.~\eqref{eq:Tk-binning} treats the sum over Fourier modes in each bin as a continuous integral. This is only approximate for a finite fundamental frequency $k_f=2\pi/L$ (given the discrete Fourier lattice), and can source a non-trivial error at very low $k$. Following \citep{Bakx:2025pop}, we quantify the error with the discreteness weight
\beq\label{eq: disc-weight}
  w \equiv \frac{\hat T_{0,\rm disc}}{\hat T_{0,\rm int}}\,,
\eeq
which is the ratio of the model averaged over the discrete lattice tetrahedra in the bin to the continuum bin average of Eq.~\eqref{eq:Tk-binning}. The bispectrum weight is defined in the same way. Since $w$ is a ratio of two averages of the same model, it depends only weakly on cosmology and bias. We therefore compute it once on the \textsc{AbacusSummit} binning for the real-space tree-level model, with $b_1=1.69$ and $\{b_2,b_{\mathcal{G}_2}\}$ from \citep{Lazeyras:2015lgp,Eggemeier:2020umu}.

For the bispectrum, the discrete average is a direct sum over all triangles in the bin. A trispectrum bin can contain up to $\sim10^{11}$ tetrahedra, so there we estimate the average using a Monte Carlo scheme, drawing at most $10^5$ lattice tetrahedra per bin. Going from $4\times10^4$ to $10^5$ samples changes the weight of a typical bin by about $0.15\%$. For the $108$ low $k$ bins where we can enumerate every tetrahedron, the Monte Carlo estimate agrees with the exact result to better than $0.5\%$, so the sampling noise is negligible.

Fig.~\ref{fig:discreteness} shows the result. The weights are very close to unity, with median $w_B=1.000$ and $w_T=0.998$. Only one of the $95$ bispectrum bins and $50$ of the $2578$ trispectrum bins depart from unity by more than $2\%$. All of them have at least one short side, with $k\le0.045\hMpc$, and most have a side at $0.025$ or $0.035\hMpc$, where the shells contain few lattice modes. The largest departures reach a few percent. The shape of the configuration matters less than its scale, although collapsed tetrahedra lie slightly below unity on average, with a median of $0.995$. This is much smaller than the $\sim15\%$ found by \citep{Ivanov:2021kcd}, because our binned integral already restricts to closed configurations (as in the revised bispectrum analysis of \citep{Bakx:2025pop}). The correction sits well below the statistical error in every bin, so we set $w=1$ in the model and do not include the weights in our analyses.

\section{Covariance Derivation}
\label{app:cov}
This appendix derives the trispectrum covariance model of Sec.~\ref{sec:cov} and demonstrates that the terms omitted from the trispectrum covariance are small. We start with the Gaussian covariance, since this already fixes the diagonal normalization and explains the strong same-sides off-diagonal structure. We then discuss the cross-covariances between the power spectrum, bispectrum and trispectrum and the leading non-Gaussian correction. 

\paragraph{Estimator Normalization.}
A trispectrum bin keeps the four side lengths and one diagonal. We write
\begin{equation}
    I=(k_1,k_2,k_3,k_4,k_{12}^I),
\end{equation}
where the estimator sums over the other diagonal, $k_{14}$. In the continuum the sum over modes gives the phase-space count
\begin{equation}
    N_I^{\rm cont}
    =
    \frac{V^3}{(2\pi)^9}
    \Delta k^4
    \int_{k_{12}^I-\Delta k/2}^{k_{12}^I+\Delta k/2} dk_{12}
    \int_{k_{14}^{\rm min}(k_{12})}^{k_{14}^{\rm max}(k_{12})}
    dk_{14}\,
    W(k_{12},k_{14}) ,
    \label{eq:NI_cont}
\end{equation}
where
\begin{equation}
    W(k_{12},k_{14})
    =
    \frac{8\pi^2}{3}\,
    \frac{k_1k_2k_3k_4\,k_{12}k_{14}}
    {V_{\rm tet}(k_1,k_2,k_3,k_4,k_{12},k_{14})}
    =
    32\pi^2\,
    \frac{k_1k_2k_3k_4\,k_{12}k_{14}}
    {\sqrt{g_1}} .
    \label{eq:W_cov}
\end{equation}
The limits $k_{14}^{\rm min}$ and $k_{14}^{\rm max}$ define the range of the free diagonal at fixed sides and fixed $k_{12}$ for which the tetrahedron closes, given by Eq.~\eqref{eq:dihedral}. The third pair diagonal is fixed by closure,
\begin{equation}
    k_{13}=
    \left(k_1^2+k_2^2+k_3^2+k_4^2-k_{12}^2-k_{14}^2\right)^{1/2}.
    \label{eq:k13_cov_app}
\end{equation}
The exact count used in the estimator is denoted by $N_I$ and is defined in Eq.~\eqref{eq:trispectrum_normalization}. It is the discrete lattice version of Eq.~\eqref{eq:NI_cont}.

\paragraph{Gaussian Trispectrum Covariance.}
The Gaussian covariance is obtained by Wick-contracting the four modes of one trispectrum estimator with the four modes of the other. We first consider the same-sides block,
\begin{equation}
    I=(k_1,k_2,k_3,k_4,k_{12}^I),
    \qquad
    J=(k_1,k_2,k_3,k_4,k_{12}^J),
\end{equation}
where the four side bins are the same, but the retained diagonal bin may be different.

The only new feature, compared with the power spectrum or bispectrum, is that a tetrahedron has three internal diagonals, $k_{12}$, $k_{14}$, and $k_{13}$. Since the estimator keeps only $k_{12}$, the retained diagonal of bin $J$ can be matched, by a Wick relabelling, to any of these three diagonals of bin $I$.
It is useful to write the derivation directly in terms of mode counts. For
$X\in\{12,14,13\}$, define
\begin{equation}
    (a_X,b_X)=
    \begin{cases}
    (1,2), & X=12,\\
    (1,4), & X=14,\\
    (1,3), & X=13.
    \end{cases}
\end{equation}
The continuum approximation to the overlap count is
\begin{equation}
\begin{split}
    N_X^{\rm cont}(I,J)
    =
    V^3
    \int_{\boldsymbol p_i\in k_i,\boldsymbol p_{12}\in k_{12}^I,\boldsymbol p_X\in k_{12}^J}
    &(2\pi)^9
    \delta_D(\boldsymbol p_1+\boldsymbol p_2-\boldsymbol p_{12})
    \delta_D(\boldsymbol p_{a_X}+\boldsymbol p_{b_X}-\boldsymbol p_X)
    \\
    &\times
    \delta_D(\boldsymbol p_1+\boldsymbol p_2+\boldsymbol p_3+\boldsymbol p_4),
\end{split}
    \label{eq:NX_cont_app}
\end{equation}
for $i\in\{1,2,3,4\}$. The auxiliary vector $\boldsymbol p_X$ only imposes that the diagonal $X$ lies in the retained-diagonal bin of $J$. For example, $N_{14}^{\rm cont}(I,J)$ counts tetrahedra in bin $I$ for which $\boldsymbol p_1+\boldsymbol p_4$ lies in the shell centered on
$k_{12}^J$.

We can now see how these counts appear in the covariance. Writing the two
binned estimators with their Dirac deltas, the Gaussian part contains
\begin{equation}
\begin{split}
    C^{T,{\rm cont}}_{IJ}
    =
    \frac{(2\pi)^{12}}{N_I^{\rm cont}N_J^{\rm cont}}
    \int_{\boldsymbol p_i\in k_i,\boldsymbol p_{12}\in k_{12}^I,\boldsymbol q_i\in k_i,\boldsymbol q_{12}\in k_{12}^J}
    &
    \delta_D(\boldsymbol p_1+\boldsymbol p_2-\boldsymbol p_{12})
    \delta_D(\boldsymbol p_3+\boldsymbol p_4+\boldsymbol p_{12})
    \\
    &\times
    \delta_D(\boldsymbol q_1+\boldsymbol q_2-\boldsymbol q_{12})
    \delta_D(\boldsymbol q_3+\boldsymbol q_4+\boldsymbol q_{12})
    \\
    &\times
    \left\langle
    \prod_{a=1}^4\delta_g(\boldsymbol p_a)
    \prod_{b=1}^4\delta_g(\boldsymbol q_b)
    \right\rangle,
\end{split}
    \label{eq:Tcov_explicit_app}
\end{equation}
again for $i\in\{1,2,3,4\}$. The Gaussian contractions give the factor $P(k_1)P(k_2)P(k_3)P(k_4)$.

The bin $I$ fixes $\boldsymbol p_1+\boldsymbol p_2$ to lie in the shell centered on $k_{12}^I$. The bin $J$ fixes $\boldsymbol q_1+\boldsymbol q_2$ to lie in the shell centered on $k_{12}^J$. After a Wick contraction, this second
diagonal can be identified with one of the three pair diagonals of the tetrahedron in bin $I$: $k^I_{12}$, $k^I_{14}$ or $k^I_{13}$.

For instance, if the retained diagonal of $J$ is mapped to the hidden diagonal $k^I_{14}$ of $I$, the condition $\boldsymbol q_1+\boldsymbol q_2=\boldsymbol q_{12}$ becomes, up to a sign irrelevant for the shell average, $\boldsymbol p_{14}\in k_{12}^J$. The remaining phase-space integral is then exactly $N_{14}^{\rm cont}(I,J)$ as defined in Eq.~\eqref{eq:NX_cont_app}. The same argument gives $N_{12}^{\rm cont}(I,J)$ when the retained diagonal of $J$ lands on $k^I_{12}$, and $N_{13}^{\rm cont}(I,J)$ when it lands on $k^I_{13}$. 

Therefore the continuum Gaussian covariance is
\begin{equation}
\begin{split}
    C^{T,{\rm cont}}_{IJ}
    =
    V^2\,
    P(k_1)P(k_2)P(k_3)P(k_4)
    \frac{
    c_{12}N_{12}^{\rm cont}(I,J)
    +c_{14}N_{14}^{\rm cont}(I,J)
    +c_{13}N_{13}^{\rm cont}(I,J)
    }{
    N_I^{\rm cont}N_J^{\rm cont}
    } .
\end{split}
    \label{eq:Tcov_cont}
\end{equation}
This is the same expression as the full-shape Gaussian trispectrum covariance \cite{Spezzati_2025}, but projected onto an estimator that keeps only one diagonal. 

Replacing the continuum counts in Eq.~\eqref{eq:Tcov_cont} by the exact lattice counts gives the covariance used in the analysis, shown in Eq.~\eqref{eq:Tcov_exact}. The replacement of $N_I^{\rm cont}$ and $N_X^{\rm cont}$ by the exact counts $N_I$ and $N_X$ is only made in the covariance normalization. For the covariance, these counts set the number of independent lattice tetrahedra contributing to each bin, so using the continuum approximation would slightly overestimate the available number of modes and underestimate the variance, specially at low-$k$. An equally valid choice would be to keep the Dirac deltas with unit weight in the numerator and let them enumerate the contributing modes directly. For the theory prediction of the trispectrum, the same bin normalization appears in the numerator and denominator of the estimator over the same closed bin geometry. The leading discreteness of the total mode count therefore cancels in the ratio, and the continuum bin integral is sufficient for evaluating the model.

\begin{figure}
\centering
\includegraphics[width=\textwidth]{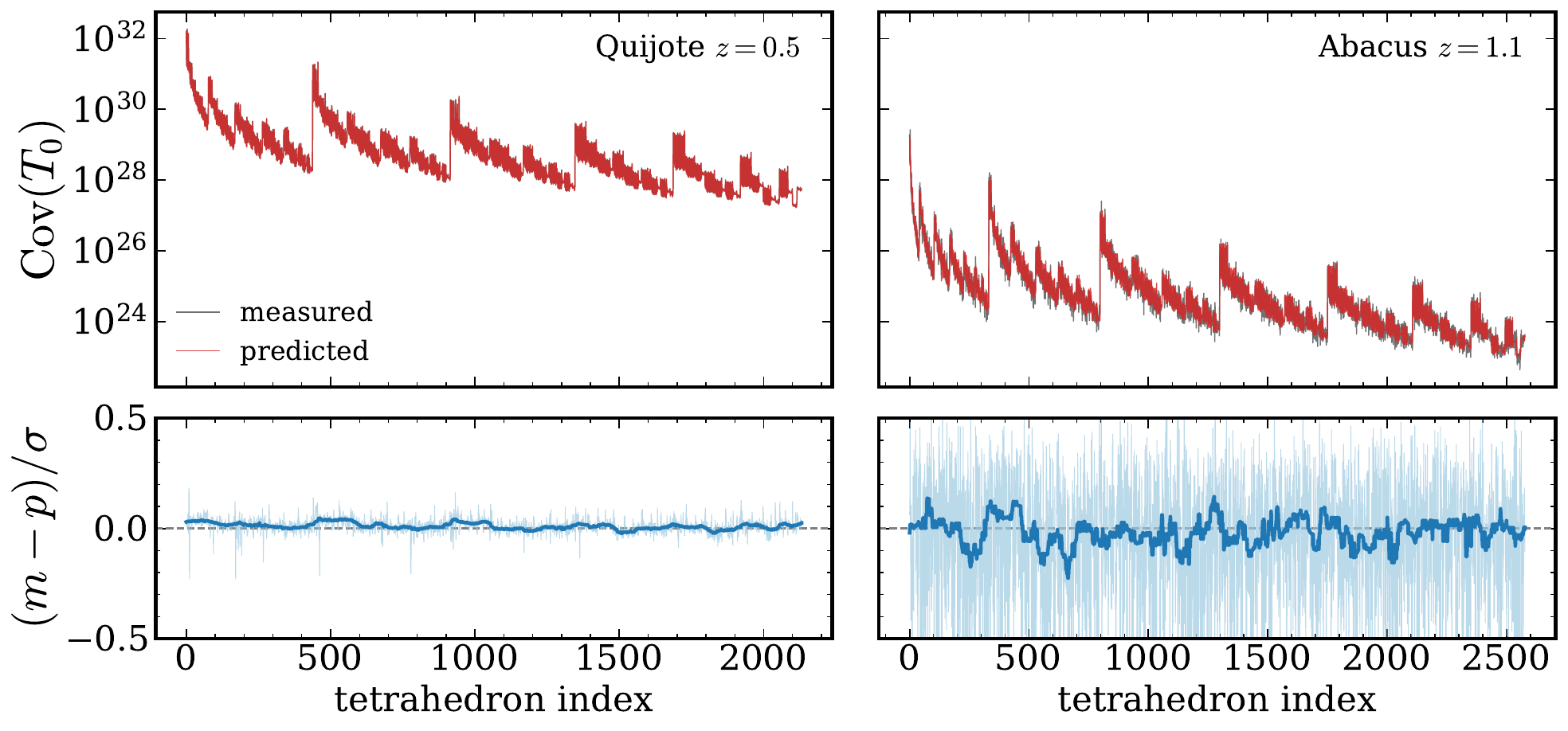}
\caption{Diagonal trispectrum covariance in redshift-space, measured (grey) versus the theory prediction from Eq.~\eqref{eq:Tcov_exact_diag_full} (red), with the fractional residual $(\widehat m-p)/\sigma$ beneath each panel. Columns: \textsc{Quijote}, $z=0.5$ (left, $15000$ realizations) and \textsc{AbacusSummit} $z=1.1$ (right, $25$ phases). The tetrahedron bins are in estimator order.}
\label{fig:Tdiag_cov}
\end{figure}

The case where $k_{12}^I=k_{12}^J$ reduces the covariance to Eq.~\eqref{eq:Tcov_exact_diag_full}, and we show in Fig.~\ref{fig:Tdiag_cov} the agreement of prediction and simulations for both \textsc{Quijote} at $z=0.5$ and \textsc{AbacusSummit} at $z=1.1$, which shows agreement over the entire $k$-range. For \textsc{Quijote}, where the sample covariance is measured precisely enough to be a meaningful reference, the residual has a median of $+1.1\%$ and the residual scatter at roughly $3\%$. For \textsc{AbacusSummit} the median residual is $-1.7\%$, again consistent with no systematic offset, while the much larger bin-to-bin scatter is accounted for by the noise of a variance estimated from only $25$ phases, for which $\sqrt{2/(N-1)}\simeq29\%$. We therefore adopt the analytic covariance throughout, noting that its diagonal is accurate to $\sim1\%$ in the median and that no residual trend with scale or configuration is apparent.

The integers $c_{12}$, $c_{14}$ and $c_{13}$ appearing in Eq.~\eqref{eq:Tcov_exact} and Eq.~\eqref{eq:Tcov_cont} are the symmetry factors for the three diagonal overlaps. These count how many Wick relabellings put the retained diagonal of bin $J$ on $k_{12}$, $k_{14}$ or $k_{13}$ of bin $I$. For four distinct side bins (\textit{i.e.}\ $k_1\neq k_2\neq k_3\neq k_4$), only one relabelling contributes,
\begin{equation}
    c_{12}=1,\qquad c_{14}=c_{13}=0 .
\end{equation}
If some of the four side bins are equal, more relabellings are allowed, which can either keep the measured diagonal as $k_{12}$ or move it to $k_{14}$ or $k_{13}$. The numbers $c_{12}$, $c_{14}$ and $c_{13}$ are fixed only by these side-bin degeneracies.

This follows a similar logic as in the bispectrum covariance. There one has the symmetry factor $s_B$ multiplying the bispectrum covariance. For example, in equilateral configurations, we have three equal sides in the triangle bin, giving $s_B=6$. For isosceles configurations, $s_B=2$, while for three distinct sides, $s_B=1$.

For the trispectrum, the diagonal covariance only depends on the part that preserves the measured diagonal, $c_{12}$. This is given by
\begin{equation}
    c_{12}=m_{12}m_{34}m ,
    \label{eq:c12_def}
\end{equation}
where
\begin{equation}
    m_{12}=
    \begin{cases}
    2, & k_1=k_2,\\
    1, & k_1\neq k_2,
    \end{cases}
    \qquad
    m_{34}=
    \begin{cases}
    2, & k_3=k_4,\\
    1, & k_3\neq k_4,
    \end{cases}
    \quad 
    m=
    \begin{cases}
    2, & \{k_1,k_2\}=\{k_3,k_4\},\\
    1, & \text{otherwise},
    \end{cases}
\end{equation}
where equalities mean equality of side bins. Thus $c_{12}=1$ if all four sides are distinct, while $c_{12}=8$ for four equal side bins.

\paragraph{$B$--$T$ Cross-Covariance.}
The largest bispectrum--trispectrum cross-covariance comes from the case in which the bispectrum triangle is one of the two faces of the trispectrum tetrahedron (see Fig~\ref{fig:bird}). For a trispectrum bin $I$, these two faces are
\begin{equation}
    (k_1,k_2,k_{12}),
    \qquad
    (k_3,k_4,k_{12}) .
\end{equation}
Consider first the covariance with $B(k_1,k_2,k_{12})$. The leading contribution contracts the two shared sides with power spectra, while the opposite face contributes the bispectrum $B(k_3,k_4,k_{12})$. This gives
\begin{equation}
    {\rm Cov}\!\left[
    B(k_1,k_2,k_{12}),
    T(k_1,k_2,k_3,k_4,k_{12})
    \right]
    =
    \frac{m_B}{N_B(k_1,k_2,k_{12})}
    P(k_1)P(k_2)
    B(k_3,k_4,k_{12}),
    \label{eq:BT_raw}
\end{equation}
where $N_B(k_1,k_2,k_{12})$ is the triangle count and $m_B$ counts the matched contractions. The Gaussian bispectrum variance is
\begin{equation}
    {\rm Var}\!\left[B(k_1,k_2,k_{12})\right]
    =
    \frac{s_B}{N_B(k_1,k_2,k_{12})}
    P(k_1)P(k_2)P(k_{12}) ,
    \label{eq:Bvar}
\end{equation}
with $s_B$ the usual bispectrum symmetry factor. For the matched-face contribution, $m_B=s_B$. Hence the same result can be written as
\begin{equation}
    {\rm Cov}\!\left[
    B(k_1,k_2,k_{12}),
    T(k_1,k_2,k_3,k_4,k_{12})
    \right]
    =
    \frac{B(k_3,k_4,k_{12})}{P(k_{12})}
    {\rm Var}\!\left[B(k_1,k_2,k_{12})\right].
    \label{eq:BTcov}
\end{equation}
The covariance with the other face is obtained by exchanging
$(k_1,k_2)$ and $(k_3,k_4)$.

In practice, we evaluate Eq.~\eqref{eq:BTcov} using the measured bispectrum and the measured bispectrum variance. This keeps the triangle count and symmetry factor fixed by the simulations. The resulting prediction matches
the measured redshift-space $B$--$T$ cross-correlations well, reproducing the matched-face correlations that reach up to $51\%$ and accounting for all $273$ cases where the correlation exceeds $20\%$, while leaving no residual correlation above $16\%$.

\begin{figure}
\centering
\includegraphics[width=\textwidth]{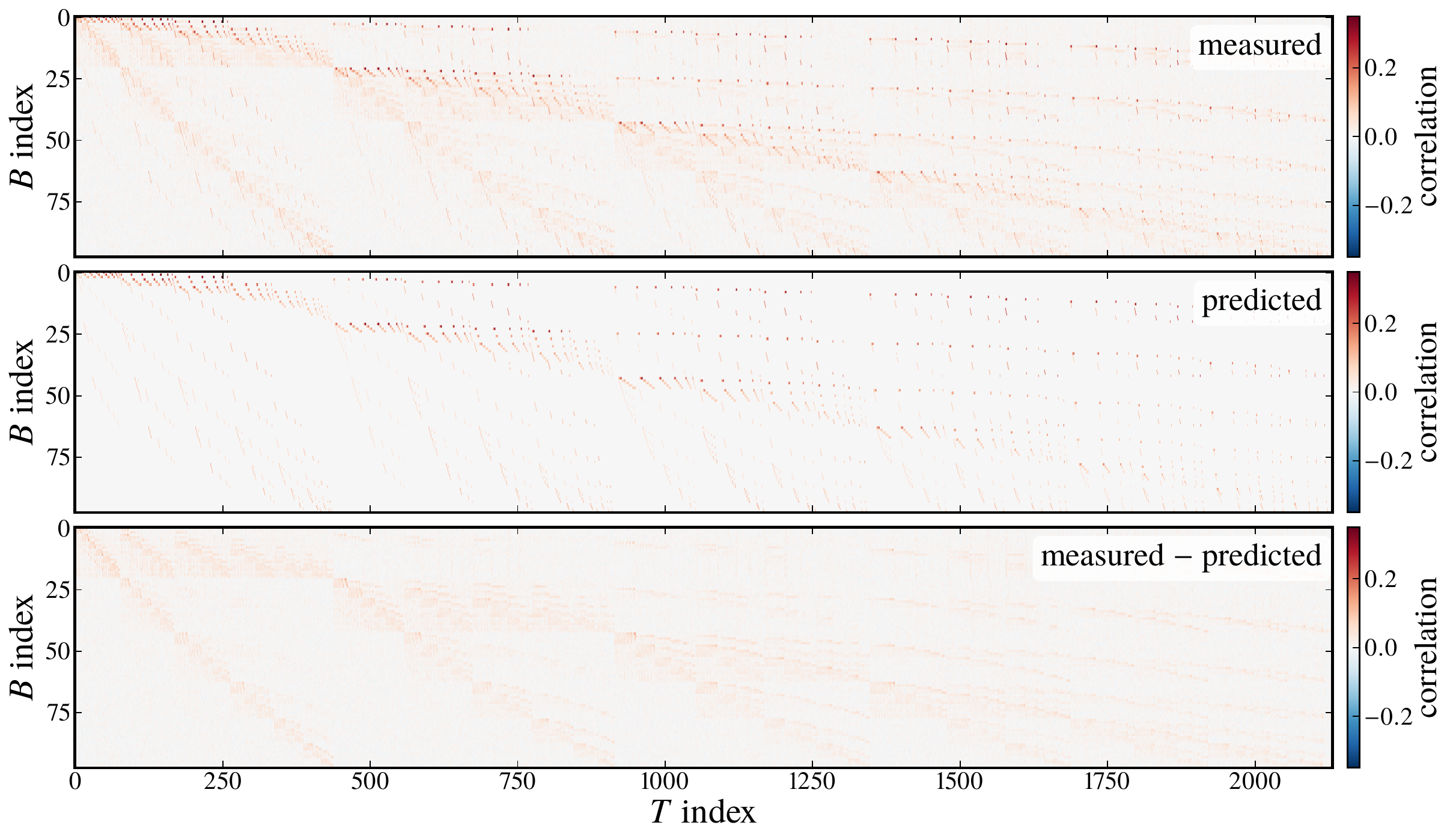}
\caption{Bispectrum--trispectrum cross-covariance measured from the
\textsc{Quijote} simulations at $z=0.5$ in redshift space, normalized to the measured diagonals and shown on the same $\pm0.35$ correlation scale as Figs.~\ref{fig:Tcov} and~\ref{fig:PTcov}. Rows are bispectrum triangles and columns are trispectrum bins. From top to bottom: measured covariance, the prediction of Eq.~\eqref{eq:BTcov}, and the residual. The prediction recovers the matched-face correlations, leaving a residual
consistent with noise.}
\label{fig:BTcov}
\end{figure}

\paragraph{$P$--$T$ Cross-Covariance.}
The power-spectrum--trispectrum cross-covariance is much smaller. Its leading terms come from contractions of the schematic form $\langle\delta^2\delta^4\rangle$, including a power spectrum multiplied by the connected trispectrum, as well as connected six-point contributions. Unlike the $B$--$T$ case, there is no matched face that produces a large, localized block of correlations.

\begin{figure}
\centering
\includegraphics[width=\textwidth]{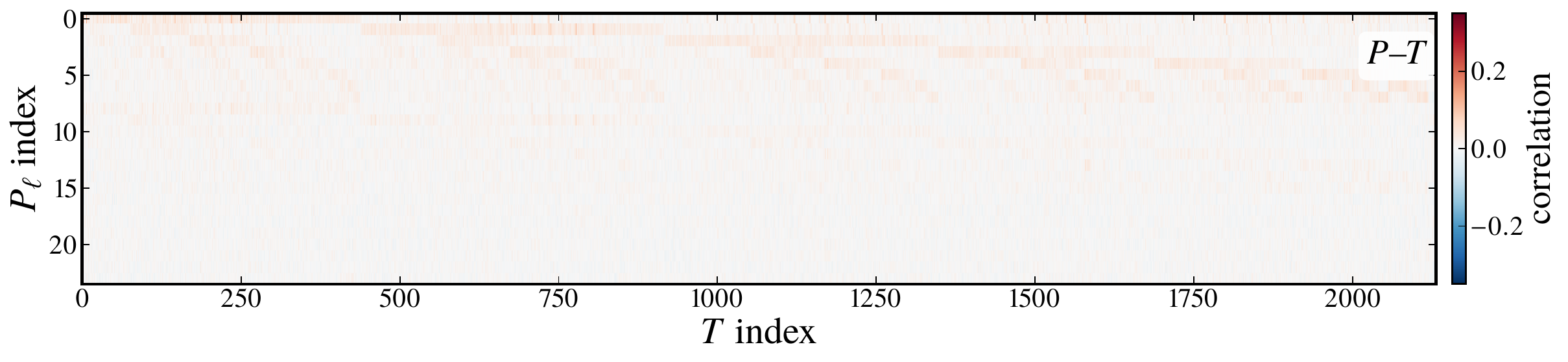}
\caption{Measured power-spectrum--trispectrum cross-covariance from the \textsc{Quijote} simulations at $z=0.5$ in redshift space, normalized to the measured diagonals. Rows are the $24$ power-spectrum multipole bins ($P_0,P_2,P_4$) and columns are the $2132$ trispectrum bins, shown on the same $\pm0.35$ correlation scale as Figs.~\ref{fig:Tcov} and~\ref{fig:BTcov}. The correlations are small, reaching a maximum of only $13\%$.}
\label{fig:PTcov}
\end{figure}

As shown in Fig~\ref{fig:PTcov}, the expectation is consistent with the measurements. For the bins used here, the redshift-space $P$--$T$ correlation stays small, reaching at most about $13\%$ on a few bins and negligible elsewhere.

\paragraph{Non-Gaussian Terms.} The Gaussian piece described above in the trispectrum covariance is only the leading contraction of the eight fields in the covariance of $T_I$ and $T_J$. To see what we are neglecting, we organize the remaining contractions by their connected pieces. We use the reduced connected polyspectra convention
\begin{equation}
    \left\langle
    \delta(\boldsymbol{k}_1)\cdots\delta(\boldsymbol{k}_n)
    \right\rangle_c
    =
    (2\pi)^3\delta_D(\boldsymbol{k}_1+\cdots+\boldsymbol{k}_n)
    P_n(\boldsymbol{k}_1,\ldots,\boldsymbol{k}_n),
    \label{eq:poly_def}
\end{equation}
with $P_2=P$, $P_3=B$, and $P_4=T$, with $P_n \sim P^{n-1}$ at tree-level. With this counting, the trispectrum covariance can be organized as
\begin{equation}
    \mathrm{Cov}(T,T)
    =
    C^{PPPP}
    +C^{TPP}
    +C^{BBP}
    +C^{T\times T}
    +C^{P_5B}
    +C^{P_6P}
    +C^{P_8}.
    \label{eq:full_cov_decomp}
\end{equation}
The schematic content and power counting are
\begin{center}
{\small
\begin{tabular}{lcc}
\toprule
term & connected partition & scaling \\
\midrule
$C^{PPPP}$ & $\langle\delta^2\rangle^4$ & $P^4$ \\
$C^{TPP}$ & $\langle\delta^4\rangle_c\langle\delta^2\rangle^2$ & $T P^2\sim P^5$ \\
$C^{BBP}$ & $\langle\delta^3\rangle_c^2\langle\delta^2\rangle$ & $B^2 P\sim P^5$ \\
$C^{T\times T}$ & $\langle\delta^4\rangle_c^2$ & $T^2\sim P^6$ \\
$C^{P_5B}$ & $\langle\delta^5\rangle_c\langle\delta^3\rangle_c$ & $P_5B\sim P^6$ \\
$C^{P_6P}$ & $\langle\delta^6\rangle_c\langle\delta^2\rangle$ & $P_6P\sim P^6$ \\
$C^{P_8}$ & $\langle\delta^8\rangle_c$ & $P_8\sim P^7$ \\
\bottomrule
\end{tabular}
}
\end{center}
This power counting is only a hierarchy for the clustering contributions. Because we do not include higher-order corrections to the bispectrum diagonal covariance, we opt to ignore the next-to-leading-order corrections ($\sim P^5$ and beyond) to the trispectrum covariance. Moreover we emphasize again that, in a real-data analysis, one would use a sample covariance measured from mocks.

\section{Fisher Forecasts and Covariance Structure}\label{app:fisher}

We use a Fisher forecast for the cosmological parameters $\{\sigma_8,h,\omega_{\rm cdm}\}$ both to quantify the information added by the trispectrum and, as the main goal of this appendix, to test how the covariance modeling propagates to the constraints, in particular the impact of the connected cross-covariances and of the off-diagonal structure within each auto-covariance.

We treat as free $\{\ln(10^{10}A_s),\omega_{\rm cdm},h\}$ and the bias parameters
$\{b_1,b_2,b_{\mathcal{G}_2}\}$ (with $n_s,\omega_b,M_\nu$ fixed) and marginalize every other parameter analytically, exactly as in the likelihood. The derivatives are evaluated with central finite differences. The Fisher matrix
$\mathbf{F}_{ab}=(\partial\boldsymbol{T}/\partial\theta_a)^{\rm T}\mathbf{C}^{-1}(\partial\boldsymbol{T}/\partial\theta_b)+\mathbf{P}_{ab}$
($\mathbf{P}$ the Gaussian prior) is constructed in the sampled basis, inverted, and the $\{\ln(10^{10}A_s),\omega_{\rm cdm},h\}$ rotated to $\{\sigma_8,\omega_{\rm cdm},h\}$ with the numerical Jacobian. 

Tab.~\ref{tab:fisher-abacus} gives the forecasted cosmological constraints for the mean of the $25$ \textsc{AbacusSummit} $z=1.1$ boxes ($200\,h^{-3}{\rm Gpc}^3$), at $\kmax^{P}=0.18$, $\kmax^{B}=\kmax^{T}=0.10\,\hMpc$, with the same analytic covariance and the same set of free and marginalized parameters used in the chains. Adding $B_0$ tightens $\omega_{\rm cdm}$ by $35\%$ and $\sigma_8$ by $16\%$ over $P_\ell$. The trispectrum then improves $\omega_{\rm cdm}$ by a further $5\%$ and $h$ by $3\%$, but leaves $\sigma_8$ almost unchanged. The gain from $T_0$ is therefore concentrated in the shape parameters rather than in the amplitude, and the same pattern is seen in the chains of Sec.~\ref{sec: results-rsd}.

\begin{table}[!htb]
\centering
\caption{Fisher forecast of the marginalized $1\sigma$ fractional errors for the mean of the $25$
\textsc{AbacusSummit} $z=1.1$ boxes, $\kmax^{P}=0.18$, $\kmax^{B}=\kmax^{T}=0.10\hMpc$, with the analytic
covariance (Gaussian $P_\ell$ including cross-$\ell$, Gaussian $B_0$, Gaussian $T_0$ with off-diagonal siblings).}
\label{tab:fisher-abacus}
{\small
\begin{tabular}{l c c c}
\hline\hline
data & $\sigma_{\sigma_8}/\sigma_8$ & $\sigma_h/h$ & $\sigma_{\omega_{\rm cdm}}/\omega_{\rm cdm}$ \\
\hline
$P_\ell$        & $0.76\%$ & $0.32\%$ & $1.23\%$ \\
$P_\ell+B_0$    & $0.64\%$ & $0.21\%$ & $0.80\%$ \\
$P_\ell+B_0+T_0$& $0.63\%$ & $0.20\%$ & $0.76\%$ \\
\hline\hline
\end{tabular}
}
\end{table}

To test the impact of the covariance form on the cosmological constraints, we use a single \textsc{Quijote} box ($1\,h^{-1}{\rm Gpc}$, $z=0.5$, $\kmax^{P}=0.14\,\hMpc$ and $\kmax^{B}=\kmax^{T}=0.10\,\hMpc$), for which we have the full
$15{,}000$-realization sample covariance. Tab.~\ref{tab:fisher-cov} compares the analytic covariance derived in Appendix~\ref{app:cov} (Gaussian $P_\ell$ with cross-$\ell$, Gaussian $B_0$, Gaussian $T_0$ with off-diagonal siblings, and no cross-statistic blocks), the full ($15{,}000$-mock, Hartlap-corrected) measured covariance and two intermediate constructions that isolate the connected cross blocks one at a time. The first keeps the full $P{+}B$ and $TT$ blocks but sets the $B\times T$ and $P\times T$ cross blocks to zero. The second keeps only the measured auto blocks and removes the $P\times B$ cross as well, so that it has the same block structure as the analytic model. The forecasts for both $P{+}B$ only and $P{+}B{+}T$ are shown. Each measured block carries the Hartlap factor appropriate to its own dimension, which matters because the block-diagonal constructions estimate far fewer elements than the full one.

\begin{table}[!htb]
\centering
\caption{Marginalized $1\sigma$ fractional errors on $\{\sigma_8,h,\omega_{\rm cdm}\}$ for a single redshift-space \textsc{Quijote} box ($z=0.5$, $\kmax^{P}=0.14\,\hMpc$ and $\kmax^{B}=\kmax^{T}=0.10\,\hMpc$) under different covariance constructions. ``Sample'' is estimated from the $15{,}000$ \textsc{Quijote} simulations with a Hartlap correction, ``sample (block-diagonal)'' contains the estimated PP, BB, PB, and TT blocks from the same simulations, while zeroing the PT and BT blocks, ``sample (auto only)'' keeps just the estimated PP, BB, and TT blocks and zeroes every cross block, and full ``analytic'' is the baseline model of
Appendix~\ref{app:cov}. }
\label{tab:fisher-cov}
{\small
\begin{tabular}{l c c c}
\hline\hline
covariance & $\sigma_{\sigma_8}/\sigma_8$ & $\sigma_h/h$ & $\sigma_{\omega_{\rm cdm}}/\omega_{\rm cdm}$ \\
\hline
$PBT$ sample                    & $9.95\%$  & $4.27\%$ & $11.70\%$ \\
$PBT$ sample (block-diagonal) & $10.29\%$ & $4.18\%$ & $11.21\%$ \\
$PBT$ sample (auto only)      & $10.16\%$ & $3.87\%$ & $9.96\%$  \\
$PBT$ analytic                                     & $9.63\%$  & $3.57\%$ & $8.32\%$  \\
\hline
$PB$ sample                                    & $11.52\%$ & $4.36\%$ & $12.22\%$ \\
$PB$ sample (auto only)                        & $11.19\%$ & $3.99\%$ & $10.64\%$ \\
$PB$ analytic                                  & $10.86\%$ & $3.73\%$ & $9.11\%$  \\
\hline\hline
\end{tabular}
}
\end{table}

We arrive at three conclusions from this test. First, the realistic covariance widens the errors relative to the Gaussian baseline. For $P{+}B$ alone, replacing the analytic covariance with the sample one inflates $\sigma(\omega_{\rm cdm})$ from $9.11\%$ to $12.22\%$, a $\sim34\%$ effect. The auto only row, at $10.64\%$, splits this almost evenly between the connected $P\times B$ cross and the off-diagonal structure of the $B_0$ auto-covariance, which the Gaussian model sets to zero. Second, the trispectrum does add extra information under the measured covariance, with $\sigma_8$ going from $11.52\%$ for $P{+}B$ to $9.95\%$ for $P{+}B{+}T$ and $\sigma(\omega_{\rm cdm})$ from $12.22\%$ to $11.70\%$. Third, the trispectrum correlation with the power spectrum and bispectrum is weak. Keeping the full measured $TT$ block while zeroing $B\times T$ and $P\times T$ changes $\sigma(\omega_{\rm cdm})$ by $4\%$ and $\sigma_8$ by $3\%$. At this volume the trispectrum can thus be treated as effectively independent from $P_\ell$ and $B_0$, which supports the covariance model used in this work. We highlight that e.g. $P\times T$ and $B\times T$ cross-covariances cannot be added without also including the corresponding off-diagonal auto-covariances, otherwise the covariance can become non-positive-definite.

Geometrically, the biggest correlations are localized in the folded (near-collinear) triangles, where the $P\times B$ cross peaks. This is the standard definition of the ``open'' triangles used for the bispectrum cross-covariance \citep{Ivanov:2021kcd}. The trispectrum analogs are the folded and equilateral tetrahedra configurations, which have the strongest $TT$ correlations. 

Overall, we find that two terms matter and that they contribute at a comparable level, the $P\times B$ cross-covariance and the off-diagonal part of the $B_0$ auto-covariance. Substituting the measured $B_0$ block into the otherwise analytic covariance recovers about two thirds of the gap to the sample covariance for $\omega_{\rm cdm}$, and most of that comes from the off-diagonal entries rather than from the $5\%$ deficit of the Gaussian variance. The analytic $T_0$ block is instead already accurate, because its sibling off-diagonals reproduce the measured ones closely and its diagonal is correct at the half percent level, and the $P\times T$ and $B\times T$ blocks are small. Improving the analytic covariance therefore means modeling the bispectrum off-diagonals, not the trispectrum ones. For a future real-data analysis, we will nevertheless use the full covariance measured from mocks, including all cross-correlations.

\section{Detection Significance}\label{app:detection}

Here we quantify how significantly the bispectrum and trispectrum are detected in each halo sample, separating the deterministic and stochastic contributions. We use the analytic block-diagonal covariance of Appendix~\ref{app:cov} for \textsc{AbacusSummit} and the full
$15{,}000$-mock measured covariance for \textsc{Quijote}, at $\kmax^{P}=0.18$ and $\kmax^{B}=\kmax^{T}=0.10\,\hMpc$.

For a statistic $X\in\{B_0,T_0\}$ the detection is the change in $\chi^2$ between modeling $X$ and setting it to zero,
\begin{equation}
 \Delta\chi^2_X=\chi^2(\text{no }X)-\chi^2(\text{full}),\qquad {\rm S/N}=\sqrt{\Delta\chi^2_X},
\end{equation}
evaluated at the joint best fit. The data vector is $\boldsymbol{T}=(P_0,P_2,P_4,B_0,T_0)$ and the bias parameters $\{b_1,b_2,b_{\mathcal{G}_2}\}$ are fixed at the $P_\ell+B_0$ best fit (cosmology held at each sample's fiducial), while the other parameters are marginalized analytically as in the
likelihood. In the ``no $X$'' hypothesis the $X$ prediction is set to zero, so $X$ enters as data with no free model or parameters, while the other statistics stay modeled and marginalized. We report every value rescaled to $V=50\,h^{-3}{\rm Gpc}^3$, a DESI-like volume. The data vector is the sample mean ($25$ phases for \textsc{AbacusSummit}, $15{,}000$
realizations for \textsc{Quijote}), for which the reduced $\chi^2$ of the full model, evaluated at the best fit and with all the free parameters marginalized, is close to unity.

A large $\Delta\chi^2$ does not by itself mean that the trispectrum is detected, since stochasticity is also present in the model. To isolate the deterministic signal we keep the stochastic amplitudes ($P_{\rm shot},A_{\rm shot}\ldots F_{\rm shot}$ for $T_0$, $P_{\rm shot},A_{\rm shot},B_{\rm shot}$ for $B_0$) marginalized in both hypotheses and remove only the deterministic piece, comparing the full model against a stochasticity only model. The resulting $\Delta\chi^2_{\rm det}$ is the signal that free shot noise parameters cannot absorb.

Tab.~\ref{tab:detection} gives the marginalized detection of $B_0$ and $T_0$ at $V=50\,h^{-3}{\rm Gpc}^3$ for the total and the deterministic value. Both statistics are detected at high significance in all three samples, and the bispectrum is stronger than the trispectrum by a factor of about $2.5$ to $3$. The deterministic contributions represent only around a third of the total signal, though they grow with number density, being largest for the dense \textsc{AbacusSummit} $z=0.5$ sample and smallest for the sparse \textsc{Quijote} one. The deterministic piece is nonetheless detected at between $75$ and $96\sigma$ for the bispectrum and $28$ to $31\sigma$ for the trispectrum. Lowering the cut to $\kmax^{B}=\kmax^{T}=0.08\,\hMpc$ reduces the numbers by roughly a factor of two and leaves these conclusions unchanged.

\begin{table}[!htb]
\centering
\caption{Marginalized detection significance ${\rm S/N}=\sqrt{\Delta\chi^2}$ of the bispectrum $B_0$ and trispectrum $T_0$ at the DESI volume $V=50\,h^{-3}{\rm Gpc}^3$ ($\kmax^{P}=0.18$, $\kmax^{B}=\kmax^{T}=0.10\,\hMpc$). The total $B_0$ and $T_0$ compares against a hypothesis in which the statistic is absent, while $B_0^{\rm det}$ and $T_0^{\rm det}$ keeps the stochastic amplitudes free and detects only the deterministic signal. \textsc{AbacusSummit} uses the analytic block-diagonal covariance and \textsc{Quijote} the full measured covariance.}
\label{tab:detection}
{\small
\begin{tabular}{l c c c c c c}
\hline\hline
sample & $b_1$ & $\bar{n}\,[h^3{\rm Mpc}^{-3}]$ & $B_0$ & $B_0^{\rm det}$ & $T_0$ & $T_0^{\rm det}$ \\
\hline
\textsc{AbacusSummit}, $z=1.1$ & $1.74$ & $3.2\times10^{-3}$ & $203$ & $82$ & $82$ & $28$ \\
\textsc{AbacusSummit}, $z=0.5$ & $1.24$ & $4.0\times10^{-3}$ & $193$ & $96$ & $76$ & $31$ \\
\textsc{Quijote}, $z=0.5$      & $1.84$ & $2.9\times10^{-4}$ & $173$ & $75$ & $59$ & $28$ \\
\hline\hline
\end{tabular}
}
\end{table}

\bibliography{short}

\providecommand{\href}[2]{#2}\begingroup\raggedright\begin{thebibliography}{100}

\bibitem{1980lssu.book.....P}
P.~J.~E. {Peebles}, \emph{{The large-scale structure of the universe}}. 1980.

\bibitem{Bernardeau:2001qr}
F.~Bernardeau, S.~Colombi, E.~Gaztanaga and R.~Scoccimarro, \emph{{Large scale structure of the universe and cosmological perturbation theory}}, \href{https://doi.org/10.1016/S0370-1573(02)00135-7}{\emph{Phys. Rept.} {\bfseries 367} (2002) 1} [\href{https://arxiv.org/abs/astro-ph/0112551}{{\ttfamily astro-ph/0112551}}].

\bibitem{Smith:2011gnl}
K.~M. Smith, S.~Ferraro and M.~LoVerde, \emph{{Halo clustering and $g_{NL}$-type primordial non-Gaussianity}}, \href{https://doi.org/10.1088/1475-7516/2012/03/032}{\emph{JCAP} {\bfseries 03} (2012) 032} [\href{https://arxiv.org/abs/1106.0503}{{\ttfamily 1106.0503}}].

\bibitem{Baumann:2012stochastic}
D.~Baumann, S.~Ferraro, D.~Green and K.~M. Smith, \emph{{Stochastic Bias from Non-Gaussian Initial Conditions}}, \href{https://doi.org/10.1088/1475-7516/2013/05/001}{\emph{JCAP} {\bfseries 05} (2013) 001} [\href{https://arxiv.org/abs/1209.2173}{{\ttfamily 1209.2173}}].

\bibitem{Ferraro:2014multifield}
S.~Ferraro and K.~M. Smith, \emph{{Using Large Scale Structure to test Multifield Inflation}}, \href{https://doi.org/10.1103/PhysRevD.91.043506}{\emph{Phys. Rev. D} {\bfseries 91} (2015) 043506} [\href{https://arxiv.org/abs/1408.3126}{{\ttfamily 1408.3126}}].

\bibitem{Desjacques:2009jb}
V.~Desjacques and U.~Seljak, \emph{{Signature of primordial non-Gaussianity of phi{\textasciicircum}3-type in the mass function and bias of dark matter haloes}}, \href{https://doi.org/10.1103/PhysRevD.81.023006}{\emph{Phys. Rev. D} {\bfseries 81} (2010) 023006} [\href{https://arxiv.org/abs/0907.2257}{{\ttfamily 0907.2257}}].

\bibitem{Xia_2011}
J.-Q. Xia, C.~Baccigalupi, S.~Matarrese, L.~Verde and M.~Viel, \emph{Constraints on primordial non-gaussianity from large scale structure probes}, \href{https://doi.org/10.1088/1475-7516/2011/08/033}{\emph{Journal of Cosmology and Astroparticle Physics} {\bfseries 2011} (2011) 033–033}.

\bibitem{Leistedt:2014zqa}
B.~Leistedt, H.~V. Peiris and N.~Roth, \emph{{Constraints on Primordial Non-Gaussianity from 800 000 Photometric Quasars}}, \href{https://doi.org/10.1103/PhysRevLett.113.221301}{\emph{Phys. Rev. Lett.} {\bfseries 113} (2014) 221301} [\href{https://arxiv.org/abs/1405.4315}{{\ttfamily 1405.4315}}].

\bibitem{Arkani-Hamed:2015bza}
N.~Arkani-Hamed and J.~Maldacena, \emph{{Cosmological Collider Physics}},  3, 2015.

\bibitem{Arkani-Hamed:2018kmz}
N.~Arkani-Hamed, D.~Baumann, H.~Lee and G.~L. Pimentel, \emph{{The Cosmological Bootstrap: Inflationary Correlators from Symmetries and Singularities}}, \href{https://doi.org/10.1007/JHEP04(2020)105}{\emph{JHEP} {\bfseries 04} (2020) 105} [\href{https://arxiv.org/abs/1811.00024}{{\ttfamily 1811.00024}}].

\bibitem{Senatore:2010jy}
L.~Senatore and M.~Zaldarriaga, \emph{{A Naturally Large Four-Point Function in Single Field Inflation}}, \href{https://doi.org/10.1088/1475-7516/2011/01/003}{\emph{JCAP} {\bfseries 01} (2011) 003} [\href{https://arxiv.org/abs/1004.1201}{{\ttfamily 1004.1201}}].

\bibitem{Chen:2009bc}
X.~Chen, B.~Hu, M.-x. Huang, G.~Shiu and Y.~Wang, \emph{{Large Primordial Trispectra in General Single Field Inflation}}, \href{https://doi.org/10.1088/1475-7516/2009/08/008}{\emph{JCAP} {\bfseries 08} (2009) 008} [\href{https://arxiv.org/abs/0905.3494}{{\ttfamily 0905.3494}}].

\bibitem{Baumann:2017jvh}
D.~Baumann, G.~Goon, H.~Lee and G.~L. Pimentel, \emph{{Partially Massless Fields During Inflation}}, \href{https://doi.org/10.1007/JHEP04(2018)140}{\emph{JHEP} {\bfseries 04} (2018) 140} [\href{https://arxiv.org/abs/1712.06624}{{\ttfamily 1712.06624}}].

\bibitem{Liu:2019fag}
T.~Liu, X.~Tong, Y.~Wang and Z.-Z. Xianyu, \emph{{Probing P and CP Violations on the Cosmological Collider}}, \href{https://doi.org/10.1007/JHEP04(2020)189}{\emph{JHEP} {\bfseries 04} (2020) 189} [\href{https://arxiv.org/abs/1909.01819}{{\ttfamily 1909.01819}}].

\bibitem{2015arXiv150200635S}
K.~M. {Smith}, L.~{Senatore} and M.~{Zaldarriaga}, \emph{{Optimal analysis of the CMB trispectrum}}, {\emph{arXiv e-prints} (2015) arXiv:1502.00635} [\href{https://arxiv.org/abs/1502.00635}{{\ttfamily 1502.00635}}].

\bibitem{Lee:2016vti}
H.~Lee, D.~Baumann and G.~L. Pimentel, \emph{{Non-Gaussianity as a Particle Detector}}, \href{https://doi.org/10.1007/JHEP12(2016)040}{\emph{JHEP} {\bfseries 12} (2016) 040} [\href{https://arxiv.org/abs/1607.03735}{{\ttfamily 1607.03735}}].

\bibitem{Chen:2009zp}
X.~Chen and Y.~Wang, \emph{{Quasi-Single Field Inflation and Non-Gaussianities}}, \href{https://doi.org/10.1088/1475-7516/2010/04/027}{\emph{JCAP} {\bfseries 04} (2010) 027} [\href{https://arxiv.org/abs/0911.3380}{{\ttfamily 0911.3380}}].

\bibitem{Philcox:2025cmbtrispectrum}
O.~H.~E. Philcox, \emph{{Searching for Inflationary Physics with the CMB Trispectrum: 1. Primordial Theory \& Optimal Estimators}},  \href{https://arxiv.org/abs/2502.04434}{{\ttfamily 2502.04434}}.

\bibitem{Philcox:2025cmbcode}
O.~H.~E. Philcox, \emph{{Searching for Inflationary Physics with the CMB Trispectrum: 2. Code \& Validation}},  \href{https://arxiv.org/abs/2502.05258}{{\ttfamily 2502.05258}}.

\bibitem{Philcox:2025cmbplanck}
O.~H.~E. Philcox, \emph{{Searching for Inflationary Physics with the CMB Trispectrum: 3. Constraints from Planck}},  \href{https://arxiv.org/abs/2502.06931}{{\ttfamily 2502.06931}}.

\bibitem{Fergusson:2010gn}
J.~R. Fergusson, D.~M. Regan and E.~P.~S. Shellard, \emph{{Optimal Trispectrum Estimators and WMAP Constraints}},  12, 2010.

\bibitem{Marzouk:2022utf}
K.~Marzouk, A.~Lewis and J.~Carron, \emph{{Constraints on \ensuremath{\tau} $_{NL}$ from Planck temperature and polarization}}, \href{https://doi.org/10.1088/1475-7516/2022/08/015}{\emph{JCAP} {\bfseries 08} (2022) 015} [\href{https://arxiv.org/abs/2205.14408}{{\ttfamily 2205.14408}}].

\bibitem{Sailer:2021yzm}
N.~Sailer, E.~Castorina, S.~Ferraro and M.~White, \emph{{Cosmology at high redshift \textemdash{} a probe of fundamental physics}}, \href{https://doi.org/10.1088/1475-7516/2021/12/049}{\emph{JCAP} {\bfseries 12} (2021) 049} [\href{https://arxiv.org/abs/2106.09713}{{\ttfamily 2106.09713}}].

\bibitem{Kalaja:2020mkq}
A.~Kalaja, P.~D. Meerburg, G.~L. Pimentel and W.~R. Coulton, \emph{{Fundamental limits on constraining primordial non-Gaussianity}}, \href{https://doi.org/10.1088/1475-7516/2021/04/050}{\emph{JCAP} {\bfseries 04} (2021) 050} [\href{https://arxiv.org/abs/2011.09461}{{\ttfamily 2011.09461}}].

\bibitem{Cabass:2022epm}
G.~Cabass, M.~M. Ivanov, O.~H.~E. Philcox, M.~Simonovic and M.~Zaldarriaga, \emph{{Constraining Single-Field Inflation with MegaMapper}},  \href{https://arxiv.org/abs/2211.14899}{{\ttfamily 2211.14899}}.

\bibitem{MoradinezhadDizgah:2020whw}
A.~Moradinezhad~Dizgah, M.~Biagetti, E.~Sefusatti, V.~Desjacques and J.~Nore\~na, \emph{{Primordial Non-Gaussianity from Biased Tracers: Likelihood Analysis of Real-Space Power Spectrum and Bispectrum}}, \href{https://doi.org/10.1088/1475-7516/2021/05/015}{\emph{JCAP} {\bfseries 05} (2021) 015} [\href{https://arxiv.org/abs/2010.14523}{{\ttfamily 2010.14523}}].

\bibitem{Baumann:2021ykm}
D.~Baumann and D.~Green, \emph{{The power of locality: primordial non-Gaussianity at the map level}}, \href{https://doi.org/10.1088/1475-7516/2022/08/061}{\emph{JCAP} {\bfseries 08} (2022) 061} [\href{https://arxiv.org/abs/2112.14645}{{\ttfamily 2112.14645}}].

\bibitem{Bertolini:2016bmt}
D.~Bertolini, K.~Schutz, M.~P. Solon and K.~M. Zurek, \emph{{The Trispectrum in the Effective Field Theory of Large Scale Structure}}, \href{https://doi.org/10.1088/1475-7516/2016/06/052}{\emph{JCAP} {\bfseries 06} (2016) 052} [\href{https://arxiv.org/abs/1604.01770}{{\ttfamily 1604.01770}}].

\bibitem{Steele:2021lnz}
T.~Steele and T.~Baldauf, \emph{{Precise Calibration of the One-Loop Trispectrum in the Effective Field Theory of Large Scale Structure}}, \href{https://doi.org/10.1103/PhysRevD.103.103518}{\emph{Phys. Rev. D} {\bfseries 103} (2021) 103518} [\href{https://arxiv.org/abs/2101.10289}{{\ttfamily 2101.10289}}].

\bibitem{Bakx:2026galaxybias}
T.~Bakx, M.~Garny, H.~Rubira and Z.~Vlah, \emph{{Galaxy bias renormalization: Two-loop Power Spectrum, One-loop Trispectrum and Bispectrum}},  \href{https://arxiv.org/abs/2606.31280}{{\ttfamily 2606.31280}}.

\bibitem{Bakx:2025cvu}
T.~Bakx, M.~Garny, H.~Rubira and Z.~Vlah, \emph{{Two-loop renormalization and running of galaxy bias}}, \href{https://doi.org/10.1088/1475-7516/2026/03/056}{\emph{JCAP} {\bfseries 03} (2026) 056} [\href{https://arxiv.org/abs/2507.13905}{{\ttfamily 2507.13905}}].

\bibitem{Leonard:2024xvy}
W.~O. Leonard, Z.~Slepian and J.~Hou, \emph{{A model for the redshift-space galaxy 4-point correlation function}}, \href{https://doi.org/10.1088/1475-7516/2025/01/090}{\emph{JCAP} {\bfseries 01} (2025) 090} [\href{https://arxiv.org/abs/2402.15510}{{\ttfamily 2402.15510}}].

\bibitem{Giri:2023modulation}
U.~Giri, M.~M{\"u}nchmeyer and K.~M. Smith, \emph{{Constraining $f_{NL}$ using the Large-Scale Modulation of Small-Scale Statistics}}, {\emph{Phys. Rev. D} {\bfseries 112} (2025) 123544} [\href{https://arxiv.org/abs/2305.03070}{{\ttfamily 2305.03070}}].

\bibitem{Goldstein_2024}
S.~Goldstein, O.~H.~E. Philcox, J.~C. Hill and L.~Hui, \emph{{Massive-ish Particles from Small-ish Scales: Non-Perturbative Techniques for Cosmological Collider Physics from Large-Scale Structure Surveys}},  \href{https://arxiv.org/abs/2407.08731}{{\ttfamily 2407.08731}}.

\bibitem{Goldstein:2026covariance}
S.~Goldstein, K.~M. Smith, U.~Giri and M.~M{\"u}nchmeyer, \emph{{If at First You Don't Succeed, Trispectrum: I. Estimating the Matter Power Spectrum Covariance with Higher-Order Statistics}},  \href{https://arxiv.org/abs/2604.18581}{{\ttfamily 2604.18581}}.

\bibitem{Gualdi:2020mattertrispectrum}
D.~Gualdi, S.~Novell-Masot, H.~Gil-Marin and L.~Verde, \emph{{Matter trispectrum: theoretical modelling and comparison to N-body simulations}}, \href{https://doi.org/10.1088/1475-7516/2021/01/015}{\emph{JCAP} {\bfseries 01} (2021) 015} [\href{https://arxiv.org/abs/2009.02290}{{\ttfamily 2009.02290}}].

\bibitem{Gualdi:2021joint}
D.~Gualdi, H.~Gil-Marin and L.~Verde, \emph{{Joint analysis of anisotropic power spectrum, bispectrum and trispectrum: application to N-body simulations}}, \href{https://doi.org/10.1088/1475-7516/2021/07/008}{\emph{JCAP} {\bfseries 07} (2021) 008} [\href{https://arxiv.org/abs/2104.03976}{{\ttfamily 2104.03976}}].

\bibitem{Philcox:2021hbm}
O.~H.~E. Philcox, J.~Hou and Z.~Slepian, \emph{{A First Detection of the Connected 4-Point Correlation Function of Galaxies Using the BOSS CMASS Sample}},  \href{https://arxiv.org/abs/2108.01670}{{\ttfamily 2108.01670}}.

\bibitem{Gualdi:2022kwz}
D.~Gualdi and L.~Verde, \emph{{Integrated trispectrum detection from BOSS DR12 NGC CMASS}}, \href{https://doi.org/10.1088/1475-7516/2022/09/050}{\emph{JCAP} {\bfseries 09} (2022) 050} [\href{https://arxiv.org/abs/2201.06932}{{\ttfamily 2201.06932}}].

\bibitem{Hou:2025desi4pcf}
J.~Hou, R.~N. Cahn et~al., \emph{{Study of the Connected Four-Point Correlation Function of Galaxies from DESI Data Release 1 Luminous Red Galaxy Sample}}, \href{https://doi.org/10.1103/1wyy-758f}{\emph{Phys. Rev. D} {\bfseries 112} (2025) 122005} [\href{https://arxiv.org/abs/2508.09070}{{\ttfamily 2508.09070}}].

\bibitem{Cahn:2021ltp}
R.~N. Cahn, Z.~Slepian and J.~Hou, \emph{{A Test for Cosmological Parity Violation Using the 3D Distribution of Galaxies}},  \href{https://arxiv.org/abs/2110.12004}{{\ttfamily 2110.12004}}.

\bibitem{Hou:2022parityodd}
J.~Hou, Z.~Slepian and R.~N. Cahn, \emph{{Measurement of Parity-Odd Modes in the Large-Scale 4-Point Correlation Function of SDSS BOSS DR12 CMASS and LOWZ Galaxies}}, \href{https://doi.org/10.1093/mnras/stad1062}{\emph{Mon. Not. Roy. Astron. Soc.} (2023) } [\href{https://arxiv.org/abs/2206.03625}{{\ttfamily 2206.03625}}].

\bibitem{Philcox:2022parityodd}
O.~H.~E. Philcox, \emph{{Probing Parity-Violation with the Four-Point Correlation Function of BOSS Galaxies}}, \href{https://doi.org/10.1103/PhysRevD.106.063501}{\emph{Phys. Rev. D} {\bfseries 106} (2022) 063501} [\href{https://arxiv.org/abs/2206.04227}{{\ttfamily 2206.04227}}].

\bibitem{Cabass:2022colliding}
G.~Cabass, M.~M. Ivanov and O.~H.~E. Philcox, \emph{{Colliding Ghosts: Constraining Inflation with the Parity-Odd Galaxy Four-Point Function}}, \href{https://doi.org/10.1103/PhysRevD.107.023523}{\emph{Phys. Rev. D} {\bfseries 107} (2023) 023523} [\href{https://arxiv.org/abs/2210.16320}{{\ttfamily 2210.16320}}].

\bibitem{Bao:2025onc}
Y.~Bao, L.-T. Wang, Z.-Z. Xianyu and Y.-M. Zhong, \emph{{Anatomy of parity-violating trispectra in galaxy surveys}}, \href{https://doi.org/10.1103/wzrq-8r8m}{\emph{Phys. Rev. D} {\bfseries 112} (2025) 103536} [\href{https://arxiv.org/abs/2504.02931}{{\ttfamily 2504.02931}}].

\bibitem{Cabass:2022rhr}
G.~Cabass, S.~Jazayeri, E.~Pajer and D.~Stefanyszyn, \emph{{Parity violation in the scalar trispectrum: no-go theorems and yes-go examples}},  \href{https://arxiv.org/abs/2210.02907}{{\ttfamily 2210.02907}}.

\bibitem{Coulton_2024}
W.~R. Coulton, O.~H.~E. Philcox and F.~Villaescusa-Navarro, \emph{{Signatures of a Parity-Violating Universe}},  \href{https://arxiv.org/abs/2306.11782}{{\ttfamily 2306.11782}}.

\bibitem{Philcox:2024mmz}
O.~H.~E. Philcox and J.~Ereza, \emph{{Could Sample Variance be Responsible for the Parity-Violating Signal Seen in the BOSS Galaxy Survey?}},  \href{https://arxiv.org/abs/2401.09523}{{\ttfamily 2401.09523}}.

\bibitem{Krolewski:2024parity}
A.~Krolewski, S.~May, K.~M. Smith and H.~Hopkins, \emph{{No evidence for parity violation in BOSS}}, \href{https://doi.org/10.1088/1475-7516/2024/08/044}{\emph{JCAP} {\bfseries 08} (2024) 044} [\href{https://arxiv.org/abs/2407.03397}{{\ttfamily 2407.03397}}].

\bibitem{Adari:2024vkf}
P.~Adari and A.~Slosar, \emph{{Searching for parity violation in SDSS DR16 Lyman-{\ensuremath{\alpha}} forest data}}, \href{https://doi.org/10.1103/PhysRevD.110.103534}{\emph{Phys. Rev. D} {\bfseries 110} (2024) 103534} [\href{https://arxiv.org/abs/2405.04660}{{\ttfamily 2405.04660}}].

\bibitem{Slepian:2025kbb}
Z.~Slepian et~al., \emph{{Measurement of Parity-Violating Modes of the Dark Energy Spectroscopic Instrument (DESI) Year 1 Luminous Red Galaxies' 4-Point Correlation Function}},  \href{https://arxiv.org/abs/2508.09133}{{\ttfamily 2508.09133}}.

\bibitem{Hou:2025cey}
J.~Hou et~al., \emph{{Parity-odd four-point correlation function from the DESI data release 1 luminous red galaxy sample}}, \href{https://doi.org/10.1103/2dmd-hyt1}{\emph{Phys. Rev. D} {\bfseries 113} (2026) 103502} [\href{https://arxiv.org/abs/2512.20132}{{\ttfamily 2512.20132}}].

\bibitem{Gao:2026zaz}
Z.~Gao, M.~S. Cagliari, A.~Moradinezhad~Dizgah and Z.~Vlah, \emph{{Testing parity with composite-field spectra of BOSS and DESI luminous red galaxies}},  \href{https://arxiv.org/abs/2604.06021}{{\ttfamily 2604.06021}}.

\bibitem{Aghamousa:2016zmz}
{\scshape DESI} collaboration, A.~Aghamousa et~al., \emph{{The DESI Experiment Part I: Science,Targeting, and Survey Design}},  \href{https://arxiv.org/abs/1611.00036}{{\ttfamily 1611.00036}}.

\bibitem{Euclid:2024yrr}
{\scshape Euclid} collaboration, Y.~Mellier et~al., \emph{{Euclid - I. Overview of the Euclid mission}}, \href{https://doi.org/10.1051/0004-6361/202450810}{\emph{Astron. Astrophys.} {\bfseries 697} (2025) A1} [\href{https://arxiv.org/abs/2405.13491}{{\ttfamily 2405.13491}}].

\bibitem{Nguyen:2024yth}
N.-M. Nguyen, F.~Schmidt, B.~Tucci, M.~Reinecke and A.~Kosti{\'c}, \emph{{How Much Information Can Be Extracted from Galaxy Clustering at the Field Level?}}, \href{https://doi.org/10.1103/PhysRevLett.133.221006}{\emph{Phys. Rev. Lett.} {\bfseries 133} (2024) 221006} [\href{https://arxiv.org/abs/2403.03220}{{\ttfamily 2403.03220}}].

\bibitem{Spezzati_2025}
F.~Spezzati, M.~Marinucci and M.~Simonovi{\'c}, \emph{{Equivalence of the field-level inference and conventional analyses on large scales}}, \href{https://doi.org/10.1103/yhg1-c63b}{\emph{Phys. Rev. D} {\bfseries 113} (2026) 063546} [\href{https://arxiv.org/abs/2507.05378}{{\ttfamily 2507.05378}}].

\bibitem{Akitsu:2025boy}
K.~Akitsu, M.~Simonovi{\'c}, S.-F. Chen, G.~Cabass and M.~Zaldarriaga, \emph{{Cosmology inference with perturbative forward modeling at the field level: a comparison with joint power spectrum and bispectrum analyses}},  \href{https://arxiv.org/abs/2509.09673}{{\ttfamily 2509.09673}}.

\bibitem{Stadler:2026ufu}
J.~Stadler, \emph{{Cosmological information from field-level analysis in redshift space}},  \href{https://arxiv.org/abs/2609.16719}{{\ttfamily 2609.16719}}.

\bibitem{Desjacques:2016bnm}
V.~Desjacques, D.~Jeong and F.~Schmidt, \emph{{Large-Scale Galaxy Bias}}, \href{https://doi.org/10.1016/j.physrep.2017.12.002}{\emph{Phys. Rept.} {\bfseries 733} (2018) 1} [\href{https://arxiv.org/abs/1611.09787}{{\ttfamily 1611.09787}}].

\bibitem{Assassi:2014fva}
V.~Assassi, D.~Baumann, D.~Green and M.~Zaldarriaga, \emph{{Renormalized Halo Bias}}, \href{https://doi.org/10.1088/1475-7516/2014/08/056}{\emph{JCAP} {\bfseries 1408} (2014) 056} [\href{https://arxiv.org/abs/1402.5916}{{\ttfamily 1402.5916}}].

\bibitem{Angulo:2015eqa}
R.~Angulo, M.~Fasiello, L.~Senatore and Z.~Vlah, \emph{{On the Statistics of Biased Tracers in the Effective Field Theory of Large Scale Structures}}, \href{https://doi.org/10.1088/1475-7516/2015/09/029, 10.1088/1475-7516/2015/9/029}{\emph{JCAP} {\bfseries 1509} (2015) 029} [\href{https://arxiv.org/abs/1503.08826}{{\ttfamily 1503.08826}}].

\bibitem{Senatore:2014eva}
L.~Senatore, \emph{{Bias in the Effective Field Theory of Large Scale Structures}}, \href{https://doi.org/10.1088/1475-7516/2015/11/007}{\emph{JCAP} {\bfseries 1511} (2015) 007} [\href{https://arxiv.org/abs/1406.7843}{{\ttfamily 1406.7843}}].

\bibitem{Mirbabayi:2014zca}
M.~Mirbabayi, F.~Schmidt and M.~Zaldarriaga, \emph{{Biased Tracers and Time Evolution}}, \href{https://doi.org/10.1088/1475-7516/2015/07/030}{\emph{JCAP} {\bfseries 1507} (2015) 030} [\href{https://arxiv.org/abs/1412.5169}{{\ttfamily 1412.5169}}].

\bibitem{Chudaykin:2020aoj}
A.~Chudaykin, M.~M. Ivanov, O.~H.~E. Philcox and M.~Simonovi\'c, \emph{{Nonlinear perturbation theory extension of the Boltzmann code CLASS}}, \href{https://doi.org/10.1103/PhysRevD.102.063533}{\emph{Phys. Rev. D} {\bfseries 102} (2020) 063533} [\href{https://arxiv.org/abs/2004.10607}{{\ttfamily 2004.10607}}].

\bibitem{Kaiser:1987qv}
N.~Kaiser, \emph{{Clustering in real space and in redshift space}}, {\emph{Mon. Not. Roy. Astron. Soc.} {\bfseries 227} (1987) 1}.

\bibitem{Gualdi:2018pyw}
D.~Gualdi, H.~Gil-Mar\'\i{}n, R.~L. Schuhmann, M.~Manera, B.~Joachimi and O.~Lahav, \emph{{Enhancing BOSS bispectrum cosmological constraints with maximal compression}}, \href{https://doi.org/10.1093/mnras/stz051}{\emph{Mon. Not. Roy. Astron. Soc.} {\bfseries 484} (2019) 3713} [\href{https://arxiv.org/abs/1806.02853}{{\ttfamily 1806.02853}}].

\bibitem{Sugiyama:2019ike}
N.~S. Sugiyama, S.~Saito, F.~Beutler and H.-J. Seo, \emph{{Perturbation theory approach to predict the covariance matrices of the galaxy power spectrum and bispectrum in redshift space}}, \href{https://doi.org/10.1093/mnras/staa1940}{\emph{Mon. Not. Roy. Astron. Soc.} {\bfseries 497} (2020) 1684} [\href{https://arxiv.org/abs/1908.06234}{{\ttfamily 1908.06234}}].

\bibitem{Ivanov:2019pdj}
M.~M. Ivanov, M.~Simonovi\'c and M.~Zaldarriaga, \emph{{Cosmological Parameters from the BOSS Galaxy Power Spectrum}}, \href{https://doi.org/10.1088/1475-7516/2020/05/042}{\emph{JCAP} {\bfseries 05} (2020) 042} [\href{https://arxiv.org/abs/1909.05277}{{\ttfamily 1909.05277}}].

\bibitem{Nishimichi:2020tvu}
T.~Nishimichi, G.~D'Amico, M.~M. Ivanov, L.~Senatore, M.~Simonovi\'c, M.~Takada et~al., \emph{{Blinded challenge for precision cosmology with large-scale structure: results from effective field theory for the redshift-space galaxy power spectrum}}, \href{https://doi.org/10.1103/PhysRevD.102.123541}{\emph{Phys. Rev. D} {\bfseries 102} (2020) 123541} [\href{https://arxiv.org/abs/2003.08277}{{\ttfamily 2003.08277}}].

\bibitem{Angulo:2014tfa}
R.~E. Angulo, S.~Foreman, M.~Schmittfull and L.~Senatore, \emph{{The One-Loop Matter Bispectrum in the Effective Field Theory of Large Scale Structures}}, \href{https://doi.org/10.1088/1475-7516/2015/10/039}{\emph{JCAP} {\bfseries 1510} (2015) 039} [\href{https://arxiv.org/abs/1406.4143}{{\ttfamily 1406.4143}}].

\bibitem{Eggemeier:2021cam}
A.~Eggemeier, R.~Scoccimarro, R.~E. Smith, M.~Crocce, A.~Pezzotta and A.~G. S\'anchez, \emph{{Testing one-loop galaxy bias: joint analysis of power spectrum and bispectrum}},  \href{https://arxiv.org/abs/2102.06902}{{\ttfamily 2102.06902}}.

\bibitem{Philcox:2022b1loop}
O.~H.~E. Philcox, M.~M. Ivanov, G.~Cabass, M.~Simonovi{\'c}, M.~Zaldarriaga and T.~Nishimichi, \emph{{Cosmology with the redshift-space galaxy bispectrum monopole at one-loop order}}, \href{https://doi.org/10.1103/PhysRevD.106.043530}{\emph{Phys. Rev. D} {\bfseries 106} (2022) 043530} [\href{https://arxiv.org/abs/2206.02800}{{\ttfamily 2206.02800}}].

\bibitem{DAmico:2022ukl}
G.~D'Amico, Y.~Donath, M.~Lewandowski, L.~Senatore and P.~Zhang, \emph{{The one-loop bispectrum of galaxies in redshift space from the Effective Field Theory of Large-Scale Structure}}, \href{https://doi.org/10.1088/1475-7516/2024/07/041}{\emph{JCAP} {\bfseries 07} (2024) 041} [\href{https://arxiv.org/abs/2211.17130}{{\ttfamily 2211.17130}}].

\bibitem{Bakx:2025pop}
T.~Bakx, M.~M. Ivanov, O.~H.~E. Philcox and Z.~Vlah, \emph{{One-Loop Galaxy Bispectrum: Consistent Theory, Efficient Analysis with COBRA, and Implications for Cosmological Parameters}},  \href{https://arxiv.org/abs/2507.22110}{{\ttfamily 2507.22110}}.

\bibitem{Rubira:2024tea}
H.~Rubira and F.~Schmidt, \emph{{The renormalization group for large-scale structure: origin of galaxy stochasticity}}, \href{https://doi.org/10.1088/1475-7516/2024/10/092}{\emph{JCAP} {\bfseries 10} (2024) } [\href{https://arxiv.org/abs/2404.16929}{{\ttfamily 2404.16929}}].

\bibitem{Rubira:2025rqo}
H.~Rubira and F.~Schmidt, \emph{{Non-Gaussian galaxy stochasticity and the noise-field formulation}}, \href{https://doi.org/10.1088/1475-7516/2026/08/023}{\emph{JCAP} {\bfseries 08} (2026) 023} [\href{https://arxiv.org/abs/2511.05484}{{\ttfamily 2511.05484}}].

\bibitem{Ivanov:2021kcd}
M.~M. Ivanov, O.~H.~E. Philcox, T.~Nishimichi, M.~Simonovi\'c, M.~Takada and M.~Zaldarriaga, \emph{{Precision analysis of the redshift-space galaxy bispectrum}}, \href{https://doi.org/10.1103/PhysRevD.105.063512}{\emph{Phys. Rev. D} {\bfseries 105} (2022) 063512} [\href{https://arxiv.org/abs/2110.10161}{{\ttfamily 2110.10161}}].

\bibitem{Jackson:2008yv}
J.~C. Jackson, \emph{{Fingers of God: A critique of Rees' theory of primoridal gravitational radiation}}, \href{https://doi.org/10.1093/mnras/156.1.1P}{\emph{Mon. Not. Roy. Astron. Soc.} {\bfseries 156} (1972) 1P} [\href{https://arxiv.org/abs/0810.3908}{{\ttfamily 0810.3908}}].

\bibitem{Perko:2016puo}
A.~Perko, L.~Senatore, E.~Jennings and R.~H. Wechsler, \emph{{Biased Tracers in Redshift Space in the EFT of Large-Scale Structure}},  \href{https://arxiv.org/abs/1610.09321}{{\ttfamily 1610.09321}}.

\bibitem{Philcox:2021kcw}
O.~H.~E. Philcox and M.~M. Ivanov, \emph{{BOSS DR12 full-shape cosmology: \ensuremath{\Lambda}CDM constraints from the large-scale galaxy power spectrum and bispectrum monopole}}, \href{https://doi.org/10.1103/PhysRevD.105.043517}{\emph{Phys. Rev. D} {\bfseries 105} (2022) 043517} [\href{https://arxiv.org/abs/2112.04515}{{\ttfamily 2112.04515}}].

\bibitem{Verde:1998zr}
L.~Verde, A.~F. Heavens, S.~Matarrese and L.~Moscardini, \emph{{Large scale bias in the universe. 2. Redshift space bispectrum}}, \href{https://doi.org/10.1046/j.1365-8711.1998.01937.x}{\emph{Mon. Not. Roy. Astron. Soc.} {\bfseries 300} (1998) 747} [\href{https://arxiv.org/abs/astro-ph/9806028}{{\ttfamily astro-ph/9806028}}].

\bibitem{Scoccimarro:1999ed}
R.~Scoccimarro, H.~M.~P. Couchman and J.~A. Frieman, \emph{{The Bispectrum as a Signature of Gravitational Instability in Redshift-Space}}, \href{https://doi.org/10.1086/307220}{\emph{Astrophys. J.} {\bfseries 517} (1999) 531} [\href{https://arxiv.org/abs/astro-ph/9808305}{{\ttfamily astro-ph/9808305}}].

\bibitem{Gil-Marin:2014sta}
H.~Gil-Marín, J.~Noreña, L.~Verde, W.~J. Percival, C.~Wagner, M.~Manera et~al., \emph{{The power spectrum and bispectrum of SDSS DR11 BOSS galaxies – I. Bias and gravity}}, \href{https://doi.org/10.1093/mnras/stv961}{\emph{Mon. Not. Roy. Astron. Soc.} {\bfseries 451} (2015) 539} [\href{https://arxiv.org/abs/1407.5668}{{\ttfamily 1407.5668}}].

\bibitem{Hashimoto:2017klo}
I.~Hashimoto, Y.~Rasera and A.~Taruya, \emph{{Precision cosmology with redshift-space bispectrum: a perturbation theory based model at one-loop order}}, \href{https://doi.org/10.1103/PhysRevD.96.043526}{\emph{Phys. Rev. D} {\bfseries 96} (2017) 043526} [\href{https://arxiv.org/abs/1705.02574}{{\ttfamily 1705.02574}}].

\bibitem{Yankelevich:2018uaz}
V.~Yankelevich and C.~Porciani, \emph{{Cosmological information in the redshift-space bispectrum}}, \href{https://doi.org/10.1093/mnras/sty3143}{\emph{Mon. Not. Roy. Astron. Soc.} {\bfseries 483} (2019) 2078} [\href{https://arxiv.org/abs/1807.07076}{{\ttfamily 1807.07076}}].

\bibitem{Baldauf:2015xfa}
T.~Baldauf, M.~Mirbabayi, M.~Simonović and M.~Zaldarriaga, \emph{{Equivalence Principle and the Baryon Acoustic Peak}}, \href{https://doi.org/10.1103/PhysRevD.92.043514}{\emph{Phys. Rev.} {\bfseries D92} (2015) 043514} [\href{https://arxiv.org/abs/1504.04366}{{\ttfamily 1504.04366}}].

\bibitem{Blas:2015qsi}
D.~Blas, M.~Garny, M.~M. Ivanov and S.~Sibiryakov, \emph{{Time-Sliced Perturbation Theory for Large Scale Structure I: General Formalism}}, \href{https://doi.org/10.1088/1475-7516/2016/07/052}{\emph{JCAP} {\bfseries 1607} (2016) 052} [\href{https://arxiv.org/abs/1512.05807}{{\ttfamily 1512.05807}}].

\bibitem{Blas:2016sfa}
D.~Blas, M.~Garny, M.~M. Ivanov and S.~Sibiryakov, \emph{{Time-Sliced Perturbation Theory II: Baryon Acoustic Oscillations and Infrared Resummation}}, \href{https://doi.org/10.1088/1475-7516/2016/07/028}{\emph{JCAP} {\bfseries 1607} (2016) 028} [\href{https://arxiv.org/abs/1605.02149}{{\ttfamily 1605.02149}}].

\bibitem{Ivanov:2018gjr}
M.~M. Ivanov and S.~Sibiryakov, \emph{{Infrared Resummation for Biased Tracers in Redshift Space}}, \href{https://doi.org/10.1088/1475-7516/2018/07/053}{\emph{JCAP} {\bfseries 1807} (2018) 053} [\href{https://arxiv.org/abs/1804.05080}{{\ttfamily 1804.05080}}].

\bibitem{Jung:2023kjh}
G.~Jung et~al., \emph{{Quijote-PNG: The Information Content of the Halo Mass Function}}, \href{https://doi.org/10.3847/1538-4357/acfe70}{\emph{Astrophys. J.} {\bfseries 957} (2023) 50}.

\bibitem{2017arXiv170903452S}
E.~{Sellentin}, A.~H. {Jaffe} and A.~F. {Heavens}, \emph{{On the use of the Edgeworth expansion in cosmology I: how to foresee and evade its pitfalls}}, {\emph{arXiv e-prints} (2017) arXiv:1709.03452} [\href{https://arxiv.org/abs/1709.03452}{{\ttfamily 1709.03452}}].

\bibitem{Namikawa:2012pe}
T.~Namikawa, D.~Hanson and R.~Takahashi, \emph{{Bias-Hardened CMB Lensing}}, \href{https://doi.org/10.1093/mnras/stt195}{\emph{Mon. Not. Roy. Astron. Soc.} {\bfseries 431} (2013) 609} [\href{https://arxiv.org/abs/1209.0091}{{\ttfamily 1209.0091}}].

\bibitem{Shen:2024vft}
D.~Shen, E.~Schaan and S.~Ferraro, \emph{{CMB lensing power spectrum without noise bias}}, \href{https://doi.org/10.1103/PhysRevD.110.043523}{\emph{Phys. Rev. D} {\bfseries 110} (2024) 043523} [\href{https://arxiv.org/abs/2402.04309}{{\ttfamily 2402.04309}}].

\bibitem{Schmidt:2020ovm}
F.~Schmidt, \emph{{An $n$-th order Lagrangian Forward Model for Large-Scale Structure}},  \href{https://arxiv.org/abs/2012.09837}{{\ttfamily 2012.09837}}.

\bibitem{Tucci:2023bag}
B.~Tucci and F.~Schmidt, \emph{{EFTofLSS meets simulation-based inference: {\ensuremath{\sigma}} $_{8}$ from biased tracers}}, \href{https://doi.org/10.1088/1475-7516/2024/05/063}{\emph{JCAP} {\bfseries 05} (2024) 063} [\href{https://arxiv.org/abs/2310.03741}{{\ttfamily 2310.03741}}].

\bibitem{Fergusson_2010}
J.~R. Fergusson, D.~M. Regan and E.~P.~S. Shellard, \emph{{Rapid Separable Analysis of Higher Order Correlators in Large Scale Structure}}, \href{https://doi.org/10.1103/PhysRevD.86.063511}{\emph{Phys. Rev. D} {\bfseries 86} (2012) 063511} [\href{https://arxiv.org/abs/1008.1730}{{\ttfamily 1008.1730}}].

\bibitem{Liberti:2014distance}
L.~Liberti, C.~Lavor, N.~Maculan and A.~Mucherino, \emph{{Euclidean Distance Geometry and Applications}}, \href{https://doi.org/10.1137/120875909}{\emph{SIAM Review} {\bfseries 56} (2014) 3} [\href{https://arxiv.org/abs/1205.0349}{{\ttfamily 1205.0349}}].

\bibitem{Paul_2024}
P.~Paul, C.~Clarkson and R.~Maartens, \emph{{The odd-parity part of the observed galaxy trispectrum}}, \href{https://doi.org/10.1088/1475-7516/2025/10/107}{\emph{JCAP} {\bfseries 10} (2025) 107} [\href{https://arxiv.org/abs/2411.10897}{{\ttfamily 2411.10897}}].

\bibitem{Alcock:1979mp}
C.~Alcock and B.~Paczynski, \emph{{An evolution free test for non-zero cosmological constant}}, \href{https://doi.org/10.1038/281358a0}{\emph{Nature} {\bfseries 281} (1979) 358}.

\bibitem{Ballinger:1996cd}
W.~Ballinger, J.~Peacock and A.~Heavens, \emph{{Measuring the cosmological constant with redshift surveys}}, \href{https://doi.org/10.1093/mnras/282.3.877}{\emph{Mon.\ Not.\ Roy.\ Astron.\ Soc.} {\bfseries 282} (1996) 877} [\href{https://arxiv.org/abs/astro-ph/9605017}{{\ttfamily astro-ph/9605017}}].

\bibitem{Song:2015gca}
Y.-S. Song, A.~Taruya and A.~Oka, \emph{{Cosmology with anisotropic galaxy clustering from the combination of power spectrum and bispectrum}}, \href{https://doi.org/10.1088/1475-7516/2015/08/007}{\emph{JCAP} {\bfseries 1508} (2015) 007} [\href{https://arxiv.org/abs/1502.03099}{{\ttfamily 1502.03099}}].

\bibitem{Philcox:2020zyp}
O.~H.~E. Philcox, M.~M. Ivanov, M.~Zaldarriaga, M.~Simonovic and M.~Schmittfull, \emph{{Fewer Mocks and Less Noise: Reducing the Dimensionality of Cosmological Observables with Subspace Projections}}, \href{https://doi.org/10.1103/PhysRevD.103.043508}{\emph{Phys. Rev. D} {\bfseries 103} (2021) 043508} [\href{https://arxiv.org/abs/2009.03311}{{\ttfamily 2009.03311}}].

\bibitem{Maus:2024sbb}
M.~Maus et~al., \emph{{A comparison of effective field theory models of redshift space galaxy power spectra for DESI 2024 and future surveys}}, \href{https://doi.org/10.1088/1475-7516/2025/01/134}{\emph{JCAP} {\bfseries 01} (2025) 134} [\href{https://arxiv.org/abs/2404.07272}{{\ttfamily 2404.07272}}].

\bibitem{Chudaykin:2025aux}
A.~Chudaykin, M.~M. Ivanov and O.~H.~E. Philcox, \emph{{Reanalyzing DESI DR1. I. {\ensuremath{\Lambda}}CDM constraints from the power spectrum and bispectrum}}, \href{https://doi.org/10.1103/qsnt-dppc}{\emph{Phys. Rev. D} {\bfseries 113} (2026) 063502} [\href{https://arxiv.org/abs/2507.13433}{{\ttfamily 2507.13433}}].

\bibitem{Tsedrik:2025hmj}
M.~Tsedrik, P.~Carrilho and C.~Moretti, \emph{{The simple way to measure evolving dark energy without prior-volume effects}},  \href{https://arxiv.org/abs/2509.09562}{{\ttfamily 2509.09562}}.

\bibitem{Lazeyras:2015lgp}
T.~Lazeyras, C.~Wagner, T.~Baldauf and F.~Schmidt, \emph{{Precision measurement of the local bias of dark matter halos}}, \href{https://doi.org/10.1088/1475-7516/2016/02/018}{\emph{JCAP} {\bfseries 1602} (2016) 018} [\href{https://arxiv.org/abs/1511.01096}{{\ttfamily 1511.01096}}].

\bibitem{Eggemeier:2020umu}
A.~Eggemeier, R.~Scoccimarro, M.~Crocce, A.~Pezzotta and A.~G. S\'anchez, \emph{{Testing one-loop galaxy bias: Power spectrum}}, \href{https://doi.org/10.1103/PhysRevD.102.103530}{\emph{Phys. Rev. D} {\bfseries 102} (2020) 103530} [\href{https://arxiv.org/abs/2006.09729}{{\ttfamily 2006.09729}}].

\bibitem{Lazeyras:2017hxw}
T.~Lazeyras and F.~Schmidt, \emph{{Beyond LIMD bias: a measurement of the complete set of third-order halo bias parameters}}, \href{https://doi.org/10.1088/1475-7516/2018/09/008}{\emph{JCAP} {\bfseries 1809} (2018) 008} [\href{https://arxiv.org/abs/1712.07531}{{\ttfamily 1712.07531}}].

\bibitem{Garrison:2021lfa}
L.~H. Garrison, D.~J. Eisenstein, D.~Ferrer, N.~A. Maksimova and P.~A. Pinto, \emph{{The abacus cosmological N-body code}}, \href{https://doi.org/10.1093/mnras/stab2482}{\emph{Mon. Not. Roy. Astron. Soc.} {\bfseries 508} (2021) 575} [\href{https://arxiv.org/abs/2110.11392}{{\ttfamily 2110.11392}}].

\bibitem{Maksimova:2021distinct}
N.~A. Maksimova, L.~H. Garrison, D.~J. Eisenstein, B.~Hadzhiyska, S.~Bose and T.~P. Satterthwaite, \emph{{AbacusSummit: a massive set of high-accuracy, high-resolution N-body simulations}}, \href{https://doi.org/10.1093/mnras/stab2484}{\emph{Mon. Not. Roy. Astron. Soc.} {\bfseries 508} (2021) 4017} [\href{https://arxiv.org/abs/2110.11398}{{\ttfamily 2110.11398}}].

\bibitem{Villaescusa-Navarro:2019fdia}
F.~Villaescusa-Navarro et~al., \emph{{The Quijote Simulations}}, \href{https://doi.org/10.3847/1538-4365/ab9d82}{\emph{Astrophys. J. Suppl.} {\bfseries 250} (2020) 2} [\href{https://arxiv.org/abs/1909.05273}{{\ttfamily 1909.05273}}].

\bibitem{Hadzhiyska:2021zbd}
B.~Hadzhiyska, D.~Eisenstein, S.~Bose, L.~H. Garrison and N.~Maksimova, \emph{{compaso: A new halo finder for competitive assignment to spherical overdensities}}, \href{https://doi.org/10.1093/mnras/stab2980}{\emph{Mon. Not. Roy. Astron. Soc.} {\bfseries 509} (2021) 501} [\href{https://arxiv.org/abs/2110.11408}{{\ttfamily 2110.11408}}].

\bibitem{Barnett_2019}
A.~H. {Barnett}, J.~{Magland} and L.~{af Klinteberg}, \emph{{A Parallel Nonuniform Fast Fourier Transform Library Based on an ``Exponential of Semicircle'' Kernel}}, \href{https://doi.org/10.1137/18M120885X}{\emph{SIAM Journal on Scientific Computing} {\bfseries 41} (2019) C479} [\href{https://arxiv.org/abs/1808.06736}{{\ttfamily 1808.06736}}].

\bibitem{Philcox:2024rqr}
O.~H.~E. Philcox and T.~Fl{\"o}ss, \emph{{PolyBin3D: A suite of optimal and efficient power spectrum, bispectrum, and trispectrum estimators for large-scale structure}}, \href{https://doi.org/10.1103/PhysRevD.110.043528}{\emph{Phys. Rev. D} {\bfseries 110} (2024) 043528} [\href{https://arxiv.org/abs/2404.07249}{{\ttfamily 2404.07249}}].

\bibitem{Alam:2016hwk}
{\scshape BOSS} collaboration, S.~Alam et~al., \emph{{The clustering of galaxies in the completed SDSS-III Baryon Oscillation Spectroscopic Survey: cosmological analysis of the DR12 galaxy sample}}, \href{https://doi.org/10.1093/mnras/stx721}{\emph{Mon. Not. Roy. Astron. Soc.} {\bfseries 470} (2017) 2617} [\href{https://arxiv.org/abs/1607.03155}{{\ttfamily 1607.03155}}].

\bibitem{Davis:1985rj}
M.~Davis, G.~Efstathiou, C.~S. Frenk and S.~D.~M. White, \emph{{The Evolution of Large Scale Structure in a Universe Dominated by Cold Dark Matter}}, \href{https://doi.org/10.1086/163168}{\emph{Astrophys. J.} {\bfseries 292} (1985) 371}.

\bibitem{Audren:2012vy}
B.~Audren, J.~Lesgourgues, S.~Bird, M.~G. Haehnelt and M.~Viel, \emph{{Neutrino masses and cosmological parameters from a Euclid-like survey: Markov Chain Monte Carlo forecasts including theoretical errors}}, \href{https://doi.org/10.1088/1475-7516/2013/01/026}{\emph{JCAP} {\bfseries 1301} (2013) 026} [\href{https://arxiv.org/abs/1210.2194}{{\ttfamily 1210.2194}}].

\bibitem{Brinckmann:2018cvx}
T.~Brinckmann and J.~Lesgourgues, \emph{{MontePython 3: boosted MCMC sampler and other features}}, \href{https://doi.org/10.1016/j.dark.2018.100260}{\emph{Phys. Dark Univ.} {\bfseries 24} (2019) 100260} [\href{https://arxiv.org/abs/1804.07261}{{\ttfamily 1804.07261}}].

\bibitem{Bakx:2024zgu}
T.~Bakx, N.~E. Chisari and Z.~Vlah, \emph{{Optimal Factorization of Cosmological Large-Scale Structure Observables}}, \href{https://doi.org/10.1103/PhysRevLett.134.191002}{\emph{Phys. Rev. Lett.} {\bfseries 134} (2025) 191002}.

\bibitem{Abidi:2018eyd}
M.~M. Abidi and T.~Baldauf, \emph{{Cubic Halo Bias in Eulerian and Lagrangian Space}}, \href{https://doi.org/10.1088/1475-7516/2018/07/029}{\emph{JCAP} {\bfseries 1807} (2018) 029} [\href{https://arxiv.org/abs/1802.07622}{{\ttfamily 1802.07622}}].

\bibitem{Philcox:2020vbm}
O.~H.~E. Philcox, \emph{{Cosmology without window functions: Quadratic estimators for the galaxy power spectrum}}, \href{https://doi.org/10.1103/PhysRevD.103.103504}{\emph{Phys. Rev. D} {\bfseries 103} (2021) 103504} [\href{https://arxiv.org/abs/2012.09389}{{\ttfamily 2012.09389}}].

\bibitem{Philcox:2021ukg}
O.~H.~E. Philcox, \emph{{Cosmology Without Window Functions: Cubic Estimators for the Galaxy Bispectrum}},  \href{https://arxiv.org/abs/2107.06287}{{\ttfamily 2107.06287}}.

\bibitem{Hotinli:2025tul}
S.~C. Hotinli, K.~M. Smith and S.~Ferraro, \emph{{Velocity Reconstruction from KSZ: Measuring $f_{NL}$ with ACT and DESILS}},  \href{https://arxiv.org/abs/2506.21657}{{\ttfamily 2506.21657}}.

\bibitem{Beyond-2pt:2024mqz}
{\scshape Beyond-2pt} collaboration, E.~Krause et~al., \emph{{A Parameter-Masked Mock Data Challenge for Beyond-Two-Point Galaxy Clustering Statistics}}, \href{https://doi.org/10.3847/1538-4357/ad781d}{\emph{Astrophys. J.} {\bfseries 990} (2025) 99} [\href{https://arxiv.org/abs/2405.02252}{{\ttfamily 2405.02252}}].

\end{thebibliography}\endgroup
\bibliographystyle{JHEP}

\end{document}